\documentclass[useamsfonts]{pasj01}
\usepackage{natbib}
\usepackage{aas_macros}
\usepackage{url}
\usepackage{bm}
\usepackage{xcolor}
\usepackage{multirow}
\usepackage{tabularx}

\newcommand{\simgt}{\lower.5ex\hbox{$\; \buildrel > \over \sim \;$}}
\newcommand{\simlt}{\lower.5ex\hbox{$\; \buildrel < \over \sim \;$}}

\begin{document}
\SetRunningHead{T. Hamana et al.}{Weak lensing clusters from HSC survey first-year data}
\Received{2020/4/1}
\Accepted{2020/6/29}
\Published{2020}

\title{Weak lensing clusters from HSC survey first-year data:
  Mitigating the dilution effect of foreground and cluster member galaxies}
%
%
%
\author{Takashi \textsc{Hamana}\altaffilmark{1,2}}
\author{Masato \textsc{Shirasaki}\altaffilmark{1}}
\author{Yen-Ting \textsc{Lin}\altaffilmark{3}}
\altaffiltext{1}{National Astronomical Observatory of Japan, Mitaka, Tokyo 181-8588, Japan}
\altaffiltext{2}{The Graduate University for Advanced Studies, SOKENDAI,
  Mitaka, Tokyo 181-8588, Japan}
\altaffiltext{3}{Academia Sinica Institute of Astronomy and Astrophysics, P.O. Box 23–141, Taipei 10617, Taiwan}
%

\KeyWords{galaxies: clusters: general --- cosmology: observations --- dark matter --- large-scale structure of universe }

\maketitle

\begin{abstract}
We present a weak lensing cluster search using Hyper Suprime-Cam Subaru
Strategic Program (HSC survey) first-year data.
We pay special attention to the dilution effect of
cluster member and foreground galaxies on weak lensing
signals from clusters of galaxies;
we adopt the globally normalized weak lensing estimator which is least
affected by cluster member galaxies, and we select source galaxies by
using photometric redshift information to mitigate the effect of
foreground galaxies. 
We produce six samples of source galaxies with different low-$z$ galaxy
cuts, construct weak lensing mass maps for each of the source sample, and
search for high peaks in the mass maps that cover an effective survey
area of $\sim$120 deg$^2$.
We combine six catalogs of high peaks into a sample of cluster
candidates which contains 124 high peaks with
signal-to-noise ratios greater than five.
We cross-match the peak sample with the public optical cluster catalog
constructed from the same HSC survey data to identify cluster
counterparts of the peaks.
We find that 107 out of 124 peaks have matched clusters within 5 arcmin
from peak positions. 
Among them, we define a sub-sample of 64 secure clusters that we use to
examine dilution effects on our weak lensing
cluster search.
We find that source samples
with the low-$z$ galaxy cuts mitigate the dilution effect on
weak lensing signals of high-$z$ clusters ($z \gtsim 0.3$),
and thus combining multiple peak catalogs from different source samples
improves the efficiency of weak lensing cluster searches.
\end{abstract}

%
%
\section{Introduction}
\label{sec:intro}

Clusters of galaxies have been playing important roles in the modern
cosmology:
Their abundance and evolution have been used to place constraints on
cosmological parameters \citep{2011ARA&A..49..409A}, and
their baryonic components (galaxies and hot intra-cluster gas)
have been used to study physical processes of hierarchical structure formation in
the universe \citep{2012ARA&A..50..353K}.
In those studies, a large sample of clusters of galaxies is
the fundamental data, which has been constructed by identifying their tracers such
as optical galaxy concentrations, X-ray emissions, Sunyaev-Zel'dovich
effect (SZE), and dark matter concentrations via the weak lensing technique
\citep{2019SSRv..215...25P}.
Since all cluster mass-observable relations have scatters, sample
completeness in terms of the cluster mass, which is the principal
quantity to link an observation to a theory, varies from method to method.
Weak lensing cluster finding is unique in that it uses the matter
concentration as the tracer regardless of physical state of baryonic
components, enabling one to locate under-luminous (in optical/X-ray/SZE)
clusters.

Observationally, there are two conflicting difficulties in constructing a
sizable cluster sample with weak lensing in a practical time scale; a
wide survey area to locate 
rare objects, and a deep imaging to achieve a sufficient number density of source galaxies.
Thanks to the development of wide-field optical cameras with
dedicated wide field surveys, weak lensing cluster finding has made
rapid progress in the last two decades,  \citep[see Table 1 of][and
  references therein]{2018PASJ...70S..27M}. 
Recently, \citet{2018PASJ...70S..27M} have conducted a weak lensing
cluster search in a $\sim$160 deg$^2$ area of Hyper Suprime-Cam
Subaru Strategic Program \citep[hereafter, HSC
  survey,][]{2018PASJ...70S...4A} first-year data
\citep{2018PASJ...70S...8A,2018PASJ...70S..25M}, and
have reported a detection of 65 peaks\footnote{In this paper, we use the
  term "peak" to mean a local maximum on a weak lensing mass map with
  its height exceeding a given threshold (see Section
  \ref{sec:peak-finding} for details).
  We adopt the signal-to-noise ratio ($SN$) of mass maps (see Section
  \ref{sec:mass-reconstruction} for its definition) to define the
  threshold, because the noise in the mass map originating from
  intrinsic galaxy shapes is well characterized by the random Gaussian field
  \citep{2000MNRAS.313..524V}, and thus it gives a rough estimate of a
  significance of a peak.}
with signal-to-noise ratio ($SN$)
greater than 4.7 in weak lensing mass maps. 
They have cross-matched the peaks with optical cluster catalogs and found
that 63 out of 65 peaks had optical counterparts, demonstrating that a
wide field survey with a sufficient depth (for their case $i= 24.5$ mag)
is indeed able to yield a sizable and high purity cluster sample.

In the near future, the size of weak lensing cluster sample will
become much larger as many more wide-area weak lensing-oriented surveys will
come:
The final survey area of HSC survey is 1400 deg$^2$ (more
than eight times of the first-year data), and Legacy Survey of Space and
Time \citep[LSST,][]{2019ApJ...873..111I} and Euclid survey
\citep{2012SPIE.8442E..0TL,2018cosp...42E2761R} will cover a large
portion of the sky with a sufficient depth.
It is thus worth improving methods of weak lensing cluster finding
by making best use of multi-band dataset that on-going/future
surveys take.
This is exactly the purpose of this paper.

In this paper, we focus on the dilution effect that we briefly explain
below:
Weak lensing effect by clusters distorts shapes of background galaxies in
a coherent manner.
Since the shape distortions by weak lensing (lensing shear), which are
generally smaller than intrinsic ellipticities of galaxies, can not be
extracted from individual galaxies, a lensing analysis
necessarily involves averaging of shear estimators among a sample of
galaxies to derive lensing shears and to suppress the noise from
intrinsic ellipticities (called the shape noise).
If a galaxy sample used for a weak lensing analysis contains not only
background lensed galaxies but also foreground and/or cluster member
galaxies which are not affected by cluster lensing and thus have no
lensing signal, the latter acts as
contaminants in weak lensing analyses and {\it dilutes} the lensing signals by
clusters \citep[see][for observational studies of the dilution effects
  in analyses of cluster lensing]{2005ApJ...619L.143B,2007ApJ...668..643L,2007MNRAS.379..317H,2007ApJ...663..717M,2008ApJ...684..177U,2010PASJ...62..811O}.
In most of recent lensing analyses of individual clusters
with known redshifts, source galaxies are selected using
multi-band galaxy photometry data so that the contamination of foreground and
cluster member galaxies is minimized \citep[see][and references therein]{2018PASJ...70...30M}.
However, in weak lensing cluster findings, redshifts of clusters are unknown
in advance, and thus a galaxy sample was commonly selected by a
simple magnitude-cut on a single band photometry \citep[for
  example,][]{2002ApJ...580L..97M,2015PASJ...67...34H}.
Such a galaxy sample inevitably contains foreground/cluster member
galaxies and suffers from the dilution effect.

Weak lensing cluster finding is based on peak heights in mass maps.
The detection threshold is set by the peak height $SN$ considering the
trade-off between completeness and purity
(lowering the threshold $SN$ leads to a larger number of cluster
detections at the cost of a higher false detection rate). 
However, the peak heights of cluster lensing are indeed affected by the
dilution effect. 
Its direct impact is the decline in numbers of cluster detections.
Another impact is on theoretical models of weak lensing
mass map peaks;
incorporating its effect into theoretical models
requires a realistic modeling of the dilution effect which is most
likely dependent on cluster mass, redshift, and galaxy selection
criteria (for example, the detection band, magnitude-cut, and size-cut). 
Therefore it is fundamentally important to understand actual dilution
effects on weak lensing mass maps on a case-by-case basis.

The purpose of this paper is two-fold:
The first is to develop a weak lensing cluster finding method that
mitigates the dilution effects by incorporating photometric redshift
information of galaxies. 
We apply it to the HSC survey first-year data in which both the weak lensing
shape catalog and photometric redshift data are publicly available
\citep{2018PASJ...70S..25M,2018PASJ...70S...9T}. 
We present a sample of weak lensing peaks located by our finding
method.
We identify their counterpart clusters by cross-matching with the optical
cluster catalog \citep{2018PASJ...70S..20O}.
Using the derived weak lensing cluster sample, we examine the
dilution effects on actual weak lensing mass maps in an empirical
manner, which is our second purpose.

The structure of this paper is as follows. 
In Section~\ref{sec:data}, we briefly summarize the HSC survey
first-year shear catalog and the photometric redshift data
used in this study. 
In Section~\ref{sec:kappa-peaks}, we describe the methods to
generate a sample of weak lensing peaks, including the selection of
source galaxies, the method to reconstruct weak lensing mass maps, and
the peak finding algorithm.
In Section~\ref{sec:cross-matching}, we cross-match
the weak lensing peaks with a sample of optical
clusters to identify their cluster counterparts.
Then we examine fundamental properties of weak lensing clusters detected
by our method.
In Section~\ref{sec:dilution_effects}, we examine the dilution effects
of foreground and cluster member galaxies on our weak lensing peaks
in an empirical manner using actual source galaxy samples and empirical
models.
Finally, we summarize and discuss our results in Section~\ref{sec:summary}.
In Appendix~\ref{sec:cross_matching}, we present results of
cross-matching of our sample of weak lensing peaks with selected
catalogs of known clusters.
In Appendix \ref{sec:neighboring_peaks}, we describe systems of
neighboring peaks in our peak sample.
In Appendix \ref{sec:cluster_mass}, we present results of the cluster
mass estimate of the weak lensing peak sample based on a model fitting
to weak lensing shear profiles. 
In Appendix \ref{sec:local_estimator}, we compare the globally
normalized $SN$ estimator, which is adopted in this study, with the
locally normalized $SN$ estimator adopted in some previous studies
\citep[for example,][]{2015PASJ...67...34H}.

Throughout this paper, unless otherwise stated,
we adopt the cosmological model with  
the cold dark matter (CDM) density $\Omega_{\rm cdm}=0.233$, the baryon
density $\Omega_{\rm b}=0.046$, the matter density
$\Omega_{\rm m}=\Omega_{\rm cdm} + \Omega_{\rm b} = 0.279$, 
the cosmological constant $\Omega_\Lambda=0.721$, the spectral index
$n_s=0.97$, the normalization of the matter fluctuation
$\sigma_8=0.82$, and the Hubble parameter $h=0.7$, which are the
best-fit cosmological parameters in the Wilkinson Microwave Anisotropy
Probe (WMAP) 9-year results \citep{2013ApJS..208...19H}.

%
%
\section{HSC survey data}
\label{sec:data}

In this section, we briefly describe those aspects of the HSC survey first
year products that are directly relevant to this study, see the following
references for full details: 
\citet{2018PASJ...70S...4A} for an overview of the HSC survey and 
survey design, \citet{2018PASJ...70S...8A} for the first public 
data release,
\citet{2018PASJ...70S...1M,2018PASJ...70S...2K,2018PASJ...70...66K,2018PASJ...70S...3F}
for the performance of 
the HSC instrument itself, \citet{2018PASJ...70S...5B} for the optical
imaging data processing pipeline used for the first-year data, 
\citet{2018PASJ...70S..25M} for the first-year shape catalog, 
\citet{2018MNRAS.481.3170M} for the
calibration of galaxy shape measurements with image simulations,
\citet{2019PASJ..tmp..106A} for the public data release of the
first-year shape catalog, 
and \citet{2018PASJ...70S...9T} for photometric redshifts derived for
the first-year data. 

%
%
\subsection{HSC first-year shape catalog}
\label{sec:shape-catalog}

We use the HSC first-year shape catalog \citep{2018PASJ...70S..25M}, in
which the shapes of galaxies are estimated on the $i$-band coadded image
adopting the re-Gaussianization PSF correction method 
\citep{2003MNRAS.343..459H}.
Only galaxies that pass given selection criteria are included in the catalog.
Among others, the four major criteria, which are relevant to the
following analyses, for galaxies to be selected are,
\begin{enumerate}
\renewcommand{\labelenumi}{(\arabic{enumi})}
\item {\it full-color and full-depth cut}: the object should be located
  in regions reaching approximately full survey depth in all five
  ($grizy$) broad bands, 
\item {\it magnitude cut}: the $i$-band cmodel magnitude (corrected for
  extinction) should be brighter than 24.5 AB mag, 
\item {\it resolution cut}: the galaxy size normalized by the PSF size,
  which varies from position to position on coadded images depending on
  observational condition, 
  defined by the re-Gaussianization method, should be larger than a given
  threshold of {\tt ishape\_hsm\_regauss\_resolution $\ge$ 0.3}, 
\item {\it bright object mask cut}: the object should 
not be located within the bright object masks.
\end{enumerate}
 See Table~4 of
\citet{2018PASJ...70S..25M} for the full description of the selection 
criteria.

The HSC shape catalog contains all the basic parameters needed to
perform weak lensing analyses in this study.
The following five sets of parameters for each galaxy are directly 
relevant to this study \citep[see][for a detail description of each
  item]{2018PASJ...70S..25M}; (1) the two-component distortion, 
$\bm{e}=(e_1,e_2)$, which represents the shape of each galaxy image, 
(2) shape weight, $w$, (3) intrinsic shape dispersion per component,
$e_{\mbox{rms}}$, (4) multiplicative bias, $m$, and (5) additive bias,
$(c_1, c_2)$.

%
%
\subsection{Photometric redshifts}
\label{sec:photo-z}

Using the HSC five-band photometry, photometric redshift (hereafter
photo-$z$) was estimated with six independent codes, described in
detail in \citet{2018PASJ...70S...9T}. 
In this study, we adopt {\tt Ephor AB} photo-$z$ data which were derived
from the PSF-matched aperture photometry (called
the {\tt afterburner} photometry) using a neural network
code,  {\tt Ephor}\footnote{https://hsc-release.mtk.nao.ac.jp/doc/index.php/photometric-redshifts/.}.
The data-set contains not only the point estimate but also the
probability distribution function of the redshift for each galaxy, that
we use to select source galaxies (see Section~\ref{sec:source-galaxy}).

%
%
\section{Weak lensing mass maps and high $SN$ peaks}
\label{sec:kappa-peaks}
In this section, we describe our procedure for constructing a sample of
high $SN$ peaks located in weak lensing mass maps.

\subsection{Source galaxy selection}
\label{sec:source-galaxy}

%
%
\begin{figure}
\begin{center}
 \includegraphics[width=82mm]{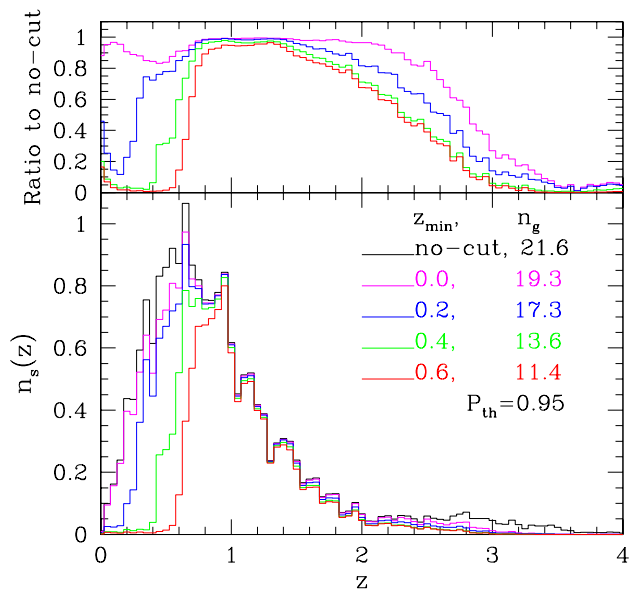}
\end{center}
\caption{{\it Bottom panel}: Estimates of redshift distribution of the
  source samples computed by summing up the redshift probability
  distribution, $P(z)$, over selected source galaxies [see equation (\ref{eq:ns})].
  The normalization is taken so that $\int dz n_s(z)=1$ for the
  ``no-cut'' case (i.e. the full galaxy sample) shown in black
  histogram.
  {\it Top panel}: Ratio of the redshift distribution for a source sample
  to that of ``no-cut'' case.
  \label{wsumpdf_zmax3.0_zmin}}
\end{figure}

%
%
\begin{table*}
  \caption{Summary of source galaxy samples:
    Total area of data-region (see Section~\ref{sec:mass-reconstruction} for
    its definition), 
    The effective number density of source galaxies
    defined by equation (1) of \citet{2012MNRAS.427..146H}
    ($\bar{n}_{g}$),  the averaged shape noise 
    ($\langle \sigma_{\rm shape}^2 \rangle^{1/2}$), and numbers of peaks
    with $SN \ge 5$.
    The last column lists the numbers of merged peaks with
    $z_{\rm opt}=z_{\rm min}$ (i.e., a peak's $SN_{\rm max}$ comes from that
    source sample) with numbers in the
    parentheses being those not existing in $z_{\rm min}=0$ sample with
    $SN(z_{\rm min}=0)\ge5$.
    \label{table:kappa_map}}
\begin{center}
  \begin{tabular}{lccccc}
    \hline
    $z_{\rm min}$ & Area & $\bar{n}_{g}$ & $\langle \sigma_{\rm shape}^2 \rangle^{1/2}$ &
    $N_{\rm peak}$ & $N_{\rm peak}$[merged] at $z_{\rm opt}$\\
    {} & [deg$^{-2}$] & [arcmin$^{-2}$] & {} & $SN\ge 5$ & $SN_{\rm max}>5$ \\
    \hline
    0.0 & 120.01 & 19.3 & 0.0158 & 68 & 24 (-) \\
    0.2 & 119.51 & 17.2 & 0.0167 & 71 & 14 (3) \\
    0.3 & 118.90 & 15.2 & 0.0179 & 70 & 18 (9) \\
    0.4 & 118.08 & 13.6 & 0.0190 & 75 & 22 (13) \\
    0.5 & 117.50 & 12.7 & 0.0198 & 73 & 15 (8) \\
    0.6 & 116.63 & 11.4 & 0.0209 & 69 & 31 (23) \\
    \hline
  \end{tabular}
\end{center}
\begin{tabnote}
  {}
\end{tabnote}
\end{table*}

We use the photo-$z$ information to select source galaxies which are used in
constructing weak lensing mass maps (detailed in the next subsection).
We adopt the {\it P-cut} method proposed by \citet{2014MNRAS.444..147O}
that uses the full probability distribution function of redshift,
denoted by $P(z)$, for each galaxy estimated by the {\tt ephor} method; 
we define samples of source galaxies that satisfy
\begin{equation}
\label{eq:Pcut}  
P_{int}\equiv \int_{z_{\rm min}}^{z_{\rm max}} P(z)~dz > P_{th},
\end{equation}
with the threshold integrated probability of $P_{th} = 0.95$.
Our main aim here is to mitigate the dilution effects of foreground and
cluster member galaxies, and thus a choice of $z_{\rm max}$ is
not crucial as long as it does not so much reduce the number
density of source galaxies. In this study, we take $z_{\rm max}=3$.
Since we do not know redshifts of clusters to be located in mass maps in
advance, we take multiple choices of $z_{\rm min}$; to be specific, we take
$z_{\rm min}=0$, 0.2, 0.3, 0.4, 0.5, and 0.6.

The summation of $P(z)$ over selected galaxies gives a reasonably
reliable estimate of redshift distribution of the source
sample\footnote{Notice that the stacking photo-$z$ $P(z)$ is not a
  mathematically sound way to infer the true redshift distribution
  \citep[see Section 5.2 of ][]{2019PASJ...71...43H}.}.
Taking the lensing weight ($w_i$) into account, we have
\begin{equation}
\label{eq:ns}
n_s(z)=\sum_i w_i P_i(z).
\end{equation}
The effective redshift distributions derived by this method are shown in 
Fig.~\ref{wsumpdf_zmax3.0_zmin} in comparison with the full galaxy sample.
It is seen in the Figure that the {\it P-cut} method works well to
suppress the probability that source samples include galaxies being at
outside the given redshift ranges. 
The mean source galaxy number densities for each sample
are summarized in Table \ref{table:kappa_map}.

%
%
\begin{table}
\caption{The effective survey area of each field. This is for the source
  sample with $z_{\rm min}=0$. Total areas of other source sample are
  summarized in Table~\ref{table:kappa_map}.}
\label{table:fields}
\begin{tabular}{lccc}
\hline
Field name & Data-region area$^{a}$ [deg$^2$] \\
\hline
XMM      & 26.30 \\
GAMA09H  & 28.52 \\
WIDE12H  & 11.45 \\
GAMA15H  & 27.50 \\
HECTOMAP & 9.48 \\
VVDS    & 16.77 \\
\hline
total   & 120.01 \\
\hline
\end{tabular}\\
$^{a}$ Area after removing regions affected by bright objects (masked-region) and
edge-region in unit of degree$^2$. See Section
  \ref{sec:mass-reconstruction} for the definitions of those regions. 
\end{table}

\subsection{Weak lensing mass reconstruction}
\label{sec:mass-reconstruction}

The weak lensing mass map which is the smoothed
lensing convergence field ($\kappa$) is evaluated from the tangential
shear data by \citep{1996MNRAS.283..837S}
\begin{equation}
\label{eq:shear2kap}
{\cal K }(\bm{\theta}) 
=\int d^2\bm{\phi}~ \gamma_t(\bm{\phi}:\bm{\theta}) Q(|\bm{\phi}|),
\end{equation}
where $\gamma_t(\bm{\phi}:\bm{\theta})$ is the tangential component of
the shear at position $\bm{\phi}$ relative to the point $\bm{\theta}$,
and $Q$ is the filter function for which we adopt the truncated Gaussian
function (for $\kappa$ field) \citep{2012MNRAS.425.2287H},
\begin{equation}
\label{eq:Q}
Q(\theta) 
={1 \over {\pi \theta^2}}
\left[1-\left(1+{{\theta^2}\over {\theta_G^2}}\right)
\exp\left(-{{\theta^2}\over {\theta_G^2}}\right)\right],
\end{equation}
for $\theta < \theta_o$ and $Q=0$ elsewhere.
The filter parameters should be chosen so that signals (high peaks in weak
lensing mass maps) from expected target clusters
(i.e. $M_{\rm vir}>10^{14}h^{-1}M_\odot$ at $0.1<z<0.6$) become largest
\citep[see][]{2004MNRAS.350..893H}. 
We take $\theta_G=1.5$ arcmin and $\theta_o=15$ arcmin.

In our actual computation, ${\cal K }$ is evaluated on regular
grid points with a grid spacing of 0.15~arcmin.
Since galaxy positions are given in the sky coordinates, we use the
tangent plane projection to define the grid.
On and around regions where no source galaxy is available due to imaging
data being affected by bright stars or large nearby galaxies, ${\cal K}$
may not be accurately evaluated. 
We define ``data-region'', ``masked-region'' and ``edge-region'' by
using the distribution of source galaxies as follows:
First, for each grid point, we check if there is a galaxy within 0.75
arcmin (about three times the mean galaxy separation) from the grid
point. If there is no galaxy, then the grid point is 
flagged as ``no-galaxy''. After performing the procedure for all the
grid points, all the ``no-galaxy'' grid points plus all the grid points
within 0.75 arcmin from all the ``no-galaxy'' grid points are defined as
the ``masked-region''.
All the masked-regions are excluded from our weak lensing analysis.
All the grid points located within 1.5 arcmin (we take this value by setting
it equal to $\theta_G$) from
any of masked-region grid points are defined as the ``edge-region''.
All the rest of grid points are defined as the ``data-region''.
Since the sky distribution of galaxies differs among different source
samples, we carry out this procedure for every source sample.
The total survey areas (data-region) of each source sample are
summarized in Table \ref{table:kappa_map}, and areas of six fields for
$z_{\rm min}=0$ sample are summarized in Table \ref{table:fields}.
The difference in the total areas among different source samples is not
large but 3 percent at largest.
The total areas of the edge-region are $\sim 30$ degree$^2$, accounting
for $\sim 20$ percent of the data- plus edge-region. 

On grid points, ${\cal K }$ is evaluated using equation
(\ref{eq:shear2kap}), but the integral in that equation is replaced
with a summation over galaxies;
\begin{equation}
\label{eq:shear2kap_sum}
{\cal K }(\bm{\theta}) 
={1\over {\bar{n}_g}} \sum_i \hat{\gamma}_{t,i} Q(|\bm{\phi_i}|),
\end{equation}
where the summation is taken over galaxies within $\theta_o$ from
a grid point at $\bm{\theta}$, $\hat{\gamma}_{t,i}$ is an estimate of 
tangential shear of $i$-th galaxy at the angular position 
$\bm{\phi_i}$ from the grid point, and $\bar{n}_g$ is
the mean galaxy number density (see Section~\ref{sec:global-kap} and
Appendix \ref{sec:local_estimator} for discussion on our
choice of the {\it global normalization}, and see also
\citealp{2011ApJ...735..119S} for a related study). 
The noise on mass maps coming from intrinsic shapes of
galaxies is evaluated on each grid point \citep{1996MNRAS.283..837S},
\begin{equation}
\label{eq:sigma_shape}
\sigma_{\rm shape}^2(\bm{\theta}) 
={1\over {2 \bar{n}_g^2}} \sum_i \hat{\gamma}_{i}^2 Q^2(|\bm{\phi_i}|).
\end{equation}
We define the signal-to-noise ratio ($SN$) of weak lensing mass map by
\begin{equation}
\label{eq:sn}  
SN(\bm{\theta})=
{{{\cal K }(\bm{\theta})}
\over {\langle \sigma_{\rm shape}^2 \rangle^{1/2}}},
\end{equation}
where $\langle \sigma_{\rm shape}^2 \rangle$ is the mean value over all
the grids in the data-region.

Taking the lensing weight, which we normalized so
  that the total weight equals the total number of galaxies (i.e.,
  $\sum_i w_i = N_g$), and measurement biases into account,
equation (\ref{eq:shear2kap_sum}) is modified to 
\citep{2018PASJ...70S..25M}, 
\begin{equation}
\label{eq:shear2kap_weight}
{\cal K }(\bm{\theta}) =
{1 \over {\bar{n}_g}}
{{\sum_i w_i ( e_{t,i}/2{\cal{R}} -\hat{c}_{t}) Q(|\bm{\phi_i}|)}
\over
{1+\hat{m}}},
\end{equation}
where $e_{t}$ is the tangential component of distortion taken from the HSC
shape catalog.
Sample averaged multiplicative bias, responsibity factor and additive
bias are given as follows,
\begin{equation}
\label{eq:hatm}
\hat{m}=
{{\sum_i w_i m_i}
  \over
      {\sum_i w_i}},
\end{equation}
\begin{equation}
\label{eq:resp}
{\cal{R}}=
1-{{\sum_i w_i e_{\mbox{rms},i}^2}
  \over
      {\sum_i w_i}},
\end{equation}
and
\begin{equation}
\label{eq:hatc}
\hat{c}_t=
{{\sum_i w_i c_{t,i}}
  \over
      {\sum_i w_i}},
\end{equation}
where $c_{t,i}$ is the tangential component the additive bias for each
galaxy.
Similarly, the expression for the shape noise, equation
(\ref{eq:sigma_shape}), is modified to,
\begin{equation}
\label{eq:sigma_shape_weight}
\sigma_{\rm shape}^2(\bm{\theta}) =
{1\over {2 \bar{n}_g^2}}
{{\sum_i w_i^2 ( e_{t,i}/2{\cal{R}} -\hat{c}_{t})^2 Q^2(|\bm{\phi_i}|)}
\over
{(1+\hat{m})^2}}.
\end{equation}
%

\subsection{Peak finding and merging multiple peak catalogs}
\label{sec:peak-finding}

We first apply the weak lensing mass reconstruction to each sample of source
galaxies.
We define a peak in the generated mass maps as the grid point with $SN$
value being higher than all the surrounding eight grid points.
We first select peaks with $SN\ge 4$ located in the data-region.
If there is a pair of peaks with separation smaller than
$\sqrt{2}\times \theta_G \simeq 2.1$ arcmin, the lower $SN$ peak of the
pair is discarded to avoid multiple peaks from a single cluster (due to,
for example, substructures of clusters).

The numbers of peaks with $SN\ge 5$ for six source samples are
summarized in Table \ref{table:kappa_map}.
Note that only peaks located in the data-region are included in the peak
catalogs. 
In the same Table, the mean shape noise values measured from each
sample are summarized, which scale with the galaxy number density
approximately as
$\langle \sigma_{\rm shape}^2 \rangle \propto \bar{n}_{g}^{-1}$ as
expected \citep{1996MNRAS.283..837S}.
It should be noticed that although the shape noise becomes larger for
higher $z_{\rm min}$ samples, the number of peak detection does not
always decrease. This may indicate that our source sample selection with
a low-$z$ cut indeed mitigates the dilution effects, that we will go
into detail in Section~\ref{sec:dilution_effects}. 

We combine the six catalogs of high peaks ($SN\ge 4$) from different
source samples by matching peak
positions to a tolerance of $2\times \theta_G =3$ arcmin.
Most of peaks have multiple matches.
Matched peaks from different source samples are merged and are
considered as peaks from the same cluster, and the highest $SN$ among
matched peaks is taken as its peak $SN$ that we denote $SN_{\rm max}$
and we define its source sample's $z_{\rm min}$ as $z_{\rm opt}$.
There are 124 merged peaks with $SN_{\rm max}\ge 5$, which we take as
our primary sample of cluster candidates.
In Table \ref{table:peaklist}, basic information of those 124 merged
peaks are summarized.

%
\begin{longtable}{lrrrrrcccl}
  \caption{Summary of weak lensing merged peaks. First column is the merged peak ID,
  the second to fifth columns are for information of weak lensing peaks (see Section \ref{sec:peak-finding}),
  the sixth to ninth columns are for information of matched CAMIRA-HSC clusters (see Section \ref{sec:cross-matching}), 
  and the last columns is for matched known clusters in \citet{2018A&A...620A...5A} (XXL clusters, see Appendix \ref{sec:xxl}),
  \citet{1989ApJS...70....1A} (Abell clusters), and \citet[][cited by M18 in this table]{2018PASJ...70S..27M} (weak lensing peaks, see Appendix \ref{sec:M18}).
  \label{table:peaklist}}
  \hline
  \multicolumn{5}{l}{Weak lensing} & \multicolumn{4}{l}{CAMIRA-HSC} & Note \\
  ID & $SN_{\rm max}$ & $z_{\rm opt}$ & RA & Dec & ID & $z_{cl}$ & $N_{\rm mem}$ & $\theta_{\rm sep}$ & {} \\      
  {} & {} & {} & \multicolumn{2}{c}{J2000.0 [$\degree$]} & {} & {} & {} & [$\arcmin$] & {} \\
  \hline
  \endhead
   HWL16a-001 &   5.19 &  0.0 &    30.3800 & $    -5.5078 $ &     - &       - &      - &      - & M18~rank~29 \\ 
   HWL16a-002 &   7.23 &  0.6 &    30.4273 & $    -5.0219 $ &    31 &   0.809 &   21.4 &    1.7 & XLSSC~114 ($z=0.234$) \\
   {} &   {} &  {} &    {} & {} &    {} &   {} &   {} &    {} & M18~rank~13 \\
   HWL16a-003 &   5.22 &  0.5 &    31.2073 & $    -3.0587 $ &    65 &   0.553 &   29.2 &    1.3 & {} \\ 
   HWL16a-004 &   5.87 &  0.6 &    31.3577 & $    -5.7178 $ &    74 &   0.290 &   43.5 &    0.9 & XLSSC~106 ($z=0.300$) \\ 
             {} &     {} &   {} &         {} & $         {} $ &    71 &   0.697 &   24.1 &    3.8 & {} \\ 
   HWL16a-005 &   5.24 &  0.5 &    31.4584 & $    -3.3714 $ &    76 &   0.167 &   21.6 &    1.3 & {} \\ 
   HWL16a-006 &   5.46 &  0.3 &    32.0082 & $    -3.4128 $ &    94 &   0.204 &   15.7 &    2.7 & {} \\ 
   HWL16a-007 &   6.53 &  0.3 &    33.1112 & $    -5.6214 $ &   149 &   0.287 &   64.0 &    1.7 & XLSSC~111 ($z=0.300$) \\ 
   HWL16a-008 &   5.46 &  0.3 &    33.1188 & $    -5.5365 $ &   149 &   0.287 &   64.0 &    4.0 & XLSSC~117 ($z=0.298$) \\
   {} &   {} &  {} &    {} & {} &    {} &   {} &   {} &    {} & M18~rank~58 \\
   HWL16a-009 &   6.11 &  0.0 &    33.3625 & $    -2.9126 $ &   165 &   0.150 &   40.4 &    2.4 & M18~rank~16 \\ 
   HWL16a-010 &   5.97 &  0.2 &    33.4777 & $    -2.8852 $ &   174 &   0.274 &   16.1 &    2.8 & M18~rank~63 \\ 
             {} &     {} &   {} &         {} & $         {} $ &   176 &   1.018 &   29.6 &    4.2 & {} \\ 
   HWL16a-011 &   5.12 &  0.6 &    33.8206 & $    -2.7029 $ &     - &       - &      - &      - & {} \\ 
   HWL16a-012 &   6.77 &  0.6 &    35.4434 & $    -3.7668 $ &   252 &   0.430 &   68.3 &    0.3 &  XLSSC~006 ($z=0.429$)\\ 
   HWL16a-013 &   5.31 &  0.6 &    36.1229 & $    -4.2378 $ &   285 &   0.264 &   15.1 &    0.9 & XLSSC~044 ($z=0.263$) \\ 
   HWL16a-014 &   6.38 &  0.2 &    36.3758 & $    -4.2496 $ &   293 &   0.155 &   18.1 &    0.7 & XLSSC~041 ($z=0.142$) \\
   {} &   {} &  {} &    {} & {} &    {} &   {} &   {} &    {} & M18~rank~36 \\
   HWL16a-015 &   6.04 &  0.6 &    37.0512 & $    -5.5929 $ &     - &       - &      - &      - & {} \\ 
   HWL16a-016 &   7.95 &  0.4 &    37.3963 & $    -3.6121 $ &   324 &   0.312 &   57.1 &    0.4 & M18~rank~9 \\ 
   HWL16a-017 &   5.45 &  0.3 &    37.5572 & $    -5.6526 $ &   327 &   0.500 &   22.3 &    1.8 & XLSSC~169 ($z=0.498$) \\ 
   HWL16a-018 &   5.26 &  0.2 &    37.7796 & $    -5.5840 $ &     - &       - &      - &      - & {} \\ 
   HWL16a-019 &   5.08 &  0.0 &    37.8096 & $    -4.4738 $ &   338 &   1.011 &   16.3 &    5.0 & {} \\ 
   HWL16a-020 &   9.68 &  0.3 &    37.9163 & $    -4.8799 $ &   343 &   0.187 &  116.8 &    0.4 & XLSSC~091 ($z=0.186$) \\
   {} &   {} &  {} &    {} & {} &   {} &   {} &  {} &    {} & Abell~362 ($z=0.184$) \\
   {} &   {} &  {} &    {} & {} &    {} &   {} &   {} &    {} & M18~rank~2 \\
   HWL16a-021 &   6.06 &  0.2 &    38.1182 & $    -4.7890 $ &   355 &   0.276 &   33.1 &    3.5 & XLSSC~151 ($z=0.189$) \\
   {} &   {} &  {} &    {} & {} &   {} &   {} &   {} &    {} & XLSSC~152 ($z=0.205$) \\
   {} &   {} &  {} &    {} & {} &    {} &   {} &   {} &    {} & M18~rank~28 \\
   HWL16a-022 &   6.59 &  0.3 &    38.1580 & $    -4.7513 $ &   355 &   0.276 &   33.1 &    0.4 & XLSSC~151 ($z=0.189$) \\ 
   HWL16a-023 &   5.30 &  0.3 &    38.3915 & $    -5.5027 $ &   362 &   0.420 &   46.9 &    0.4 & XLSSC~105 ($z=0.432$) \\ 
   HWL16a-024 &   5.11 &  0.5 &   129.3206 & $     1.6069 $ &   401 &   0.360 &   36.5 &    0.8 & {} \\ 
   HWL16a-025 &   5.64 &  0.0 &   130.3706 & $     0.4379 $ &   433 &   0.413 &   19.0 &    0.4 & M18~rank~22 \\ 
             {} &     {} &   {} &         {} & $         {} $ &   430 &   0.454 &   24.4 &    4.0 & {} \\ 
             {} &     {} &   {} &         {} & $         {} $ &   435 &   0.215 &   21.3 &    4.1 & {} \\ 
   HWL16a-026 &   7.93 &  0.5 &   130.5895 & $     1.6473 $ &   438 &   0.424 &   78.9 &    0.4 & {} \\ 
   HWL16a-027 &   5.11 &  0.4 &   131.0585 & $     1.0644 $ &   449 &   0.323 &   29.1 &    1.5 & {} \\ 
             {} &     {} &   {} &         {} & $         {} $ &   448 &   0.661 &   16.8 &    1.8 & {} \\ 
   HWL16a-028 &   6.98 &  0.6 &   133.1296 & $     0.4041 $ &   505 &   0.270 &   43.2 &    1.6 & M18~rank~18 \\ 
   HWL16a-029 &   5.27 &  0.6 &   135.9841 & $     1.4088 $ &   591 &   0.817 &   16.1 &    2.0 & {} \\ 
             {} &     {} &   {} &         {} & $         {} $ &   592 &   0.665 &   15.6 &    2.8 & {} \\ 
   HWL16a-030 &   5.16 &  0.3 &   136.0810 & $     0.5865 $ &   594 &   0.399 &   16.1 &    2.7 & {} \\ 
             {} &     {} &   {} &         {} & $         {} $ &   593 &   0.303 &   26.8 &    4.9 & {} \\ 
   HWL16a-031 &   5.18 &  0.2 &   136.8921 & $    -0.0580 $ &   628 &   1.062 &   21.4 &    0.4 & {} \\ 
   HWL16a-032 &   8.31 &  0.0 &   138.4612 & $    -0.7631 $ &   686 &   0.285 &   36.1 &    1.0 & M18~rank~4 \\ 
   HWL16a-033 &   5.55 &  0.0 &   138.5051 & $     1.6655 $ &   691 &   0.380 &   15.6 &    4.6 & M18~rank~20 \\ 
   HWL16a-034 &   7.72 &  0.4 &   139.0387 & $    -0.3966 $ &   716 &   0.315 &   78.5 &    1.4 & Abell~776 ($z=0.336$) \\
   {} &   {} &  {} &    {} & {} &    {} &   {} &   {} &    {} & M18~rank~8 \\
   HWL16a-035 &   5.29 &  0.4 &   139.3198 & $     0.9985 $ &   733 &   0.344 &   24.7 &    1.5 & {} \\ 
   HWL16a-036 &   5.99 &  0.6 &   139.3405 & $     3.8281 $ &   734 &   0.420 &   22.9 &    0.4 & {} \\ 
   HWL16a-037 &   5.18 &  0.6 &   140.0954 & $     1.5748 $ &   770 &   0.697 &   55.0 &    0.3 & {} \\ 
   HWL16a-038 &   5.03 &  0.3 &   140.1431 & $     0.7907 $ &   771 &   0.463 &   25.2 &    1.3 & {} \\ 
   HWL16a-039 &   5.88 &  0.0 &   140.4154 & $    -0.2491 $ &   784 &   0.310 &   29.1 &    1.7 & M18~rank~39 \\ 
   HWL16a-040 &   5.97 &  0.6 &   140.5592 & $    -0.1323 $ &   793 &   0.794 &   17.5 &    0.9 & {} \\ 
   HWL16a-041 &   6.01 &  0.3 &   140.6790 & $     2.1327 $ &   798 &   0.194 &   24.8 &    0.2 & M18~rank~30 \\ 
   HWL16a-042 &   5.53 &  0.0 &   177.1051 & $    -0.6610 $ &   864 &   0.419 &   25.9 &    3.5 & M18~rank~19 \\ 
             {} &     {} &   {} &         {} & $         {} $ &   861 &   0.898 &   15.9 &    4.3 & {} \\ 
   HWL16a-043 &   5.49 &  0.6 &   177.1322 & $     0.0088 $ &   860 &   1.080 &   27.6 &    5.0 & {} \\ 
   HWL16a-044 &   6.14 &  0.6 &   177.2646 & $     0.2836 $ &   870 &   0.260 &   16.5 &    4.1 & {} \\ 
   HWL16a-045 &   5.26 &  0.4 &   177.2946 & $     0.3636 $ &   870 &   0.260 &   16.5 &    1.1 & {} \\ 
   HWL16a-046 &   8.07 &  0.5 &   177.5842 & $    -0.6009 $ &   878 &   0.135 &   51.8 &    0.6 & Abell~1392 ($z=0.139$) \\
   {} &   {} &  {} &    {} & {} &    {} &   {} &   {} &    {} & M18~rank~10 \\
   HWL16a-047 &   7.05 &  0.4 &   178.0615 & $     0.5187 $ &   892 &   0.472 &   61.0 &    0.3 & M18~rank~17 \\ 
   HWL16a-048 &   6.11 &  0.4 &   178.0989 & $    -0.5111 $ &   893 &   0.311 &   15.7 &    2.2 & M18~rank~32 \\ 
   HWL16a-049 &   5.22 &  0.4 &   178.6288 & $    -0.1237 $ &   909 &   0.246 &   30.6 &    3.4 & {} \\ 
             {} &     {} &   {} &         {} & $         {} $ &   916 &   0.548 &   19.0 &    3.6 & {} \\ 
             {} &     {} &   {} &         {} & $         {} $ &   912 &   0.892 &   17.7 &    5.0 & {} \\ 
   HWL16a-050 &   5.05 &  0.4 &   178.8288 & $     0.8712 $ &   922 &   0.481 &   21.9 &    1.1 & {} \\ 
   HWL16a-051 &   7.75 &  0.0 &   179.0517 & $    -0.3490 $ &   928 &   0.254 &   66.9 &    1.5 & M18~rank~5 \\ 
   HWL16a-052 &   5.25 &  0.4 &   179.6138 & $    -0.0412 $ &   942 &   0.252 &   29.5 &    1.7 & {} \\ 
   HWL16a-053 &   6.36 &  0.3 &   180.4286 & $    -0.1839 $ &   966 &   0.167 &   45.8 &    1.2 & Abell~1445 ($z=0.169$) \\
   {} &   {} &  {} &    {} & {} &    {} &   {} &   {} &    {} & M18~rank~12 \\
   HWL16a-054 &   5.01 &  0.6 &   180.4536 & $    -0.4986 $ &   968 &   0.322 &   24.5 &    0.7 & {} \\ 
             {} &     {} &   {} &         {} & $         {} $ &   967 &   0.162 &   24.0 &    0.9 & {} \\ 
   HWL16a-055 &   5.39 &  0.4 &   180.6834 & $     0.9709 $ &   978 &   0.568 &   33.7 &    0.7 & {} \\ 
             {} &     {} &   {} &         {} & $         {} $ &   982 &   0.434 &   23.2 &    3.3 & {} \\ 
   HWL16a-056 &   6.88 &  0.6 &   181.3878 & $    -0.6432 $ &   994 &   0.470 &   40.7 &    0.7 & {} \\ 
   HWL16a-057 &   5.32 &  0.4 &   210.7874 & $    -0.3084 $ &  1037 &   0.450 &   35.1 &    0.0 & {} \\ 
   HWL16a-058 &   5.25 &  0.6 &   211.2955 & $    -0.1472 $ &  1046 &   0.248 &   27.5 &    0.7 & {} \\ 
   HWL16a-059 &   5.52 &  0.5 &   211.7872 & $    -0.2717 $ &  1057 &   0.561 &   48.3 &    1.3 & {} \\ 
   HWL16a-060 &   7.33 &  0.4 &   211.9925 & $    -0.4857 $ &  1062 &   0.469 &   34.0 &    0.9 & M18~rank~35 \\ 
   HWL16a-061 &   5.26 &  0.4 &   212.3195 & $    -0.1997 $ &     - &       - &      - &      - & {} \\ 
   HWL16a-062 &   5.42 &  0.2 &   213.6054 & $    -0.3669 $ &  1103 &   0.144 &   60.0 &    1.1 & M18~rank~54 \\ 
             {} &     {} &   {} &         {} & $         {} $ &  1105 &   1.024 &   15.4 &    3.6 & {} \\ 
   HWL16a-063 &   5.07 &  0.4 &   213.7248 & $    -0.3423 $ &  1105 &   1.024 &   15.4 &    4.4 & {} \\ 
   HWL16a-064 &   5.44 &  0.0 &   213.7770 & $    -0.4892 $ &  1112 &   0.144 &   38.8 &    0.5 & Abell~1882 ($z=0.137$) \\
   {} &   {} &  {} &    {} & {} &    {} &   {} &   {} &    {} & M18~rank~26 \\
   HWL16a-065 &   7.16 &  0.2 &   213.8891 & $    -0.0527 $ &     - &       - &      - &      - & M18~rank~23 \\ 
   HWL16a-066 &   5.63 &  0.5 &   214.6762 & $    -0.0506 $ &     - &       - &      - &      - & {} \\ 
   HWL16a-067 &   5.13 &  0.0 &   214.8009 & $     0.2418 $ &     - &       - &      - &      - & {} \\ 
   HWL16a-068 &   5.31 &  0.0 &   215.0330 & $     0.9984 $ &  1155 &   0.322 &   16.6 &    2.8 & {} \\ 
             {} &     {} &   {} &         {} & $         {} $ &  1157 &   0.515 &   46.3 &    3.9 & {} \\ 
   HWL16a-069 &   7.18 &  0.6 &   215.0729 & $     0.9557 $ &  1157 &   0.515 &   46.3 &    0.5 &  M18~rank~33\\ 
             {} &     {} &   {} &         {} & $         {} $ &  1155 &   0.322 &   16.6 &    1.2 & {} \\ 
             {} &     {} &   {} &         {} & $         {} $ &  1161 &   0.168 &   15.3 &    4.7 & {} \\ 
   HWL16a-070 &   6.22 &  0.6 &   215.2574 & $     0.3665 $ &  1164 &   0.645 &   47.6 &    1.4 & {} \\ 
   HWL16a-071 &   5.38 &  0.5 &   215.9165 & $     0.4491 $ &  1191 &   0.534 &   29.4 &    0.5 & {} \\ 
   HWL16a-072 &   5.11 &  0.6 &   216.0089 & $     0.1418 $ &  1195 &   0.539 &   39.1 &    0.8 & {} \\ 
             {} &     {} &   {} &         {} & $         {} $ &  1192 &   0.319 &   17.8 &    2.9 & {} \\ 
   HWL16a-073 &   7.02 &  0.5 &   216.6510 & $     0.8016 $ &  1220 &   0.604 &   18.8 &    3.8 & M18~rank~62 \\ 
   HWL16a-074 &   5.01 &  0.0 &   216.6535 & $    -0.0903 $ &     - &       - &      - &      - & {} \\ 
   HWL16a-075 &   5.65 &  0.3 &   216.6760 & $     0.1643 $ &  1222 &   0.531 &   25.3 &    2.2 & {} \\ 
   HWL16a-076 &   9.36 &  0.5 &   216.7785 & $     0.7267 $ &  1226 &   0.296 &   24.2 &    0.4 &  M18~rank~14\\ 
   HWL16a-077 &   5.46 &  0.6 &   216.8310 & $     0.9541 $ &  1231 &   0.294 &   36.6 &    2.0 & {} \\ 
   HWL16a-078 &   5.28 &  0.0 &   216.8484 & $    -0.2403 $ &  1229 &   0.164 &   19.3 &    4.2 & {} \\ 
   HWL16a-079 &   5.05 &  0.0 &   216.8659 & $    -0.1978 $ &     - &       - &      - &      - & M18~rank~40 \\ 
   HWL16a-080 &   6.14 &  0.5 &   217.6808 & $     0.8093 $ &  1244 &   0.312 &   35.4 &    0.6 &  M18~rank~47\\ 
   HWL16a-081 &   5.66 &  0.4 &   218.8457 & $    -0.3931 $ &  1273 &   0.283 &   27.3 &    1.3 & {} \\ 
   HWL16a-082 &   5.49 &  0.4 &   218.8858 & $    -1.1228 $ &  1274 &   0.756 &   15.6 &    3.5 & {} \\ 
             {} &     {} &   {} &         {} & $         {} $ &  1277 &   0.260 &   24.4 &    4.1 & {} \\ 
   HWL16a-083 &   5.82 &  0.0 &   219.2131 & $    -0.7026 $ &  1288 &   0.198 &   18.4 &    4.1 & M18~rank~25 \\ 
   HWL16a-084 &   5.42 &  0.3 &   220.0846 & $    -0.6101 $ &  1322 &   0.549 &   39.6 &    1.4 & {} \\ 
   HWL16a-085 &   5.82 &  0.0 &   220.4015 & $    -0.9068 $ &  1339 &   0.536 &   52.9 &    1.0 & M18~rank~53 \\ 
             {} &     {} &   {} &         {} & $         {} $ &  1341 &   0.884 &   17.3 &    2.5 & {} \\ 
   HWL16a-086 &   5.23 &  0.5 &   220.4589 & $    -0.8247 $ &  1344 &   0.149 &   29.3 &    1.6 & {} \\ 
             {} &     {} &   {} &         {} & $         {} $ &  1341 &   0.884 &   17.3 &    4.9 & {} \\ 
   HWL16a-087 &   5.02 &  0.0 &   220.5909 & $     0.3367 $ &  1347 &   0.166 &   21.0 &    2.6 & M18~rank~64 \\ 
   HWL16a-088 &   6.01 &  0.4 &   220.7952 & $     1.0452 $ &  1351 &   0.528 &   40.2 &    0.6 & M18~rank~48 \\ 
   HWL16a-089 &   5.71 &  0.0 &   221.0371 & $     0.1743 $ &     - &       - &      - &      - & M18~rank~11 \\ 
   HWL16a-090 &   5.39 &  0.6 &   221.1442 & $     0.2464 $ &  1363 &   0.295 &   54.2 &    1.4 & {} \\ 
   HWL16a-091 &   5.91 &  0.6 &   221.1917 & $    -0.6694 $ &  1366 &   0.523 &   36.6 &    0.4 & {} \\ 
   HWL16a-092 &   5.05 &  0.3 &   221.2090 & $     0.2015 $ &     - &       - &      - &      - & {} \\ 
   HWL16a-093 &   6.80 &  0.6 &   221.3335 & $     0.1116 $ &  1371 &   0.286 &   32.7 &    0.2 & M18~rank~57 \\ 
   HWL16a-094 &   6.49 &  0.2 &   223.0801 & $     0.1689 $ &  1417 &   0.592 &   26.1 &    0.3 & M18~rank~34 \\ 
   HWL16a-095 &   7.66 &  0.6 &   223.0929 & $    -0.9713 $ &  1418 &   0.304 &   38.4 &    0.2 & M18~rank~42 \\ 
   HWL16a-096 &   5.02 &  0.2 &   223.9242 & $    -0.3384 $ &     - &       - &      - &      - & {} \\ 
   HWL16a-097 &   5.05 &  0.0 &   224.2746 & $     0.1164 $ &  1443 &   0.220 &   25.9 &    0.6 & {} \\ 
   HWL16a-098 &   5.58 &  0.4 &   224.6567 & $     0.4858 $ &  1454 &   0.395 &   18.1 &    0.4 & {} \\ 
   HWL16a-099 &   5.76 &  0.2 &   244.4326 & $    42.5427 $ &  1530 &   0.285 &   30.8 &    1.2 & {} \\ 
             {} &     {} &   {} &         {} & $         {} $ &  1528 &   0.598 &   17.9 &    4.1 & {} \\ 
   HWL16a-100 &   5.78 &  0.0 &   245.0550 & $    42.5052 $ &  1540 &   0.141 &   27.4 &    0.9 &  M18~rank~24\\ 
             {} &     {} &   {} &         {} & $         {} $ &  1543 &   0.800 &   22.7 &    4.9 & {} \\ 
   HWL16a-101 &   8.24 &  0.3 &   245.3758 & $    42.7648 $ &  1547 &   0.152 &   33.7 &    0.6 & Abell~2183 ($z=0.136$) \\
   {} &   {} &  {} &    {} & {} &    {} &   {} &   {} &    {} & M18~rank~1 \\
   HWL16a-102 &   5.04 &  0.4 &   246.1339 & $    43.3203 $ &  1557 &   0.287 &   26.1 &    0.8 & {} \\ 
   HWL16a-103 &   5.58 &  0.0 &   246.5173 & $    43.7147 $ &  1561 &   0.260 &   17.1 &    1.1 & {} \\ 
   HWL16a-104 &   6.31 &  0.6 &   333.0522 & $    -0.1334 $ &  1622 &   0.350 &   30.9 &    0.4 &  M18~rank~44\\ 
   HWL16a-105 &   5.36 &  0.3 &   333.3515 & $    -0.2017 $ &  1628 &   0.100 &   19.7 &    3.8 & {} \\ 
             {} &     {} &   {} &         {} & $         {} $ &  1630 &   0.357 &   33.3 &    4.3 & {} \\ 
   HWL16a-106 &   5.55 &  0.3 &   333.3714 & $    -0.1542 $ &  1628 &   0.100 &   19.7 &    1.4 & M18~rank~50 \\ 
             {} &     {} &   {} &         {} & $         {} $ &  1630 &   0.357 &   33.3 &    3.9 & {} \\ 
   HWL16a-107 &   5.57 &  0.6 &   333.5929 & $     0.7956 $ &  1635 &   0.308 &   26.7 &    0.6 & {} \\ 
   HWL16a-108 &   5.14 &  0.2 &   333.6801 & $     1.1171 $ &  1637 &   0.469 &   25.2 &    2.2 & {} \\ 
             {} &     {} &   {} &         {} & $         {} $ &  1638 &   0.759 &   18.5 &    4.3 & {} \\ 
   HWL16a-109 &   5.32 &  0.5 &   333.7900 & $     1.0422 $ &  1639 &   0.702 &   26.7 &    1.2 & {} \\ 
   HWL16a-110 &   5.91 &  0.3 &   335.2140 & $     0.9704 $ &  1664 &   0.323 &   22.1 &    0.1 & {} \\ 
   HWL16a-111 &   5.43 &  0.4 &   335.4040 & $     1.3854 $ &  1670 &   0.790 &   22.8 &    2.4 & M18~rank~56 \\ 
             {} &     {} &   {} &         {} & $         {} $ &  1669 &   0.324 &   16.9 &    3.3 & {} \\ 
   HWL16a-112 &   8.34 &  0.2 &   336.0366 & $     0.3331 $ &  1683 &   0.154 &   44.3 &    0.5 & M18~rank~3 \\ 
   HWL16a-113 &   6.75 &  0.2 &   336.2291 & $    -0.3668 $ &  1688 &   0.308 &   33.1 &    0.8 & M18~rank~21 \\ 
             {} &     {} &   {} &         {} & $         {} $ &  1687 &   0.140 &   19.5 &    1.0 & {} \\ 
   HWL16a-114 &   5.63 &  0.4 &   336.4066 & $    -0.3068 $ &  1694 &   0.402 &   17.8 &    0.2 & {} \\ 
   HWL16a-115 &   6.58 &  0.5 &   336.4217 & $     1.0730 $ &  1696 &   0.281 &   49.9 &    0.8 & M18~rank~52 \\ 
   HWL16a-116 &   5.31 &  0.0 &   336.9540 & $     0.1141 $ &  1707 &   0.410 &   17.3 &    2.0 & {} \\ 
   HWL16a-117 &   6.23 &  0.6 &   337.1293 & $     1.7135 $ &  1709 &   0.338 &   31.5 &    0.6 & M18~rank~61 \\ 
   HWL16a-118 &   5.66 &  0.5 &   338.0182 & $     0.0231 $ &     - &       - &      - &      - & M18~rank~59 \\ 
   HWL16a-119 &   5.39 &  0.6 &   338.0183 & $     0.2288 $ &  1724 &   1.013 &   18.6 &    3.2 & {} \\ 
   HWL16a-120 &   5.13 &  0.6 &   338.5233 & $     1.6997 $ &  1733 &   0.246 &   16.0 &    3.8 & {} \\ 
   HWL16a-121 &   5.39 &  0.2 &   338.9150 & $     1.4837 $ &     - &       - &      - &      - & Abell~2457 ($z=0.059$) \\
   {} &   {} &  {} &    {} & {} &    {} &   {} &   {} &    {} & M18~rank~6 \\
   HWL16a-122 &   6.47 &  0.6 &   339.1320 & $     1.5266 $ &     - &       - &      - &      - & {} \\ 
   HWL16a-123 &   5.54 &  0.6 &   339.3176 & $    -0.3629 $ &     - &       - &      - &      - & {} \\ 
   HWL16a-124 &   5.65 &  0.0 &   339.7643 & $     0.6652 $ &  1748 &   0.264 &   19.4 &    1.2 & M18~rank~65 \\ 
             {} &     {} &   {} &         {} & $         {} $ &  1749 &   0.200 &   19.6 &    3.5 & {} \\ 
\hline
\end{longtable}

In the last column of Table \ref{table:kappa_map}, we present numbers of
merged peaks for each $z_{\rm opt}=z_{\rm min}$ with numbers in the
parentheses showing those that do not exist in the $z_{\rm min}=0$ sample.
We see that $z_{\rm opt}$ is distributed rather broadly with a
noticeable number at the highest $z_{\rm min}$ sample.
It is found that 56 out of 124 merged peaks have $SN(z_{\rm min}=0) < 5$
(to be specific, $SN$s of those peaks measured in mass maps from
$z_{\rm min}=0$ source sample are smaller than 5).
This may be an indication that the dilution effects indeed have non-negligible
influence on peak $SN$s in mass maps of $z_{\rm min}=0$.

%
%
\section{Cross-matching with CAMIRA-HSC clusters}
\label{sec:cross-matching}

We cross-match our merged peak catalog with
the CAMIRA \citep[Cluster-finding Algorithm based on Multi-band Identification
  of Red-sequence gAlaxies,][]{2014MNRAS.444..147O} HSC cluster sample
to identify clusters of galaxies from which weak lensing
peak signals originate.
CAMIRA-HSC cluster sample is based on the same HSC S16A data set
\citep{2018PASJ...70S..20O} used in our study, and thus covers
our survey fields uniformly except
for regions affected by blight objects.
We take this optically-selected cluster catalog as our
primary reference sample, because it covers a sufficiently wide redshift
range ($0.1 < z <1.1$) and cluster mass range (the richness
$N_{mem}>15$, where richness is defined as the effective number of
member galaxies above stellar mass greater than $10^{10.2}M_\odot$). 
For each cluster, the sky coordinates and cluster
redshifts based on
the red sequence of cluster member galaxies
are estimated \cite[see details of
 cluster finding algorithm and definitions of those
 quantities,][]{2014MNRAS.444..147O,2018PASJ...70S..20O}, that we use
in the following analysis.
See Appendix \ref{sec:cross_matching} for results of cross-matching with
other selected cluster samples.

We cross-match our merged peak catalog with CAMIRA-HSC
clusters\footnote{There are some different CAMIRA-HSC
  catalogs based on different HSC data sets. We use the HSC wide cluster
  catalog based on HSC S16A data with updated star mask called 'Arcturus'
  \citep{2018PASJ...70S..25M}. The catalog is available from
  https://www.slac.stanford.edu/{\textasciitilde}oguri/cluster/.}
with their positions to a tolerance of 5 arcmin.
We summarize the results in Table \ref{table:kappa_map}, in which the
angular separation between a peak position and a matched CAMIRA-HSC cluster
position is given ($\theta_{\rm sep}$).
Since the smoothing scale of weak lensing mass map is $\theta_G =1.5$
arcmin, the tolerance radius could be large enough to identify 
clusters of galaxies from which the weak lensing peaks originate.
However, 17 out of 124 peaks have no CAMIRA-HSC cluster
counterpart (see Appendix \ref{sec:no_camira} for some details of those
peaks).
Among the rest of 107 peaks, 25 peaks have multiple matches (mostly 
matching with two CAMIRA-HSC clusters, but 3 out of 25 peaks have three matches).
There are some possible reasons for those systems:
Some of such peaks could be due to physically interacting nearby cluster
systems, but others could be generated not from a single
system but from a line-of-sight projection of multiple clusters
\citep{2004MNRAS.350..893H}. 
In this paper, we are not going into details of such multiple-match
peaks.

%
%
\begin{figure}
\begin{center}
 \includegraphics[width=82mm]{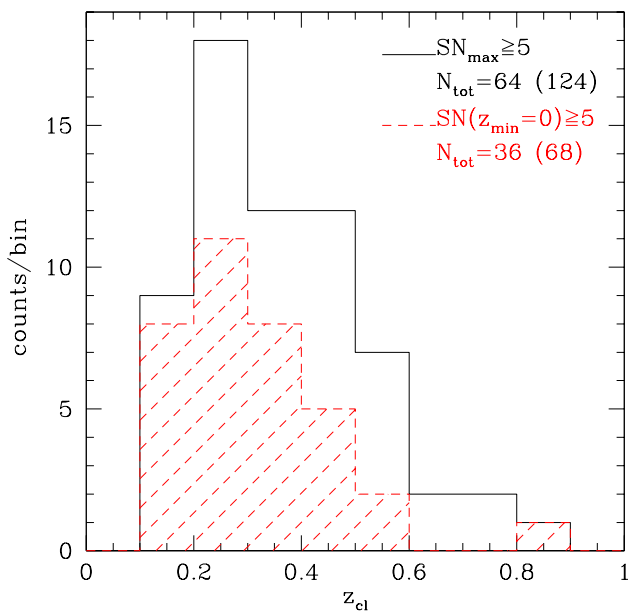}
\end{center}
\caption{Black histogram shows the redshift distribution of the weak
  lensing secure clusters (64 out of 124 merged peaks), whereas the red
  hatched histogram shows the same clusters but having $SN\ge 5$ in weak
  lensing mass maps from the source sample of 
  $z_{\rm min}=0$ (36 out of 68 peaks located in $z_{\rm min}=0$ mass maps with
  $SN\ge 5$).  
  \label{npeak_z0_z}}
\end{figure}

Among 82 peaks matching with a single CAMIRA-HSC cluster, 64 peaks have
CAMIRA-HSC cluster counterparts within 2 arcmin from peak positions.
Although it is possible that some of those peaks are affected by
line-of-sight projections of small clusters (below the richness
threshold of CAMIRA algorithm), it is highly likely that the major
lensing contribution comes from the matched CAMIRA-HSC clusters. 
We have also visually inspected those systems with HSC $riz$-color image, and 
found good correlations between weak lensing mass over-densities and
galaxy concentrations for all the cases.
We thus define those 64 peaks
as the {\it secured sample of WL clusters}, with redshift (that we
denote $z_{cl}$) taken from the
matched CAMIRA-HSC cluster, which 
we will use to investigate the dilution effects
in the following section. The redshift distribution of those secured weak lensing clusters is shown
in Figure \ref{npeak_z0_z}.
In the same plot, we also show the distribution of those weak
lensing secure clusters that have $SN\ge 5$ in weak lensing mass maps from
the source sample of $z_{\rm min}=0$.
Comparing the two distributions, we see that a large part of
clusters at $z_{cl}>0.4$ have peak $SN$s below our threshold of $SN=5$
in the mass maps of $z_{\rm min}=0$, and pass the threshold in mass
maps of $z_{\rm min}\ge 0.2$.

%
%
\begin{figure}
\begin{center}
 \includegraphics[width=82mm]{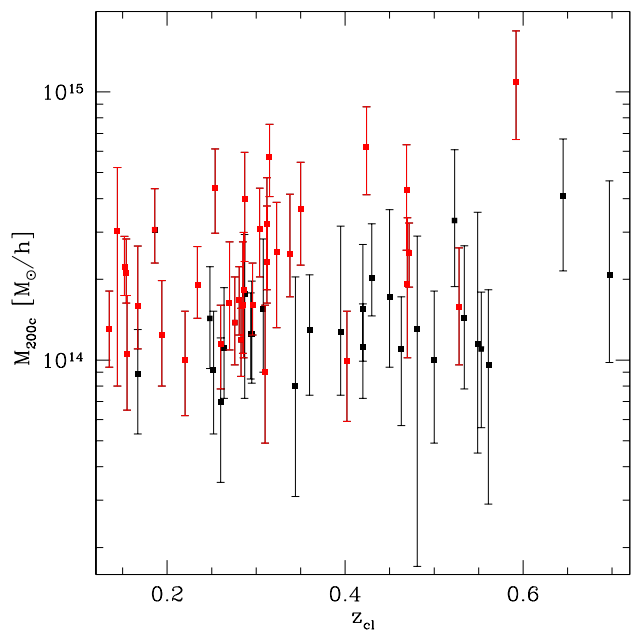}
\end{center}
\caption{Distribution of weak lensing secure clusters on the
  $M_{200c}-z_{cl}$ plane. 
  The cluster masses defined by the spherical over-density mass $M_{200c}$
  are derived by fitting the NFW model to measured weak lensing shear
  profiles based on the standard likelihood analysis (see Appendix
  \ref{sec:cluster_mass}), and 
  filled squares and error bars show the peak and 68.3\% confidence
  interval of the posterior distributions.
  Red (black) symbols are for clusters with the peak $SN\ge 5$ ($<5$) in weak
  lensing mass maps from the source sample of 
  $z_{\rm min}=0$.
  \label{z_m}}
\end{figure}

We derive the cluster masses of the weak lensing secure clusters
by fitting the NFW model to measured weak lensing shear profiles
based on the standard likelihood analysis (see Appendix
\ref{sec:cluster_mass} for details).
Derived cluster masses are plotted on the cluster mass--redshift plane 
in Figure \ref{z_m}, where
red (black) symbols are for clusters with the
peak $SN\ge 5(<5)$ in weak lensing mass maps from the source sample of
$z_{\rm min}=0$.
From this Figure, we find that clusters below the peak height threshold
($SN=5$) in the mass maps of $z_{\rm min}=0$ are mostly relatively lower
mass clusters at $z_{cl} \gtsim 0.4$.
This is a natural result of the following two facts that (1)
for a fixed cluster redshift, the peak height is higher for more
massive clusters \citep{2004MNRAS.350..893H}, and (2) the dilution effect of
foreground galaxies is stronger for higher redshift clusters and for
lower $z_{\rm min}$ galaxy samples (see the next section).

%
%
\section{Dilution effects on weak lensing peaks from clusters}
\label{sec:dilution_effects}

%
%
\begin{figure*}
\begin{center}
  \includegraphics[width=82mm]{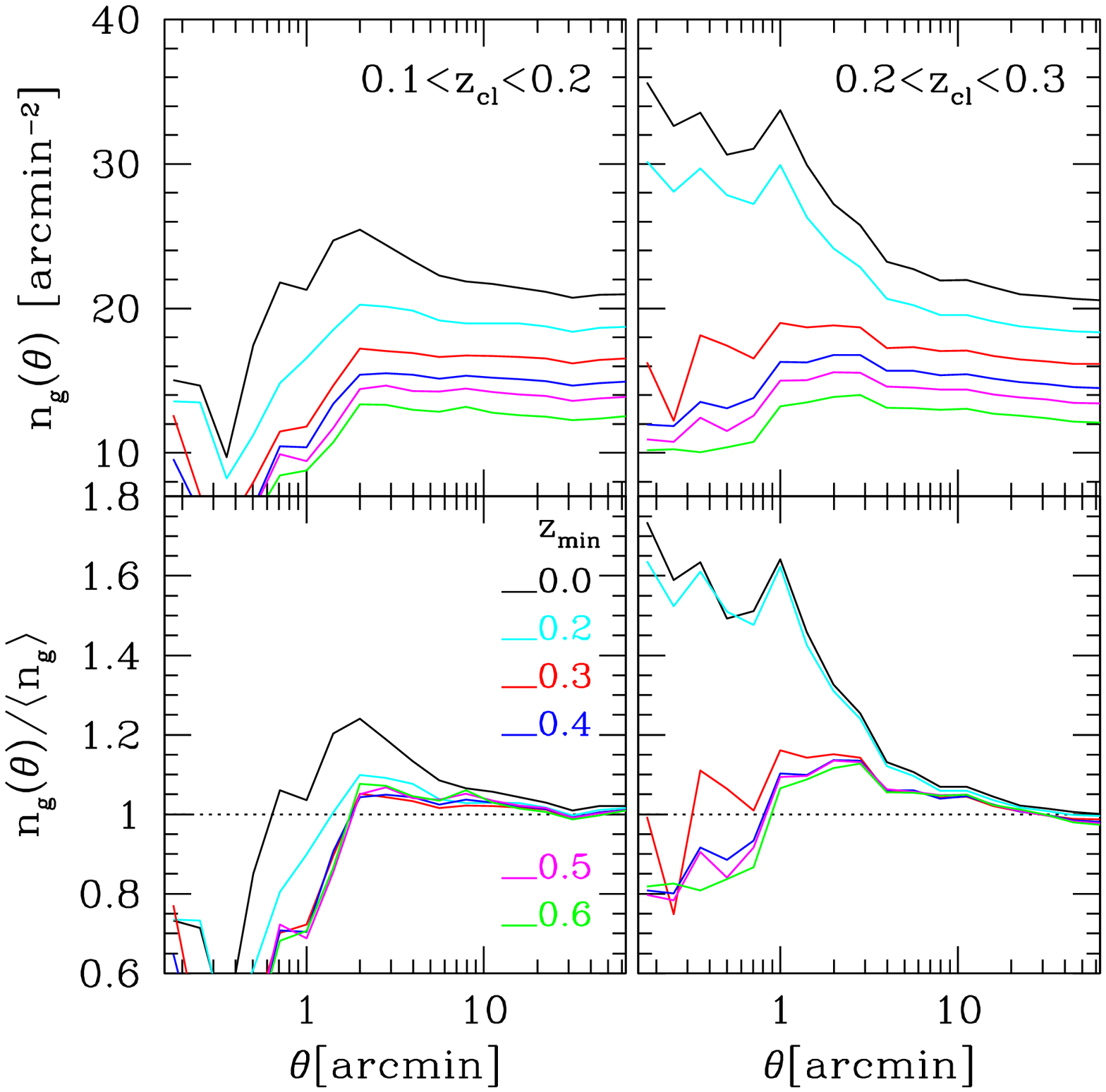}
  \hspace{1em}
  \includegraphics[width=82mm]{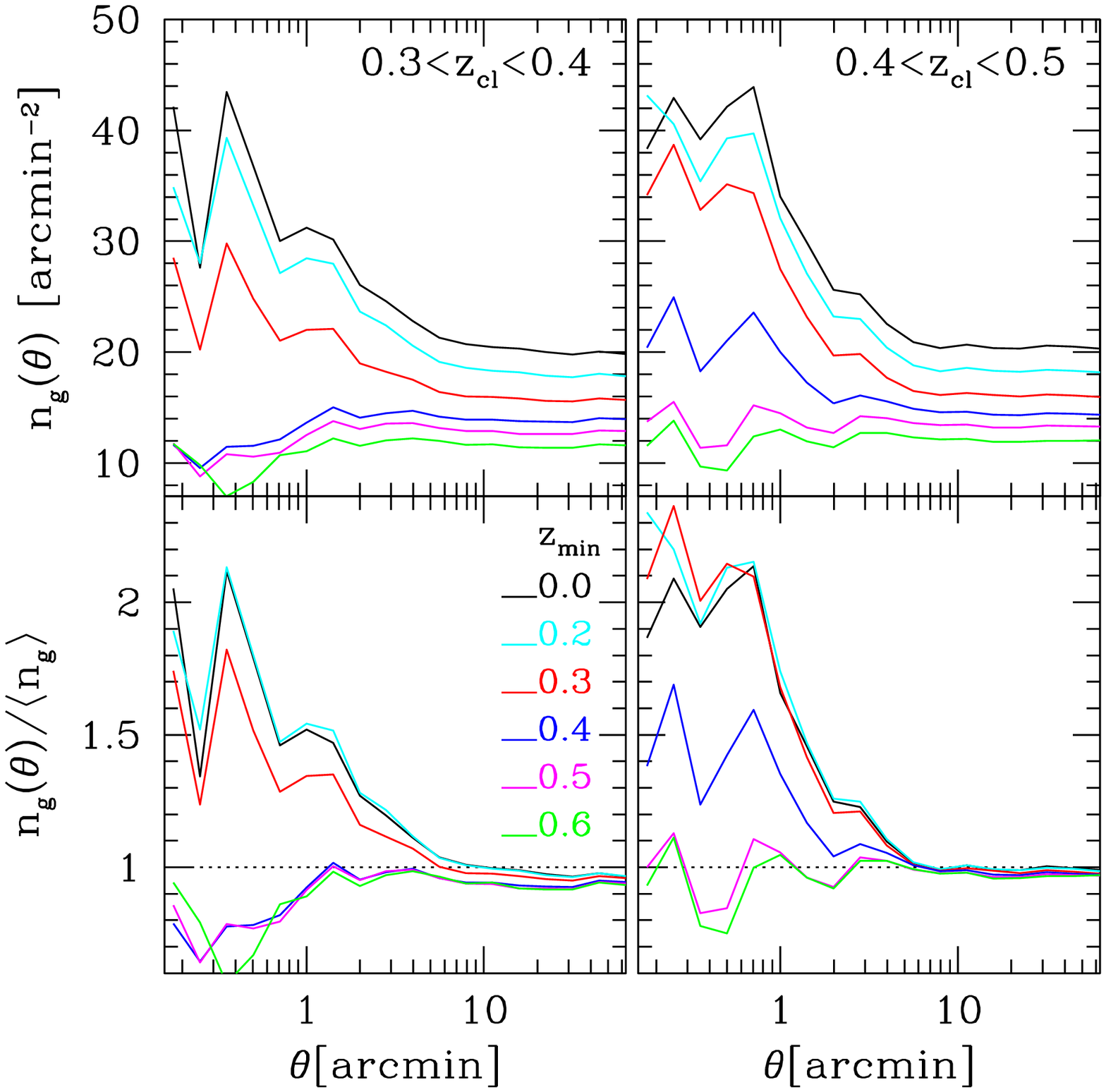}
\end{center}
\caption{{\it Top-panels}: Stacked galaxy number density profiles as a
  function of angular radius from peak positions. Samples of the weak
  lensing secure clusters (defined in Section 4) with cluster redshifts
  within a given redshift range (denoted in each panel) are
  used. Different colors are for different source galaxy samples
  characterized by $z_{\rm min}$ (see Section \ref{sec:source-galaxy}).
  {\it Bottom-panels}: The same stacked galaxy number density profiles
  shown in the top-panels but normalized by the mean number density of
  each source galaxy sample.
  \label{gdensprof_zcl}}
\end{figure*}

The dilution effects on weak lensing high peaks originating from
clusters are caused by foreground and cluster member galaxies
\citep[for observational studies of the dilution effects
  in analyses of cluster lensing, see for
  example][]{2005ApJ...619L.143B,2010PASJ...62..811O,2018PASJ...70...30M}.
Let us first make a rough estimate of proportions of those galaxies in
our source galaxy samples. 
We see from the estimated redshift distributions of source samples shown
in Figure \ref{wsumpdf_zmax3.0_zmin} that a proportion of foreground to
background galaxies depends strongly on both cluster redshifts and
source samples, and it can be more than 20 percent for high-$z$
clusters in low-$z_{\rm min}$ source samples.
We estimate the proportion of cluster member galaxies by measuring
stacked galaxy number density profiles of sub-samples
of weak lensing secure clusters selected based on cluster redshifts.
The measurement is done for every source galaxy sample and
the results are presented in Figure \ref{gdensprof_zcl} for four
redshift ranges. 
We find that, at cluster central regions, a considerable
number of cluster member galaxies are contained in source samples with
$z_{\rm min}<z_{cl}$ except for the case of the lowest cluster redshift range.
The excess mostly disappears in source samples with $z_{\rm min}>z_{cl}$. 
However, we note that the degree of the excess and
its suppression largely vary from cluster to cluster.

We have adopted two means to mitigate the dilution effects:
One is to take the {\it globally normalized} $SN$ estimator,
equation (\ref{eq:sn}) with equations (\ref{eq:shear2kap_sum}) and
(\ref{eq:sigma_shape}), and the other is to combine multiple peak
catalogs from weak lensing mass maps of source
samples with different $z_{\rm min}$.
In the following sub-sections, we will first describe the former, then
we will examine the effectiveness of the latter using actual
source galaxy samples.

Another important point seen in Figure \ref{gdensprof_zcl}
is that the deficiency of source galaxies in cluster central regions for the
lowest redshift cluster sample and for the other samples with
$z_{\rm min}>z_{cl}$. 
There are two possible causes of this:
One is the masking effect of bright cluster galaxies that screen
background galaxies behind them.
The other is the lensing magnification effect that enlarges a sky area
behind clusters resulting in a decrease in the local galaxy number
density (for more details, see \citealp{2001PhR...340..291B}; and see
\citealp{2019arXiv190902042C} for a
measurement of lensing magnification effect in the HSC data).
We are not going into further details of those two effects because it is
beyond the scope of this paper, but we examine
their influence on the peak height using empirical models in
Section~\ref{sec:defficiency_effect}. 

%
%
\subsection{The globally normalized $SN$ estimator}
\label{sec:global-kap}

Here, we explain how the globally normalized $SN$ estimator
defined by equation~(\ref{eq:sn}) 
can mitigate the dilution effect of cluster member galaxies.
We examine actual advantage of this estimator over the locally
normalized estimator in Appendix \ref{sec:local_estimator}. 

Let us assume the following simple model of a galaxy distribution which consists
of three populations;
lensed background galaxies ($bg$), unlensed foreground galaxies ($fg$),
and unlensed cluster member galaxies ($cl$), with number densities of
$n_{bg}$, $n_{fg}$, and $n_{cl}(\theta)$, respectively.
Note that we have
assumed that only $n_{cl}(\theta)$ has a non-uniform sky distribution
associated with clusters of galaxies.
As is seen in Figure \ref{gdensprof_zcl}, $n_{cl}(\theta)$ can
be comparable to $n_{bg}+n_{fg}$ at cluster central regions.
However, since the cluster population is very rare in the sky, in what
follows, we assume that the globally averaged
$n_{cl}(\theta)$ is much smaller than $n_{bg}+n_{fg}$, and we
take $\bar{n}_g=n_{bg}+n_{fg}$.
Then the globally normalized estimator,
equation~(\ref{eq:shear2kap_sum}), can be formally written by
\begin{eqnarray}
\label{eq:shear2kap_sum_G}
{\cal K }_G(\bm{\theta}) 
&=&{1\over {\bar{n}_g}} \sum_i \hat{\gamma}_{t,i} Q_i \nonumber\\
&=&{1\over {\bar{n}_g}}
\left(
\sum_{i\in bg} \hat{\gamma}_{t,i} Q_i
+\sum_{i\in fg} \hat{\gamma}_{t,i} Q_i
+\sum_{i\in cl} \hat{\gamma}_{t,i} Q_i
\right)\nonumber\\
&=&{1\over {n_{fg}+n_{bg}}}
\sum_{i\in bg} \hat{\gamma}_{t,i} Q_i,
\end{eqnarray}
where from the second to third line, we have used the fact that the foreground
and cluster member galaxies have no lensing signal.
Denoting the galaxy intrinsic ellipticity by $e^{\rm int}$ and its shear
converted one by $\hat{e}=e^{\rm int}/2\cal{R}$, the estimator
of the shape noise, equation~(\ref{eq:sigma_shape}), can be written, in
the same manner, by
\begin{eqnarray}
\label{eq:sigma_shape_G}
\sigma_{{\rm shape},G}^2(\bm{\theta}) 
&=&{1\over {2 (n_{fg}+n_{bg})^2}}\nonumber \\
&&\times \left(
\sum_{i\in bg} \hat{e}_i^2 Q_i^2
+\sum_{i\in fg} \hat{e}_i^2 Q_i^2
+\sum_{i\in cl} \hat{e}_i^2 Q_i^2
\right),
\end{eqnarray}
where we have ignored the contribution from lensing shear.
Taking the average over a survey field, we have,
\begin{eqnarray}
\label{eq:sigma_shape_Gave}
\langle \sigma_{{\rm shape},G}^2\rangle 
&\simeq& {1\over {2 (n_{fg}+n_{bg})^2}} \nonumber \\
&&\times \left(
\left\langle \sum_{i\in bg} \hat{e}_i^2 Q_i^2 \right\rangle
+\left\langle \sum_{i\in fg} \hat{e}_i^2 Q_i^2 \right\rangle
\right),
\end{eqnarray}
where we have again assumed that on global average the contribution from
the cluster member population is small and have ignored it.
Using those expressions, the globally normalized $SN$ defined by
equation~(\ref{eq:sn}), can be written by
\begin{eqnarray}
\label{eq:sn_G}
SN_G(\bm{\theta})
&=& {{{\cal K }_G(\bm{\theta})} \over
{\langle \sigma_{{\rm shape},G}^2\rangle}^{1/2}}\nonumber\\
&\simeq&
{\sqrt{2}{\sum_{bg} \hat{\gamma}_{t,i} Q_i}
\over
{\left( \bigl\langle \sum_{bg} \hat{e}_i^2 Q_i^2 \bigr\rangle
+\bigl\langle  \sum_{fg} \hat{e}_i^2 Q_i^2 \bigr\rangle\right)^{1/2}}}.
\end{eqnarray}
Note that in the above expression, there is no contribution from cluster
member population.
Therefore, the globally normalized estimator is, to a good approximation,
free from the dilution effect of the cluster member galaxies.

%
%
\subsection{Dilution effect of foreground galaxies}
\label{sec:foreground}

%
%
\begin{figure}
\begin{center}
  \includegraphics[width=82mm]{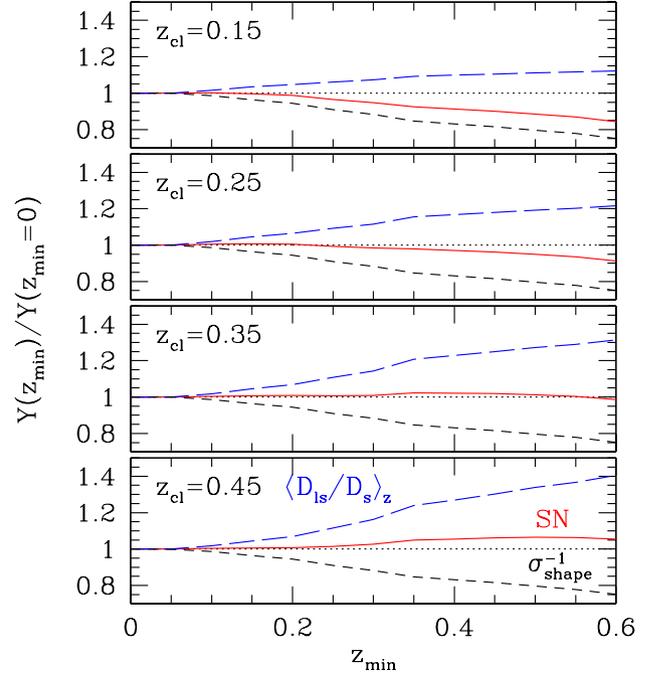}
\end{center}
\caption{Shown is dependence of any of $\langle D_{ls}/D_s\rangle_z$ (blue
  long-dashed lines), $\sigma_{\rm shape}^{-1}$ (black dashed lines) or
  $SN_{\rm peak}$ (red solid lines) on $z_{\rm min}$.
  All the quantities (signified by $Y(z_{\rm min})$) are normalized by their values
  at $z_{\rm min}=0$.
  Different panels for different cluster redshifts, $z_{cl}$, which are
  denoted in each panel.
  \label{fg_contmi}}
\end{figure}

Foreground galaxies have two effects on weak lensing peak $SN$s from clusters.
One is to dilute the lensing signal, and the other is to make the shape
noise level on mass maps smaller.
Below we will first derive relevant expressions for these effects,
and evaluate the dilution effect of foreground galaxies using the actual
redshift distributions of source galaxies.
Then, we will compare it with the real data measured using the secure
weak lensing cluster sample.

Focusing on contributions from foreground and background galaxies,
from equation (\ref{eq:shear2kap_sum_G}), the peak signal can be
approximately written by
\begin{equation}
\label{eq:kap_G_app}
{\cal K }_G(\bm{\theta}) 
={1\over {n_{fg}+n_{bg}}}
\sum_{i\in bg} \hat{\gamma}_{t,i} Q_i
\propto  {{n_{bg} \langle \hat{\gamma}_t \rangle_z}\over {n_{fg}+n_{bg}}},
\end{equation}
where $\langle \hat{\gamma}_t \rangle_z$ is the source redshift distribution weighted
mean tangential shear.
Since the source redshift dependence of the tangential shear enters only
through the distance ratio, $D_{ls}/D_s$, we can re-write equation
(\ref{eq:kap_G_app}) by
\begin{equation}
  \label{eq:dratio}
  {\cal K }_G(\bm{\theta}) \propto
\left\langle{{D_{ls}} \over {D_s}}\right\rangle_z =
{{\int_{z_{cl}}^\infty dz~n_s(z) D_{ls}(z_{cl},z)/D_{s}(z)}
  \over
 {\int_0^\infty dz~n_s(z)} 
},
\end{equation}
where $n_s(z)$ is the redshift distribution of source galaxies.

%
%
\begin{figure}
\begin{center}
  \includegraphics[width=82mm]{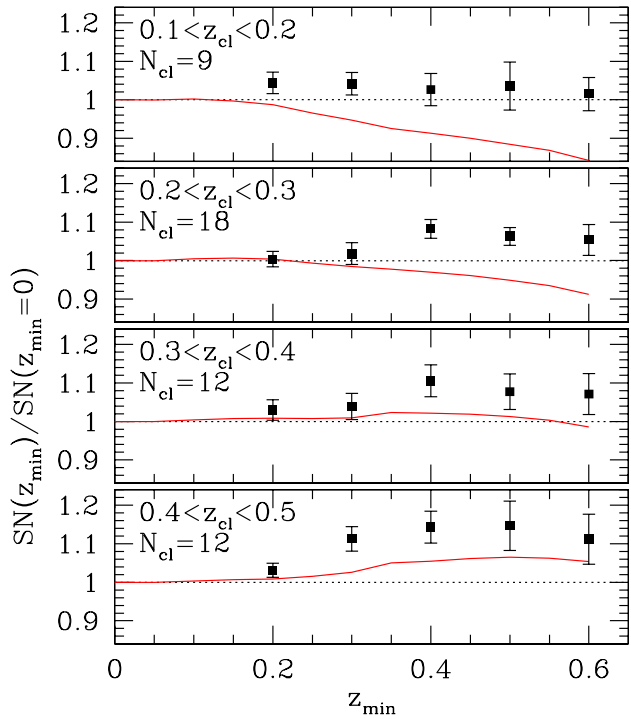}
\end{center}
\caption{Shown is the $SN(z_{\rm min})$ of weak lensing secure cluster
  normalized by its $SN(z_{\rm min}=0)$.
  Weak lensing secure clusters are divided into four
  sub-samples based on the cluster redshifts (denoted in panels).
  The horizontal axis is $z_{\rm min}$ of source galaxy samples.
  For each sub-sample and each source galaxy sample, the mean and
  its 1-$\sigma$ error among the clusters (the number of clusters in each
  sub-sample is given in each panel) are plotted.
  For comparison, the red lines show the expected $SN$ ratios plotted in
  Figure \ref{fg_contmi} (red lines) for a cluster at a central redshift
  in respective redshift ranges.
  \label{sn_zmin_stats}}
\end{figure}

In the same manner, from equation (\ref{eq:sigma_shape_Gave}) we have
\begin{equation}
\label{eq:sigma_shape_Gapp}
\langle \sigma_{{\rm shape},G}^2\rangle 
\propto
{{(n_{fg}+n_{bg})\langle\hat{e}^2\rangle}
\over {2 (n_{fg}+n_{bg})^2}} 
\propto {{\langle\hat{e}^2\rangle} \over {\bar{n}_g}},
\end{equation}
where we have ignored a possible redshift dependence of $\langle\hat{e}^2\rangle$.
This is the well known scaling relation between the shape noise and
galaxy number density \citep{1996MNRAS.283..837S}.

The weak lensing peak $SN$ from clusters is related to the source redshift
weighted distance ratio and the shape noise via
\begin{equation}
\label{eq:SN-peak}
SN \propto
{ {\langle{{D_{ls}}/ {D_s}}\rangle_z }
\over
{\langle \sigma_{{\rm shape},G}^2 \rangle ^{1/2}}}. 
\end{equation}
Since both the distance ratio and the shape noise depend on source
galaxy samples, so does the peak SN from a cluster.
In our case, the source galaxy selection is characterized by $z_{\rm min}$,
and thus we evaluate the dependence of those quantities on $z_{\rm min}$
using the redshift distributions of our source samples defined by
equation (\ref{eq:ns}).  
The results are shown in Figure \ref{fg_contmi} for cluster redshifts of
$z_{cl}=0.15$, 0.25, 0.35, and 0.45.
Findings from that figure are as follows:
The shape noise, which is not dependent on $z_{cl}$, monotonically
increases with $z_{\rm min}$ as expected. 
The source redshift weighted distance ratio increases with
$z_{\rm min}$. 
Since a fraction of foreground galaxies is larger for the higher
redshift clusters, the higher the cluster redshift, the larger the distance ratio.
Those two effects compete;
For clusters with redshifts lower than 0.3, their weak lensing peak $SN$
decreases with $z_{\rm min}$.
However, for higher redshift clusters, $SN$ stays almost constant or
slightly increases with $z_{\rm min}$.

We examine the actual dependence of peak $SN$s on $z_{\rm min}$ using the
weak lensing secure clusters.
In doing so, we divide the secure clusters into four sub-samples based
on the cluster redshift (to be specific, $0.1<z_{cl}<0.2$, $0.2<z_{cl}<0.3$,
$0.3<z_{cl}<0.4$, and $0.4<z_{cl}<0.5$).
For each sub-sample, we evaluate the mean of
$SN(z_{\rm min})/SN(z_{\rm  min}=0)$ and its standard error among sample
clusters.
The results are shown in Figure \ref{sn_zmin_stats}.
We find that the measured ratios of $SN(z_{\rm min})/SN(z_{\rm  min}=0)$ is
systematically larger than the expectations shown by the red lines
(which are same as ones plotted in Figure
\ref{fg_contmi}), especially for lower-$z$ clusters ($z_{cl}<0.3$).
The reason of this is unclear; a possible cause is the intrinsic
alignment of galaxies:
Because the major axis of galaxies surrounding a cluster tend to point
towards the cluster center due to the intrinsic alignment effects
(originating from, e.g., the gravitational tidal stretching, see for a
review \citealp{2015SSRv..193....1J}, and references therein), it reduces 
the peak $SN$ value.
Peak $SN$s of $z_{\rm min}=0$ maps are likely affected by this effect,
and consequently they are likely biased low.
If this is the case, it accounts for the systematically larger
$SN(z_{\rm min})/SN(z_{\rm min}=0)$ found in the measured results,
though this argument is rather phenomenological.
Aside from this systematic difference, the measured ratios are in 
reasonable agreement with expectations in their amplitudes and in its
increasing trend toward higher-$z$ clusters.
From the above findings, we conclude that combining multiple peak
catalogs from source samples with different $z_{\rm min}$ can mitigate
the dilution effect of foreground galaxies, especially on high-$z$
clusters.

%
%
\subsection{Impact of the source galaxy deficiency on peak heights}
\label{sec:defficiency_effect}

\begin{figure}
\begin{center}
  \includegraphics[width=82mm]{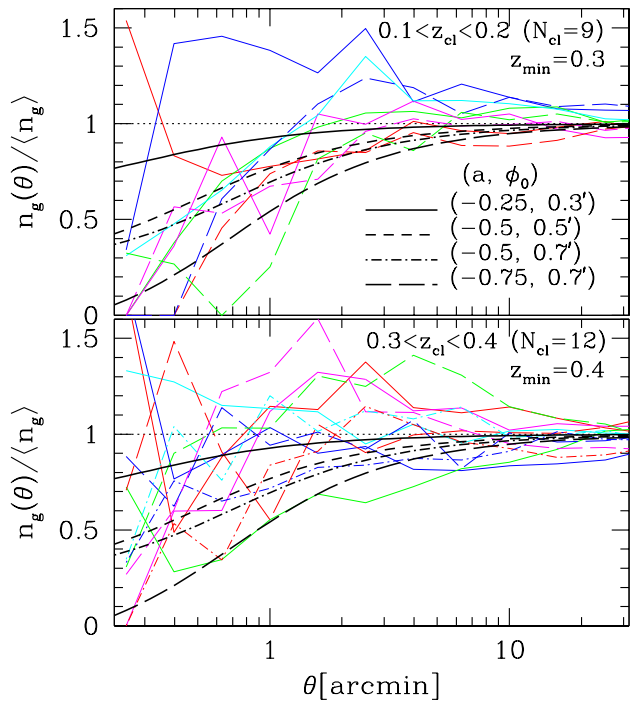}
\end{center}
\caption{Thin colored lines show azimuthally averaged galaxy number
  density normalized by the mean number density as a function of the 
  angular separation from peak positions.  
  Each colored line is for an individual cluster.
  Sub-samples of weak lensing secure clusters are shown.
  The upper/lower panel is for the clusters at
  $0.1<z_{cl}<0.2$/$0.3<z_{cl}<0.4$ with the 
  source galaxy sample of $z_{\rm min}=0.3$/$0.4$.
  The thick black lines show the parametric model,
  equation (\ref{eq:fcl_fit}) but plotted is $[1+f_m(\theta)]$,
  for four sets of parameters denoted in the upper panel.  
  \label{gdensprof_indv}}
\end{figure}

Here we examine the impact on the peak $SN$ from source galaxy deficiency
at cluster central regions seen in the stacked galaxy number density
profiles shown in Figure \ref{gdensprof_zcl}.
We note, however, that deficiency profiles vary greatly from cluster to
cluster as is shown in Figure \ref{gdensprof_indv}.

%
%

%
%
\begin{figure}
\begin{center}
  \includegraphics[width=82mm]{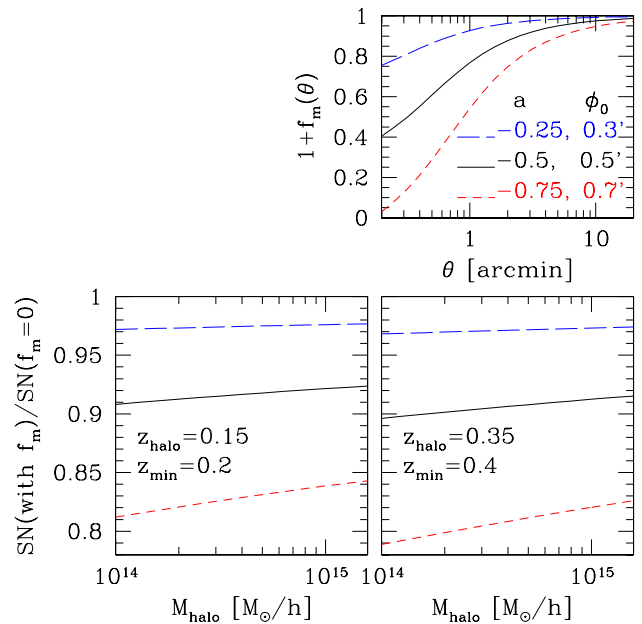}
\end{center}
\caption{{\it Top-panel}: Three models of the source galaxy deficiency
  profile are shown. See equation (\ref{eq:fcl_fit}) for the
  functional form of the model.
  {\it Bottom-panels}: Shown is ratios between expected peak $SN$s of an
  NFW halo with and without taking the source galaxy deficiency into account as a
  function of halo mass (the virial mass is taken here).  
  The left-panel is for the case $z_{\rm halo}=0.15$ with the galaxy 
  redshift distribution taken from the source sample with $z_{\rm min}=0.2$,
  whereas the right-panel is for $z_{\rm halo}=0.35$ with $z_{\rm min}=0.4$.
  Different lines are for different deficiency models shown in
  the top-panel.
  \label{fmask_ratsn}}
\end{figure}

We use the empirical models of dark matter
halo and simple model of the galaxy deficiency profiles, which we
describe below.
In the presence the source galaxy deficiency, the
theoretical expression for the lensing signal from clusters, equation
(\ref{eq:shear2kap}), is modified to
\begin{equation}
\label{eq:shear2kap_m}
{\cal K }(\bm{\theta})
=
\int d^2\bm{\phi}~
[1+f_m(\phi)]
\gamma_t(\bm{\phi}:\bm{\theta}) Q(|\bm{\phi}|),
\end{equation}
where $f_m(\phi)$ is the source galaxy deficiency profile, for which 
we adopt the following parametric function,
\begin{equation}
\label{eq:fcl_fit}
f_m(\phi) = \max\left\{
a \left( {\pi \over 2} -\tan^{-1}({\phi \over {\phi_0}}) \right),-1
\right\},
\end{equation}
where $a$ and $\phi_0$ are the amplitude and scale parameters,
respectively.
We fit the measured deficiency profiles shown in Figure
\ref{gdensprof_zcl} with this function and derive typical
values of those parameters that we take in the following analysis.
We consider three models of $f_m(\phi)$ shown in the top-panel of Figure
\ref{fmask_ratsn}:
The model with $a=-0.25$ mimics the stacked deficiency profile
of a case $0.3<z_{cl}<0.4$ with $z_{\rm min} > z_{cl}$, and the model with
$a=-0.5$ mimics ones of $0.1<z_{cl}<0.2$ (see Figure
\ref{gdensprof_zcl}), whereas the model with $a=-0.75$ represents 
extreme cases from individual clusters shown in Figure
\ref{gdensprof_indv}.
Since details of the models needed to compute equation
(\ref{eq:shear2kap_m}) are described in
\citet{2012MNRAS.425.2287H}, here we summarize the main ingredients and
relevant references:
\begin{itemize}
\item Dark matter halos of clusters are modeled by the truncated NFW
  model \citep{2009JCAP...01..015B} with
  the mass-concentration relation by
  \citet{2008MNRAS.391.1940M} and \citet{2008MNRAS.390L..64D}.
\item Redshift distributions of source galaxies of our source samples
  are estimated by equation (\ref{eq:ns}), which are dependent on $z_{\rm min}$,
  and are presented in Figure \ref{wsumpdf_zmax3.0_zmin}.
\end{itemize}
Results are presented in Figure \ref{fmask_ratsn}.
Note, however, that since the deficiency model is a crude approximation
and halo mass dependence of the source deficiency is not taken into
account, the results 
should be considered as a rough estimate of the impact of source
deficiency on a peak $SN$. 
In the bottom-panels, ratios between expected peak $SN$s with and
without taking the source deficiency into account as a function of halo
mass are shown for three deficiency models presented in the top-panel.
The left-panel is for the case $z_{\rm halo}=0.15$, whereas the right-panel is
for $z_{\rm halo}=0.35$.
We find that in both the cases, the suppression of $SN$ due to the
source deficiency is 2--5 percent for models with $a=-0.25$, 8--10
percent for $a=-0.5$ and 16--20 percent for the extreme case of
$a=-0.75$.
We thus conclude that a typical impact of source deficiency on a peak
$SN$ is a suppression of a few to $\sim10$ percent.
However it can be $\sim20$ percent for individual cases.
It is also seen in the Figure that for a given source deficiency model,
the suppression decreases with increasing halo mass.
The reason for this is that a relative
contribution to a peak $SN$ from galaxies within a fixed aperture is smaller
for more massive halos.

%
%
\section{Summary and discussions}
\label{sec:summary}

We have presented a weak lensing cluster search using HSC first-year data.
We generated six samples of source galaxies
with different $z_{\rm min}$-cuts (we took $z_{\rm min}=0$, 0.2, 0.3,
0.4, 0.5 and 0.6), and made weak lensing mass maps for each source
sample from which we searched for high peaks.
From each source sample, we detected a sample of 68--75 weak lensing
peaks with $SN\ge 5$.
We compiled the six peak samples into a sample of merged peaks.
We obtained a sample of 124 weak lensing merged peaks with $SN_{\rm max}\ge 5$
which are candidates of clusters of galaxies.

We cross-matched our peak sample with CAMIRA-HSC clusters
\citep{2018PASJ...70S..20O} to identify cluster counterparts of the
peaks.
We found that 107 out of 124 merged peaks have matched CAMIRA-HSC
clusters within 5 arcmin from the peak positions.
Among the 107 matched peaks, 25 peaks have multiple matches, which might
be generated by line-of-sight projections of multiple clusters.
Among the remaining 82 peaks matching with a single CAMIRA-HSC cluster,
64 peaks have CAMIRA-HSC cluster counterparts within 2 arcmin from peak
positions. 
We confirmed by visual inspection of HSC images and found that, for all 64
peaks, there exist good correlations between weak lensing mass
over-densities and galaxy concentrations.
We thus defined those peaks as the sample of {\it weak lensing secure
  clusters}, and used them to examine the dilution effects on our weak
lensing peak finding.

We have paid particular attention
to the dilution effect of cluster member and foreground galaxies on
weak lensing peak $SN$s, and have adopted two means to mitigate its impact,
namely the globally normalized
estimator [equations (\ref{eq:shear2kap_sum}), and
  (\ref{eq:sigma_shape})], and the source galaxy selection with
different $z_{\rm min}$-cuts using the
full probability distribution function of galaxy photo-$z$s.

We have demonstrated, using the simple model of galaxy populations
introduced in Section \ref{sec:global-kap}, that the peak $SN$ defined by
the globally normalized estimators is, to a good approximation, not
affected by the dilution effect of the cluster member galaxies.
This is in marked contract to the locally normalized $SN$ which is
indeed affected by the cluster member galaxies as demonstrated in Appendix
\ref{sec:local_estimator}.
We compared the peak heights of the globally normalized $SN_G$ with
ones of the locally normalized $SN_L$ using our weak lensing mass maps,
and found that for the peak samples with $SN_G \ge 5$, $SN_G$s
are, on average, about 10 percent larger than the corresponding $SN_L$s.

In Section \ref{sec:foreground}, we have examined the dilution effect of
foreground galaxies and have demonstrated the ability of our source
galaxy selection to mitigate it.
We used the probability distribution function of photo-$z$,
and adopted {\it P-cut} method \citep{2014MNRAS.444..147O} to remove
galaxies at $z<z_{\rm min}$ which are foreground galaxies of clusters at
$z_{cl} > z_{\rm min}$.
This galaxy selection has two competing influences on weak lensing peak
$SN$s from clusters:
One is to reduce the dilution effect of foreground galaxies, and the
other is to increase the shape noise level as the $z_{\rm min}$-cut reduces
the number density of source galaxies. 
We examined the expected impact on peak $SN$ heights from those two
factors using the estimated redshift distribution of the source samples,
and found that for high/low-$z$ clusters
($z \gtsim 0.3$/$z \lesssim 0.2$), the former/latter is 
more effective than the latter/former, leading to a gain/decline in peak
$SN$s with increasing $z_{\rm min}$.

We examined the actual dependence of peak $SN$s on the source selection
using the weak lensing secure clusters.
We measured the ratios of $SN(z_{\rm min})/SN(z_{\rm  min}=0)$ for four
sub-samples of secure clusters divided based on the cluster redshift
(shown in Figure \ref{sn_zmin_stats}).
We found that the measured results were in reasonable agreement with
the expectations in their amplitudes and in their
increasing trend toward higher-$z$ clusters, except for the systematic 
offset of about $+5$ to $+10$ percent which could be due to the intrinsic
alignment of cluster neighbor galaxies.
From the above findings, along with the fact that the number of merged
peak sample (124 for $SN_{\rm max}\ge 5$) is nearly twice of the numbers
from individual source samples (68--75 for $SN \ge 5$), we conclude that
combining multiple peak samples from source samples with different
$z_{\rm min}$ indeed improve the efficiency of weak lensing cluster
search, especially for high-$z$ clusters.

We have also examined the effect of source galaxy deficiency on weak
lensing peak heights.
The source deficiency was clearly observed in stacked galaxy number density
profiles of secure clusters at cluster central regions for the
cluster sample of $0.1<z_{cl}<0.2$ and for the other samples with
$z_{\rm min}>z_{cl}$ (Figure \ref{gdensprof_zcl}).
This can be due to the masking effect of bright cluster galaxies and/or
the lensing magnification effect.
Using a simple model consisting of spatial profiles of dark matter halo
density and source deficiency, we make predictions for the source
deficiency effect on the peak $SN$. 
We found that for realistic models of source deficiency, a peak $SN$ is
suppressed by a few to $\sim10$ percent.

Since we have focused on the dilution effect, there are some important
tasks/issues related to weak lensing cluster search which have not
been examined in this paper: The three major matters among others are:
\begin{enumerate}
\item The purity of the sample of 124 weak lensing cluster candidates:
For each of 64 weak lensing secure clusters, we have found a good
correlation between weak lensing mass over-density and galaxy
concentration, and have concluded that those weak lensing signals have
a physical relationship with the counterpart CAMIRA-HSC cluster.
The remaining 60 peaks fall into the following three categories:\\
(a) 18 cases: peaks matching with a single
CAMIRA-HSC clusters but
their separations are larger than 2 arcmin. Physical connections between
those weak lensing mass over-densities and clusters are not clear, which
are a subject of a future study.\\
(b) 25 cases: peaks having multiple CAMIRA-HSC clusters within 5
arcmin. In 23 out of the 25 cases, matched clusters of the same peak are
separated in the redshift direction by $\Delta z > 0.1$. 
Thus those peaks are likely affected by line-of-sight
projections of physically unrelated clusters, though detail
investigations of each peak are required to reveal their real nature.\\
(c) 17 cases: Peaks have no matched CAMIRA-HSC cluster, for which 
we searched for possible counterpart clusters in a known cluster
database taken from a compilation by {\tt NASA/IPAC
  Extragalactic Database} (NED\footnote{http://ned.ipac.caltech.edu/}).
The results are presented in Appendix \ref{sec:no_camira}.
In 10 out of 17 peaks, possible counterpart clusters are found (see Table
\ref{table:no_camira}).
In Figure \ref{riz_image}, we show HSC $riz$ composite images of the
remaining 7 peaks (that have no counterpart cluster found), in which good
correlations between the weak lensing mass over-density and galaxy
concentration are seen in some of those systems.
Clearly, the above information is not enough to evaluate the purity of
our sample; further followup studies combining information from other
wavelength data (for example, X-ray and Sunyaev-Zel’dovich effect) are
required. 
\item Weak lensing mass estimate of our cluster candidates: Although
cluster mass derived from weak lensing analysis can add valuable
information to our sample, an accurate determination of
cluster redshift as well as carefully taking account of line-of-sight projections
of uncorrelated objects are required to estimate weak lensing mass accurately.
We have derived weak lensing cluster masses only for weak lensing secure
clusters which have good correlations between the weak lensing mass peak
and galaxy over-density (see Appendix \ref{sec:cluster_mass}).
Since the remaining weak lensing peaks have either multiple CAMIRA-HSC cluster
counterparts or a less correlated/no CAMIRA-HSC cluster counterpart, further
detailed studies of individual systems are needed to derive their cluster
masses, which we leave for a future study. 
\item Masking effect of bright cluster galaxies and lensing
magnification effect on weak lensing peak finding:
As we discussed in the above, we have seen an observational indication of
those effects as the deficiency of source galaxies at cluster central
regions, and have examined their impact on the peak
$SN$ in Section~\ref{sec:defficiency_effect}.
Since those effects are unavoidable in weak lensing cluster search, a
further detail study of those effects is important for cosmological
applications of weak lensing selected clusters.
It is, however, beyond the scope of this
paper, and we leave it for a future study.
\end{enumerate}

Weak lensing mass maps contain a wealth of cosmological information beyond
those obtained by analyses of the cosmic shear power spectrum or
two-point correlation function \citep[see, for example,
][]{2010MNRAS.402.1049D,2011PhRvD..84d3529Y,2013PhRvD..88l3002P,2017MNRAS.466.2402S}. 
However, in this study, we showed that if a source galaxy sample is
selected by, for example, a simple magnitude-cut, the dilution effects
may alter $SN$s of high peaks in a non-negligible amount,
and thus may modify statistical properties of weak lensing mass maps.
Therefore, when one uses weak lensing mass
maps for a cosmological application,
the dilution and the source deficiency effects must be taken into account.
We note that the effects are dependent on the source sample that one
takes, and thus should be examined on a case-by-case basis.
At the same time, developing source galaxy selection methods that can
mitigate the dilution
effects is another important subject in that research field.

%
\begin{ack}
We would like to thank Masamune Oguri for useful comments on an earlier
manuscript, and Satoshi Miyazaki for useful discussions. 
We would like to thank Nick Kaiser for making the software {\tt
  imcat} publicly available, and {\tt ds9} developers for {\tt ds9}
publicly available.
We have heavily used those softwares in this study.
We would like to thank HSC data analysis software team for their effort
to develop data processing software suite, and HSC data archive team for
their effort to build and to maintain the HSC data archive system.

This work was supported in part by JSPS KAKENHI
Grant Number JP17K05457.
MS is supported by JSPS Overseas Research Fellowships.

Data analysis were in part carried out on PC cluster at Center for
Computational Astrophysics, National Astronomical Observatory of
Japan. Numerical computations were in part carried out on Cray XC30 and
XC50 at Center for Computational Astrophysics, National Astronomical
Observatory of Japan, and also on Cray XC40 at YITP in Kyoto
University.

The Hyper Suprime-Cam (HSC) collaboration includes the astronomical communities 
of Japan and Taiwan, and Princeton University.  
The HSC instrumentation and software were developed by the National Astronomical 
Observatory of Japan (NAOJ), the Kavli Institute for the Physics and Mathematics 
of the Universe (Kavli IPMU), the University of Tokyo, the High Energy Accelerator 
Research Organization (KEK), the Academia Sinica Institute for Astronomy and 
Astrophysics in Taiwan (ASIAA), and Princeton University.  
Funding was contributed by the FIRST program from Japanese Cabinet Office, 
the Ministry of Education, Culture, Sports, Science and Technology (MEXT), 
the Japan Society for the Promotion of Science (JSPS),  Japan Science
and Technology Agency (JST),  the Toray Science  Foundation, NAOJ, Kavli
IPMU, KEK, ASIAA, and Princeton University.
This paper makes use of software developed for the Large Synoptic Survey
Telescope. We thank the LSST Project for making their code available as
free software at \url{http://dm.lsst.org}

The Pan-STARRS1 Surveys (PS1) have been made possible through 
contributions of the Institute for Astronomy, the University of Hawaii, 
the Pan-STARRS Project Office, the Max-Planck Society and its 
participating institutes, the Max Planck Institute for Astronomy, 
Heidelberg and the Max Planck Institute for Extraterrestrial Physics, 
Garching, The Johns Hopkins University, Durham University, the 
University of Edinburgh, Queen's University Belfast, the 
Harvard-Smithsonian Center for Astrophysics, the Las Cumbres Observatory 
Global Telescope Network Incorporated, the National Central University 
of Taiwan, the Space Telescope Science Institute, the National 
Aeronautics and Space Administration under Grant No. NNX08AR22G issued 
through the Planetary Science Division of the NASA Science Mission 
Directorate, the National Science Foundation under Grant No. AST-1238877, 
the University of Maryland, and Eotvos Lorand
University (ELTE) and the Los Alamos National Laboratory.

Based in part on data collected at the Subaru Telescope and retrieved
from the HSC data archive system, which is operated by Subaru Telescope
and Astronomy Data Center at National Astronomical Observatory of Japan.
\end{ack}




\appendix

%
%
\section{Cross-matching with selected cluster catalogs}
\label{sec:cross_matching}

%
%
\begin{table*}
\caption{Summary of known cluster counterparts of 17 weak lensing merged
  peaks which have no CAMIRA-HSC cluster within 5 arcmin radius from the peak
  positions. 
  A known cluster database taken from a compilation by {\tt NASA/IPAC
  Extragalactic Database} (NED) was used for this counterpart search.
\label{table:no_camira}}
\begin{tabular}{llccl}
\hline
ID & Cluster name & $z_{cl}$ & $\theta_{\rm sep}{}^a$ & Ref \\
{} & {} & {} & [arcmin] & {} \\
\hline
HWL16a-001 & - & - & - & - \\ 
HWL16a-011 & - & - & - & - \\
HWL16a-015 & CFHTLS W1-2593 & 0.30 & 1.4 & {\citet{2011A&A...535A..65D}} \\
{}        & CFHTLS W1-2588 & 0.68 & 4.1 & {\citet{2011A&A...535A..65D}} \\
{}        & CFHT-W CL~J022757.5$-$053537 & 0.32 & 4.9 & \citet{2011ApJ...734...68W} \\
HWL16a-018 & CFHT-W CL~J023111.0$-$0536 & 0.67 & 1.9 & \citet{2011ApJ...734...68W} \\
{}        & CFHTLS W1-2864 & 0.60 & 2.8  & {\citet{2011A&A...535A..65D}} \\
{}        & CFHTLS W1-2588 & 0.68 & 2.9 & {\citet{2011A&A...535A..65D}} \\
{}        & CFHTLS W1-2589 & 1.00 & 3.2 & {\citet{2011A&A...535A..65D}} \\
HWL16a-061 & - & - & - & - \\ 
HWL16a-065 & SDSS CE~J213.904556$-$00.069648 & 0.29 & 1.4 & \citet{2002AJ....123.1807G} \\
{}        & WHL~J141527.6$-$000319 & 0.15 & 1.5 & \citet{2009ApJS..183..197W} \\
{}        & SDSS CE~J213.843536$-$00.001681 & 0.29 & 4.1 & \citet{2002AJ....123.1807G} \\
{}        & SDSS CE~J213.919922$+$00.023597 & 0.33 & 4.9 &  \citet{2002AJ....123.1807G} \\
HWL16a-066 & SDSS CE~J214.633743$-$00.016635 & 0.23 & 3.3 & \citet{2002AJ....123.1807G} \\
HWL16a-067 & SDSS CE~J214.788757$+$00.220532 & 0.44 & 1.5 & \citet{2002AJ....123.1807G} \\
HWL16a-074 & GMBCG~J216.67104$-$00.08426 & 0.39 & 1.1 & \citet{2010ApJS..191..254H} \\
{}        & SDSS CE~J216.649841$-$00.110289 & 0.27 & 1.2 &  \citet{2002AJ....123.1807G} \\
{}        & GMBCG~J216.63912$-$00.10900 & 0.25 & 1.4 & \citet{2010ApJS..191..254H} \\
{}        & SDSS CE~J216.635178$-$00.044207 & 0.42 & 3.0 &  \citet{2002AJ....123.1807G} \\
{}        & GMBCG~J216.67010$-$00.03407 & 0.40 & 3.5 & \citet{2010ApJS..191..254H} \\
HWL16a-079 & SDSS CE~J16.867157$-$00.209108 & 0.18 & 0.7 & \citet{2002AJ....123.1807G} \\
{}        & SDSS CE~J216.868240$-$00.171960 & 0.35 & 1.6 &  \citet{2002AJ....123.1807G} \\
{}        & SDSS CE~J216.852905$-$00.249845 & 0.18 & 3.2 &  \citet{2002AJ....123.1807G} \\
HWL16a-089 & SDSS CE~J221.044815$+$00.172764 & 0.30 & 0.2  & \citet{2002AJ....123.1807G} \\
{}        & GMBCG~J221.00862$+$00.12188 & 0.29 & 3.6  & \citet{2010ApJS..191..254H} \\
HWL16a-092 & FAC2011~CL~0061 & 0.62 & 1.2 & \citet{2011MNRAS.417.1402F}\\
{}        & GMBCG~J221.15835$+$00.19581 & 0.43 & 3.1 & \citet{2010ApJS..191..254H} \\
{}        & WHL~J144437.5$+$001402 & 0.31 & 3.7 & \citet{2009ApJS..183..197W} \\
{}        & SDSS CE~J221.230865$+$00.138749 & 0.27 & 4.0 &  \citet{2002AJ....123.1807G} \\
{}        & MaxBCG~J221.20075$+$00.12862 & 0.29 & 4.4 & \citet{2007ApJ...660..239K}\\
{}        & SDSS CE~J221.138031$+$00.233130 & 0.30 & 4.7 &  \citet{2002AJ....123.1807G} \\
HWL16a-096 & - & - & - & - \\
HWL16a-118 & - & - & - & - \\
HWL16a-121 & WHL~J223540.8$+$012906 & 0.058 & 0.3 & \citet{2009ApJS..183..197W} \\
{}        & MCXC~J2235.6$+$0128 & 0.060 & 0.8 & {\citet{2011A&A...534A.109P}} \\
{}        & ABELL 2457 & 0.059 & 1.5 & \citet{1989ApJS...70....1A} \\
HWL16a-122 & - & - & - & - \\
HWL16a-123 & - & - & - & - \\
\hline
\end{tabular}
\begin{tabnote}
  {$^a$ The angular separation between the weak lensing peak position
    and the cluster position.}
\end{tabnote}
\end{table*}

%
%
\begin{figure*}
    \begin{tabular}{c}
      \begin{minipage}{0.5\hsize}
          \includegraphics[width=66mm]{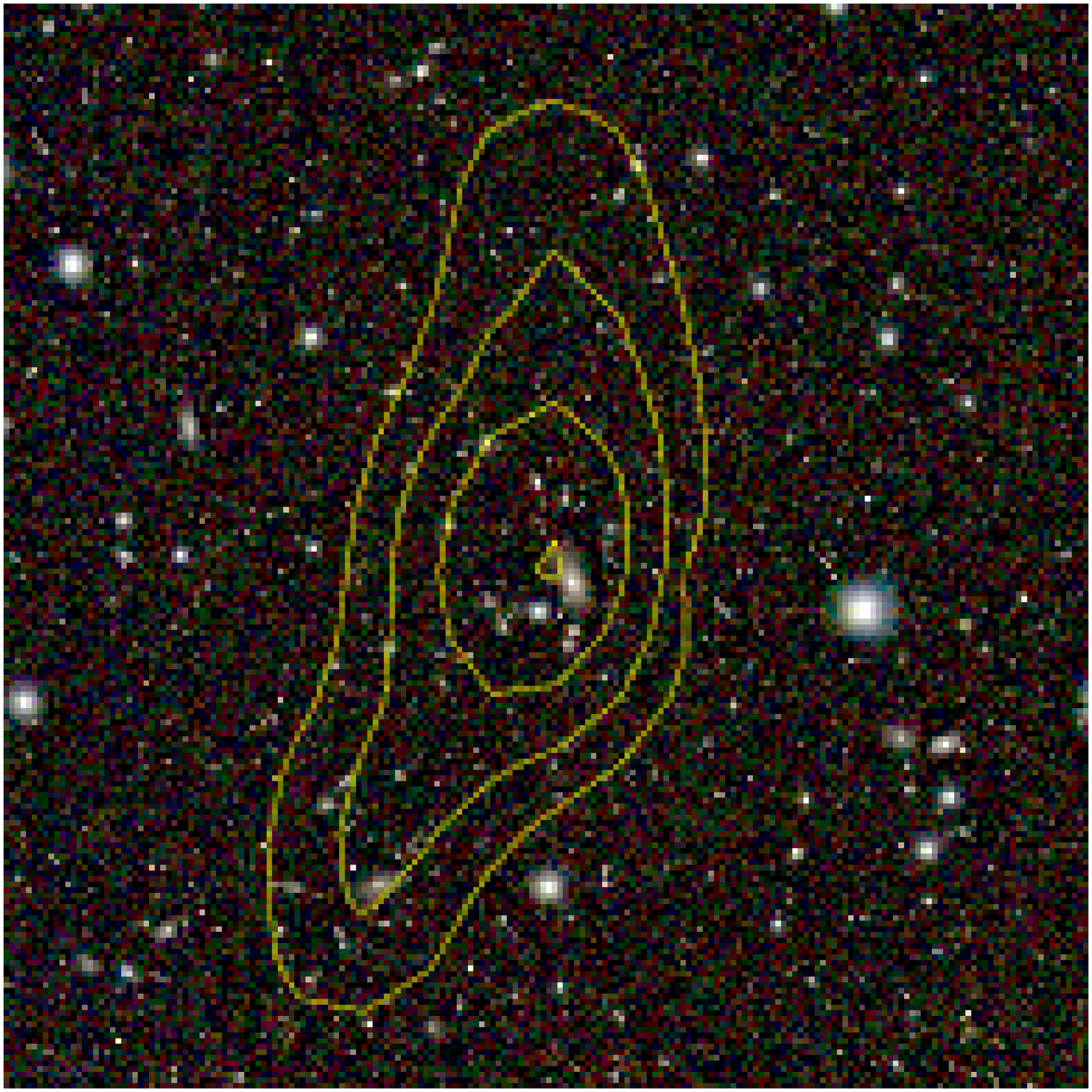}\\
          (a) HWL16a-001
      \end{minipage}
      \begin{minipage}{0.5\hsize}
          \includegraphics[width=66mm]{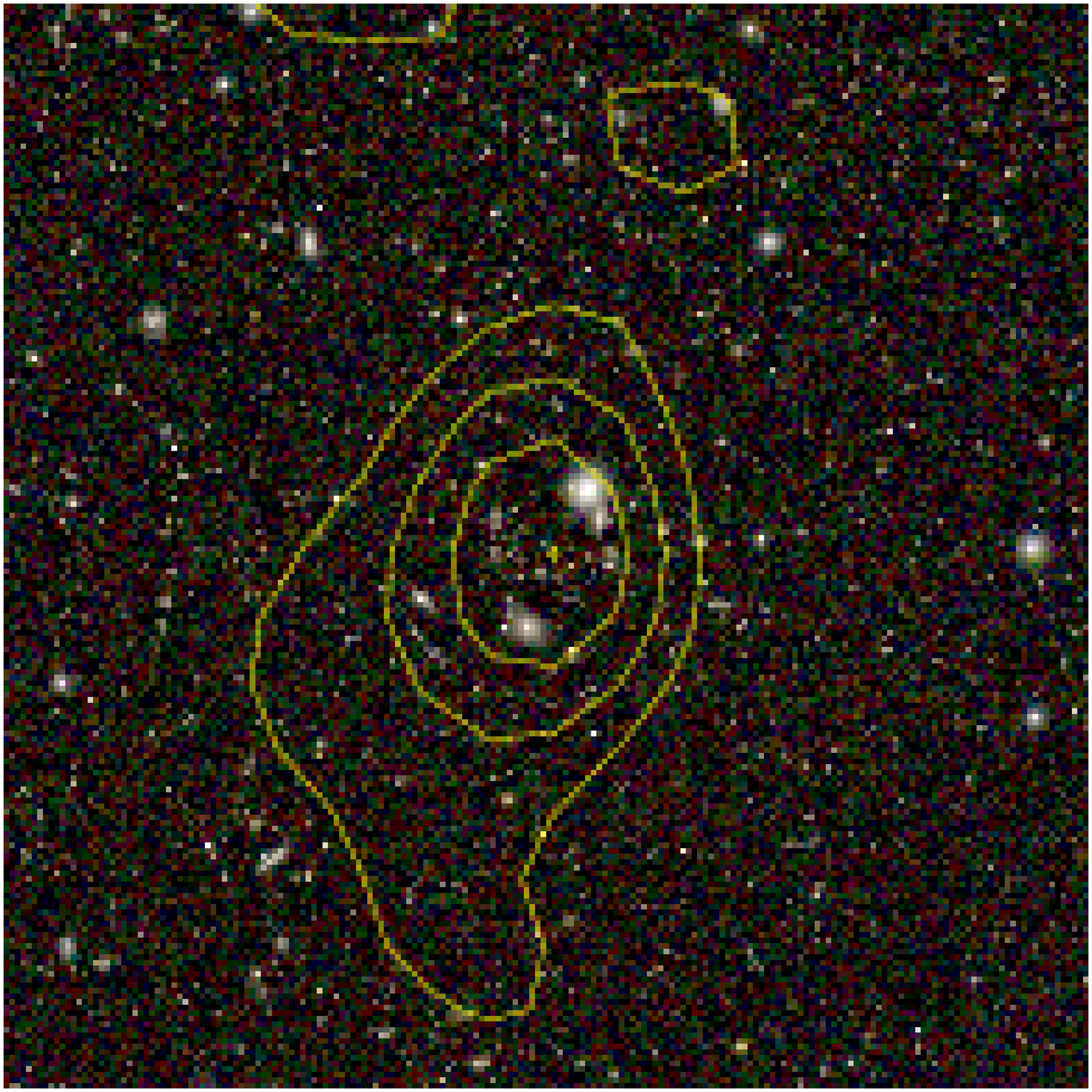}\\
          (b) HWL16a-011
      \end{minipage}\\
      \vspace{0mm}\\
      \begin{minipage}{0.5\hsize}
          \includegraphics[width=66mm]{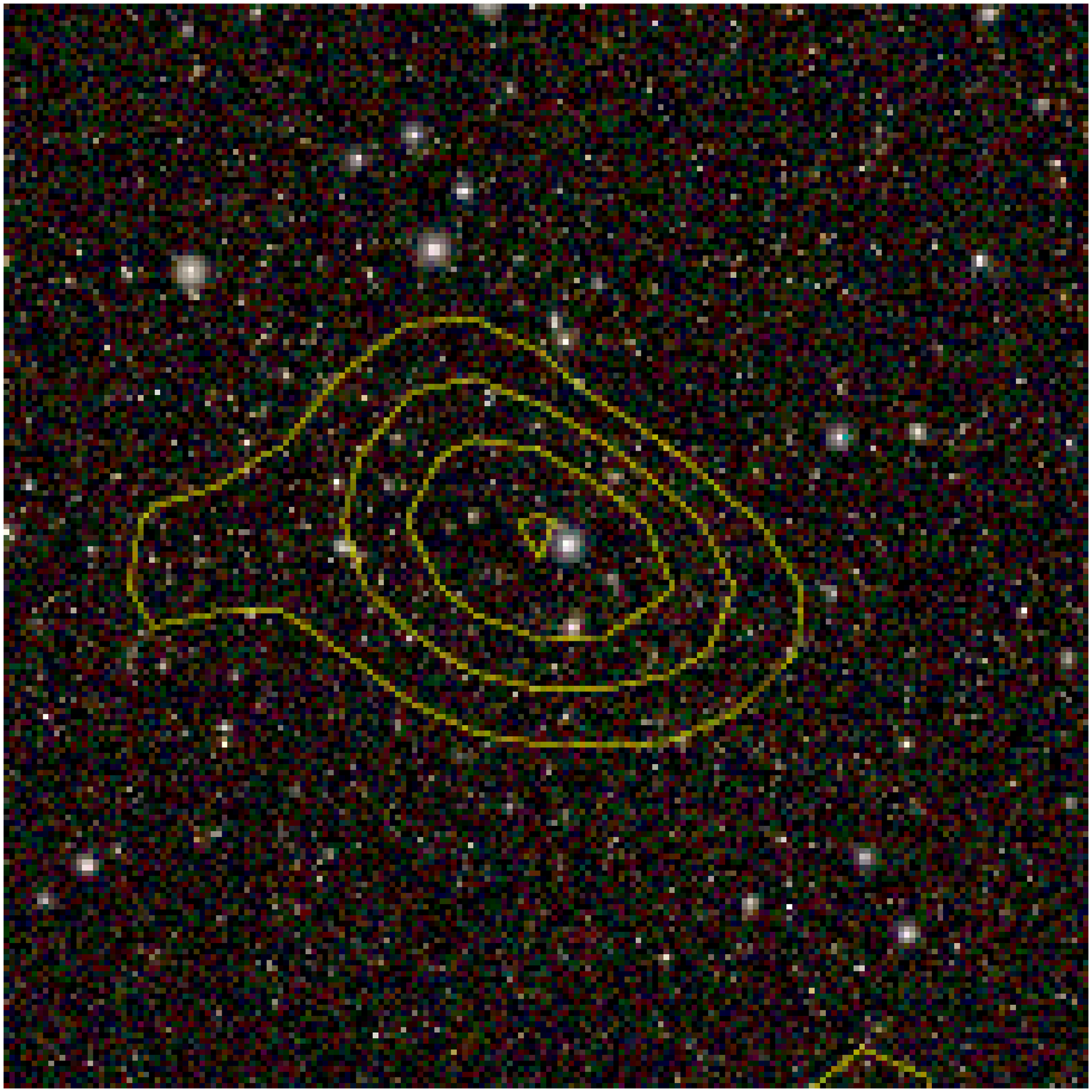}\\
          (c) HWL16a-061
      \end{minipage}
      \begin{minipage}{0.5\hsize}
          \includegraphics[width=66mm]{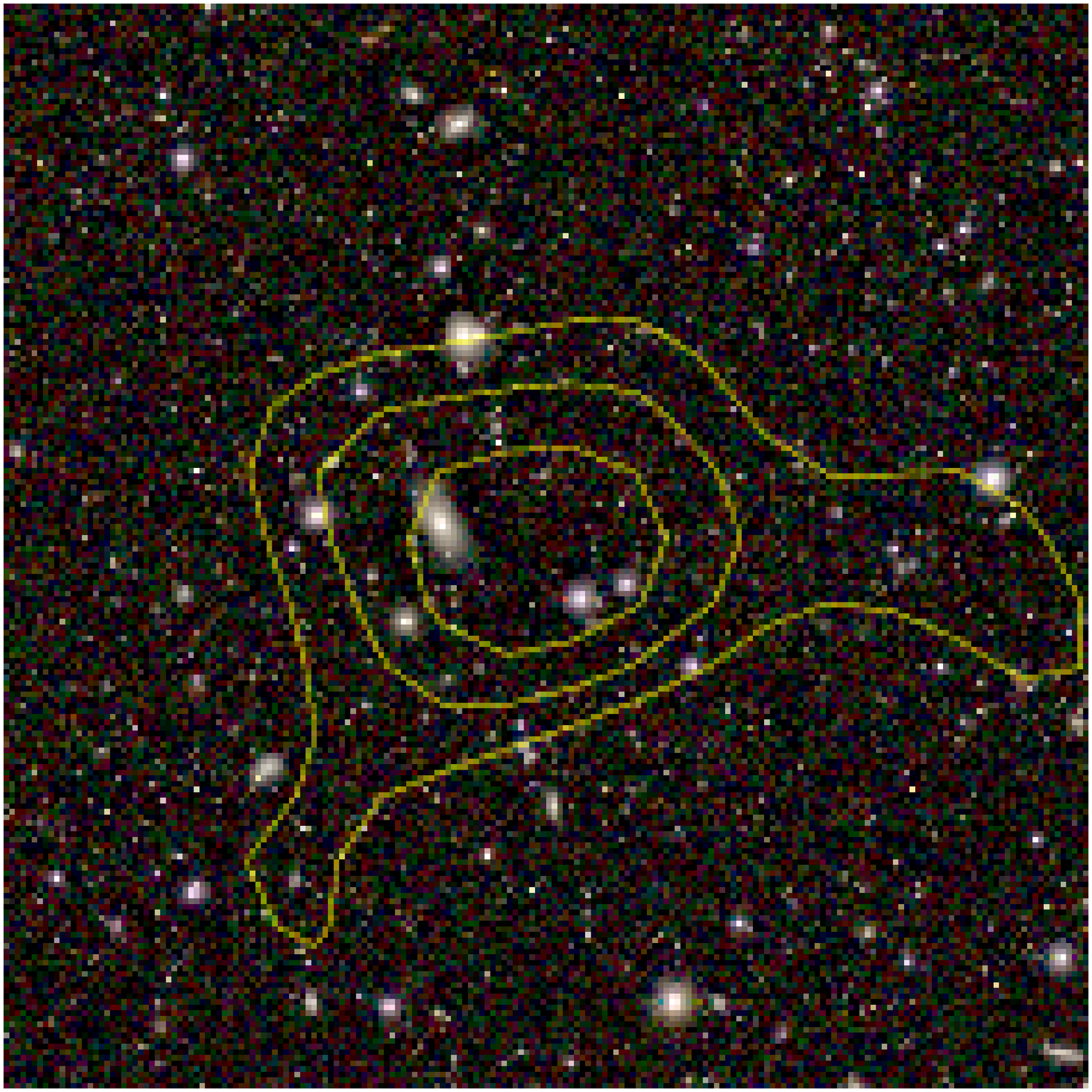}\\
          (d) HWL16a-096
      \end{minipage}\\
      \vspace{0mm}\\
      \begin{minipage}{0.5\hsize}
          \includegraphics[width=66mm]{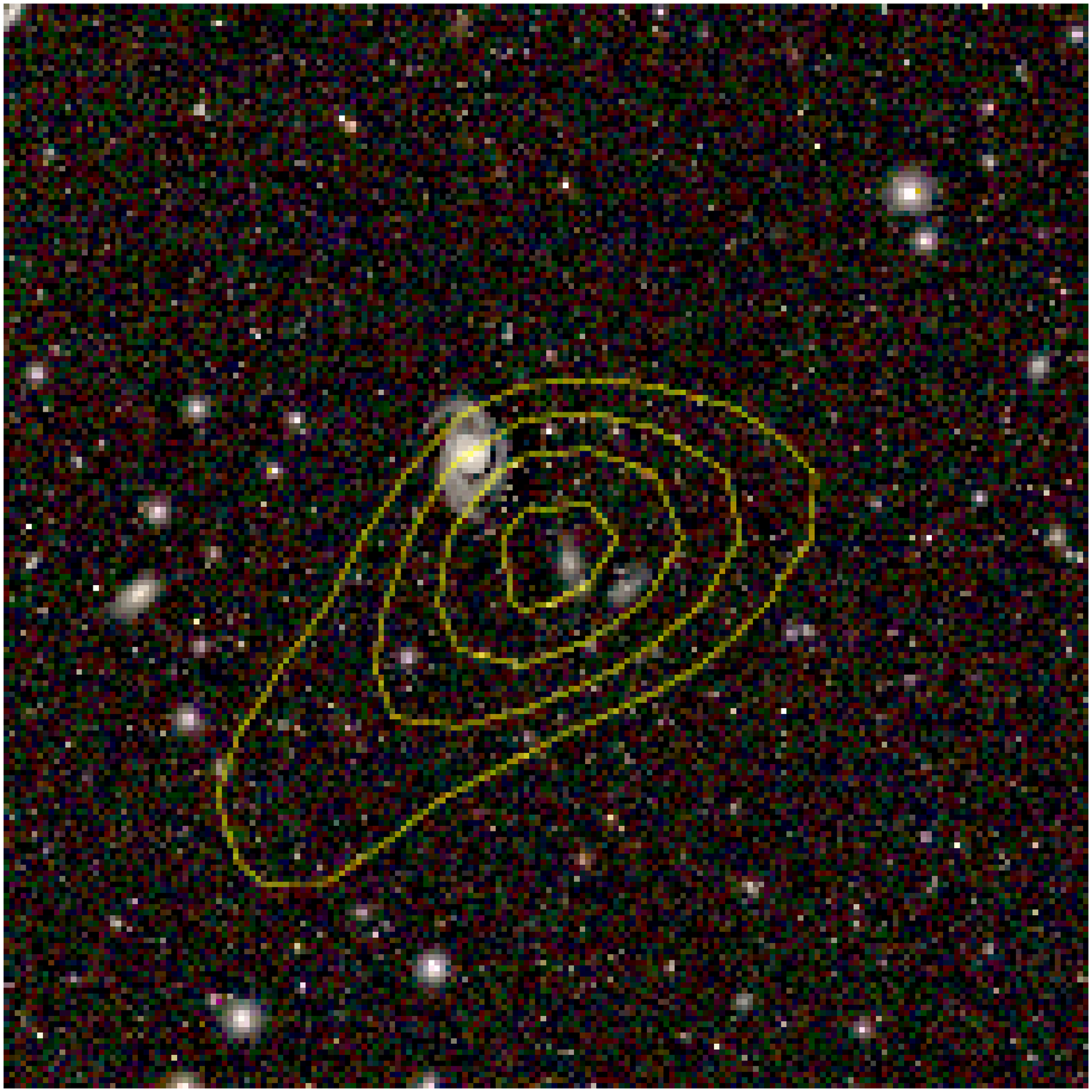}\\
          (e) HWL16a-118
      \end{minipage}
      \begin{minipage}{0.5\hsize}
          \includegraphics[width=66mm]{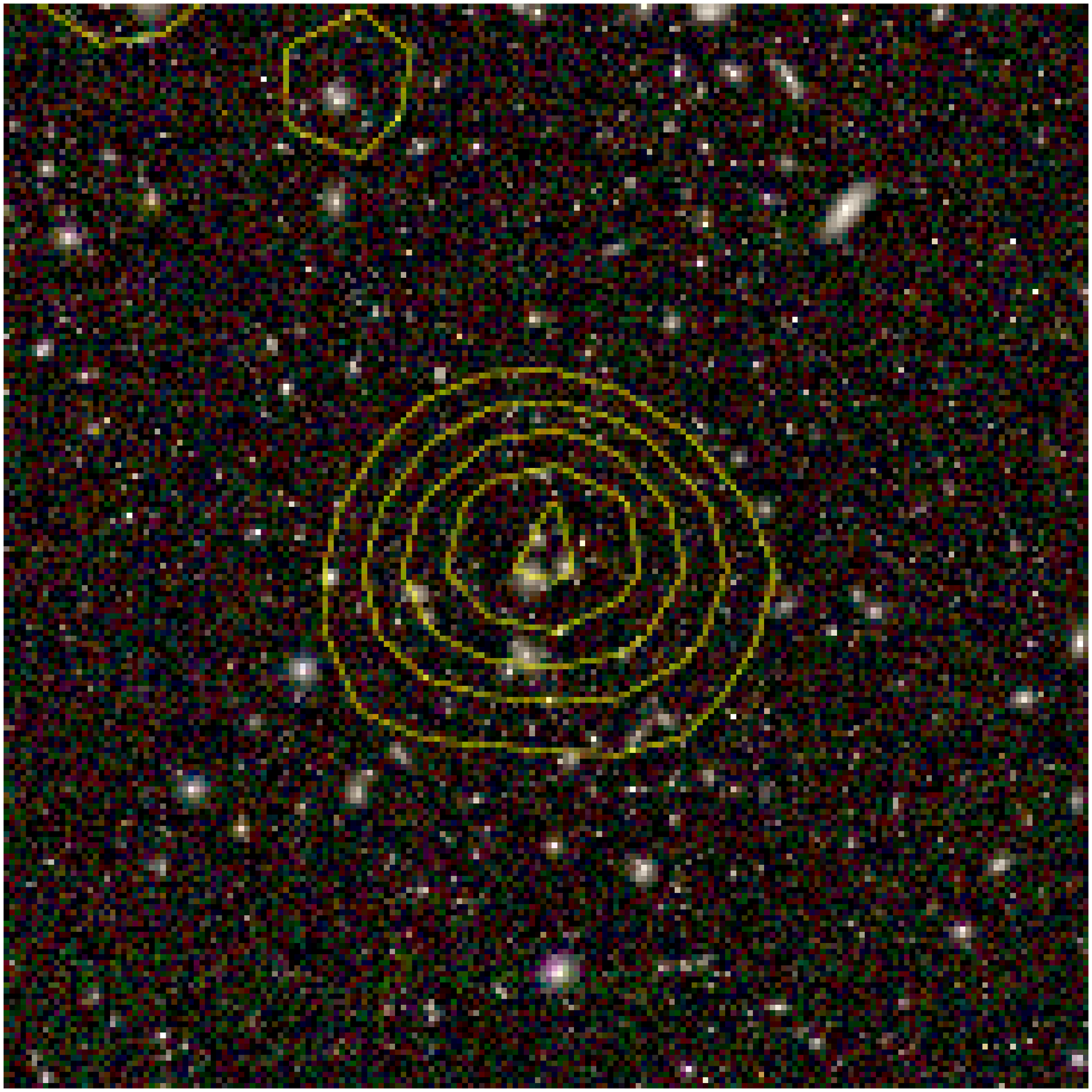}\\
          (f) HWL16a-122
      \end{minipage}\\
      \vspace{0mm}\\
    \end{tabular}
  \caption{HSC $riz$-band composite images with a side length of 10
    arcmin.
    The yellow contour shows the weak lensing $SN$, the contour lines
    start from $SN=2$ with the interval of 1.
    The first 7 panels [(a)--(g)] show weak lensing peaks having no
    CAMIRA-HSC counterpart being matched within 5 arcmin radius from the
    peak position.
    The last 7 panels [(h)--(n)] show systems of neighboring peaks
    having a common CAMIRA-HSC counterpart. Positions of
    CAMIRA-HSC cluster 
    \citep[based on HSC S16A data with updated 
  star mask,][]{2018PASJ...70S..20O} and XXL clusters
  \citep{2018A&A...620A...5A} are marked with plus symbols.
  \label{riz_image}}
\end{figure*}

%
%
\addtocounter{figure}{-1}
\begin{figure*}
  \begin{center}
    \begin{tabular}{c}
      \begin{minipage}{0.5\hsize}
          \includegraphics[width=66mm]{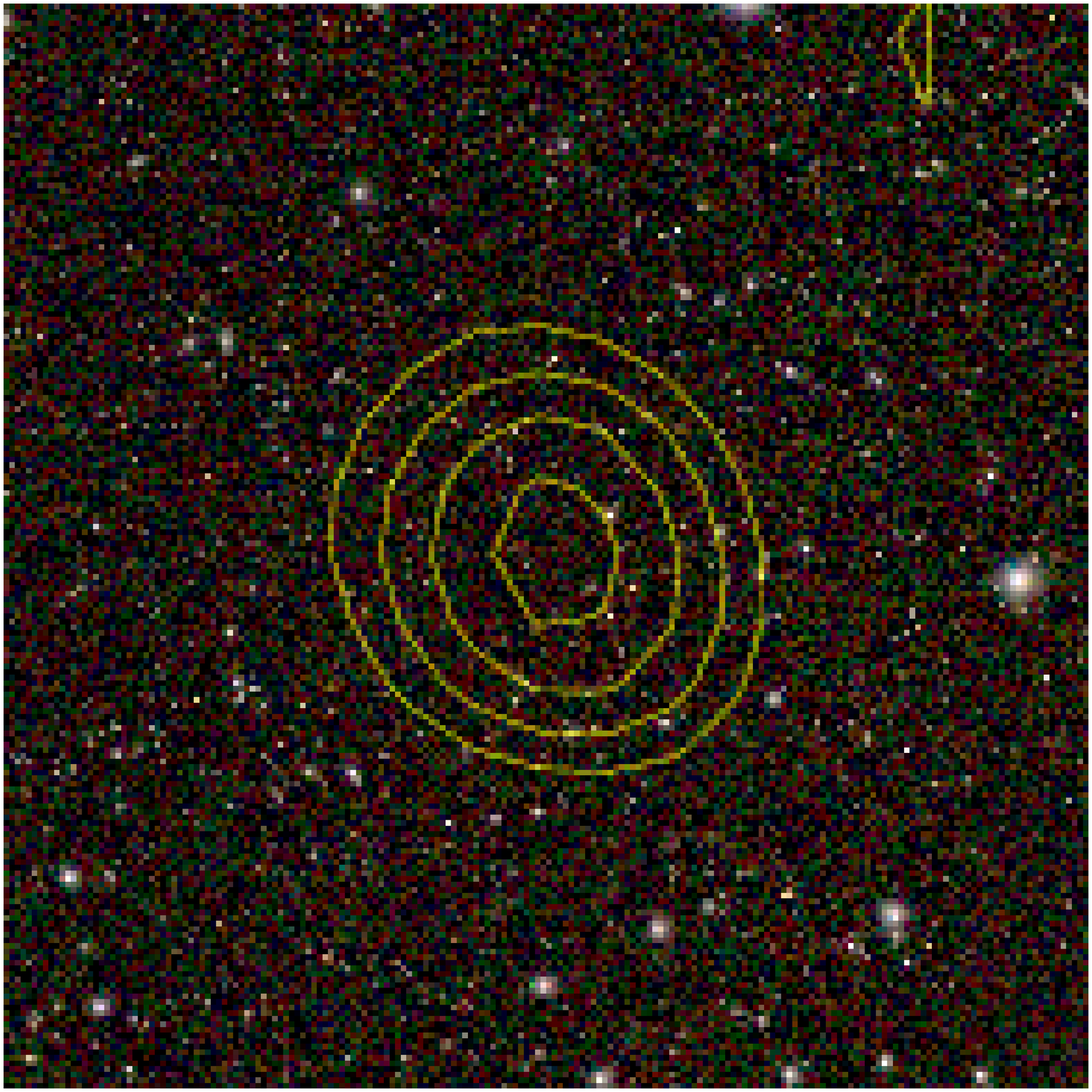}\\
          (g) HWL16a-123
      \end{minipage}
      \begin{minipage}{0.5\hsize}
          \includegraphics[width=66mm]{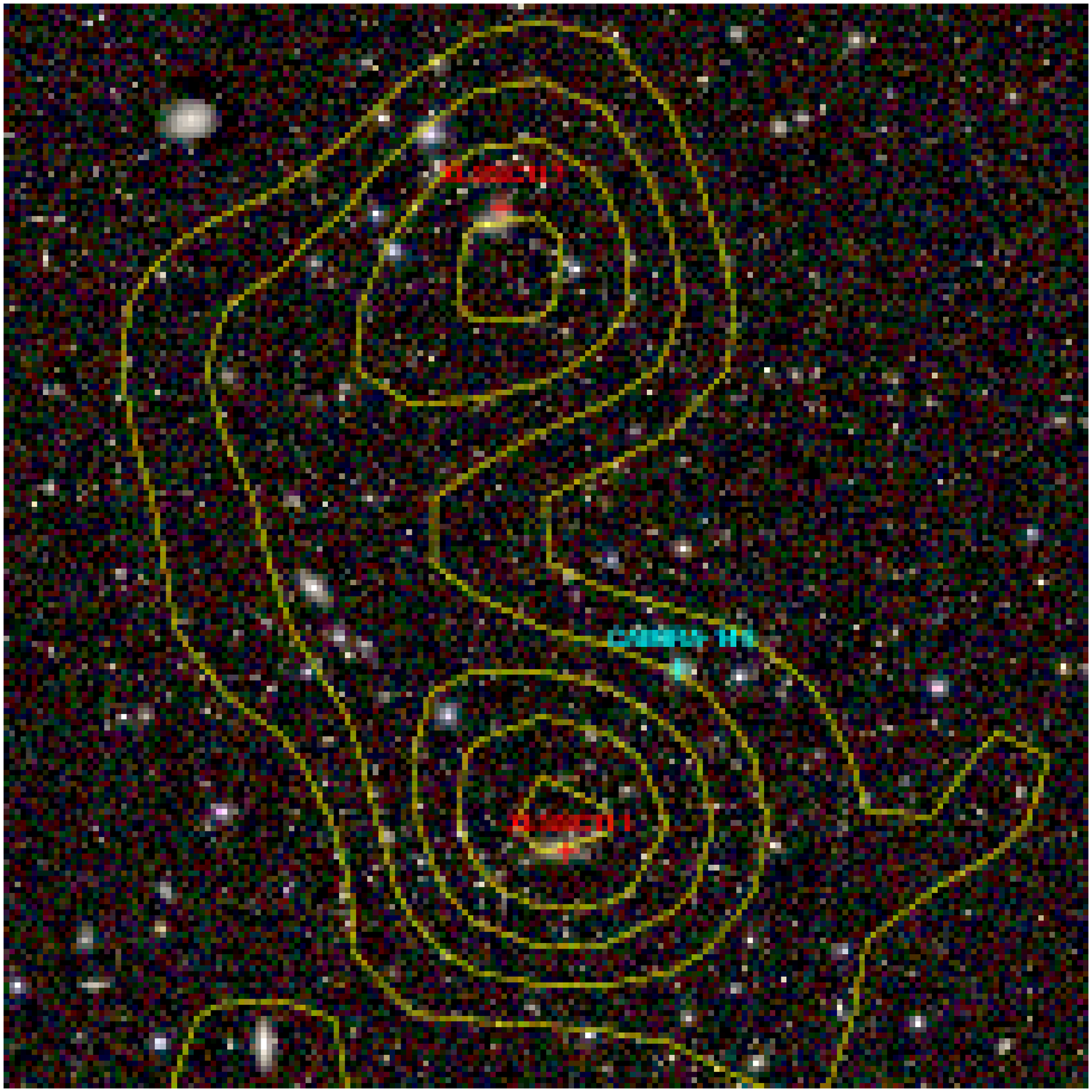}\\
          (h) HWL16a-007 (lower peak), and 008 (upper peak)
      \end{minipage}\\
      \vspace{0mm}\\
      \begin{minipage}{0.5\hsize}
          \includegraphics[width=66mm]{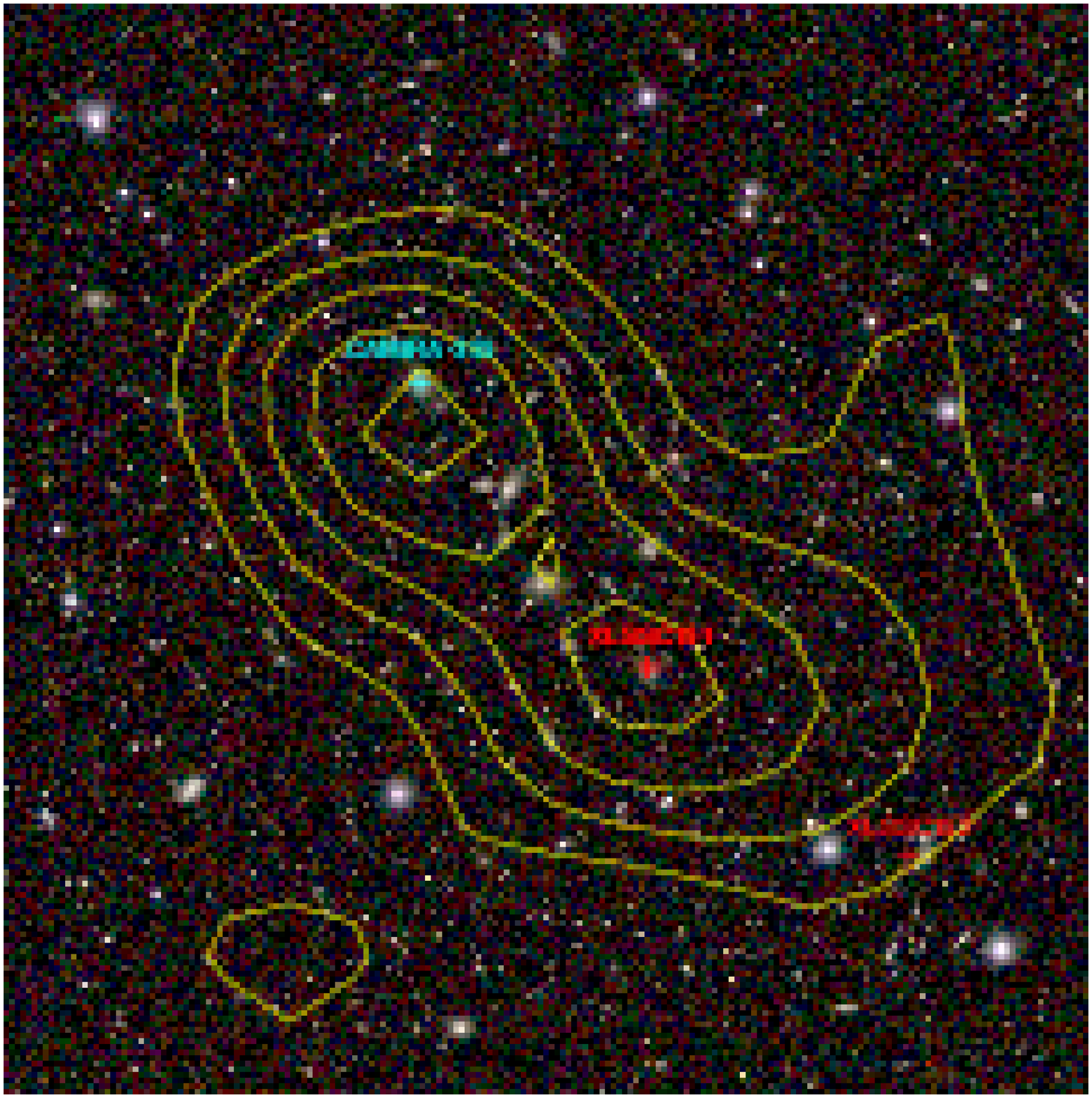}\\
          (i) HWL16a-021 (lower peak), and 022 (upper peak)
      \end{minipage}
      \begin{minipage}{0.5\hsize}
          \includegraphics[width=66mm]{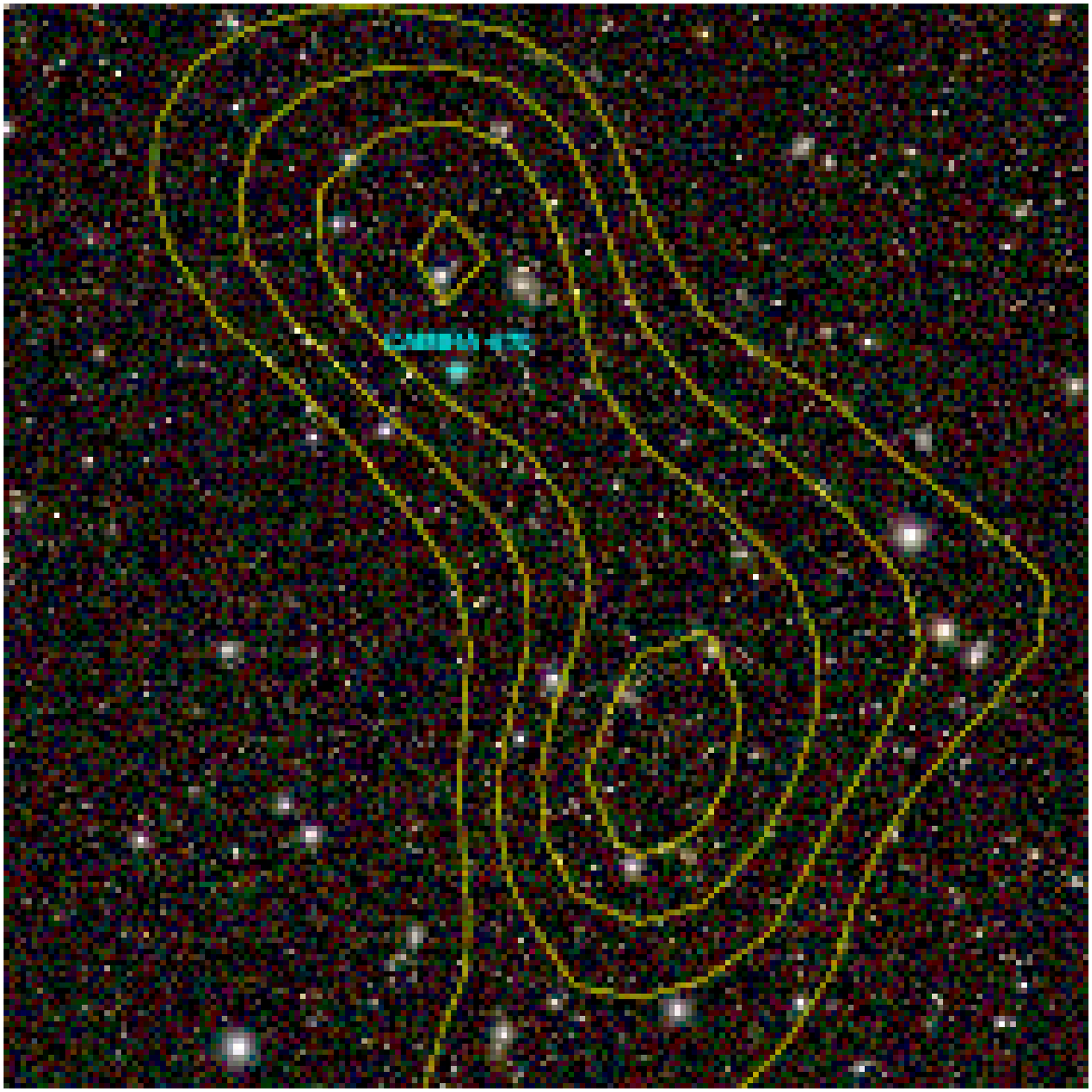}\\
          (j) HWL16a-044 (lower peak), and 045 (upper peak)
      \end{minipage}\\
      \vspace{0mm}\\
      \begin{minipage}{0.5\hsize}
          \includegraphics[width=66mm]{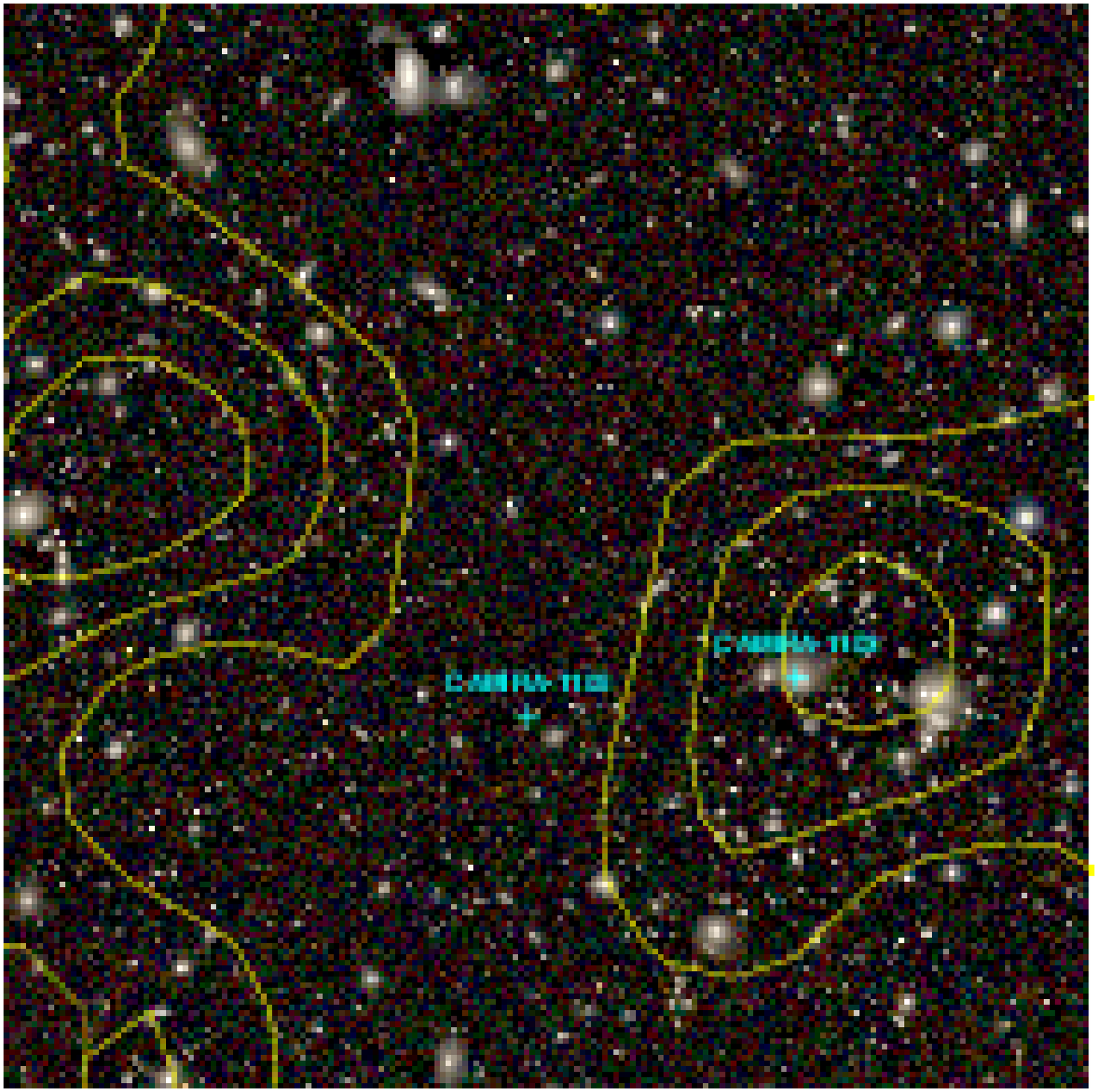}\\
          (k) HWL16a-062 (right-peak), and 063 (left peak)
      \end{minipage}
      \begin{minipage}{0.5\hsize}
          \includegraphics[width=66mm]{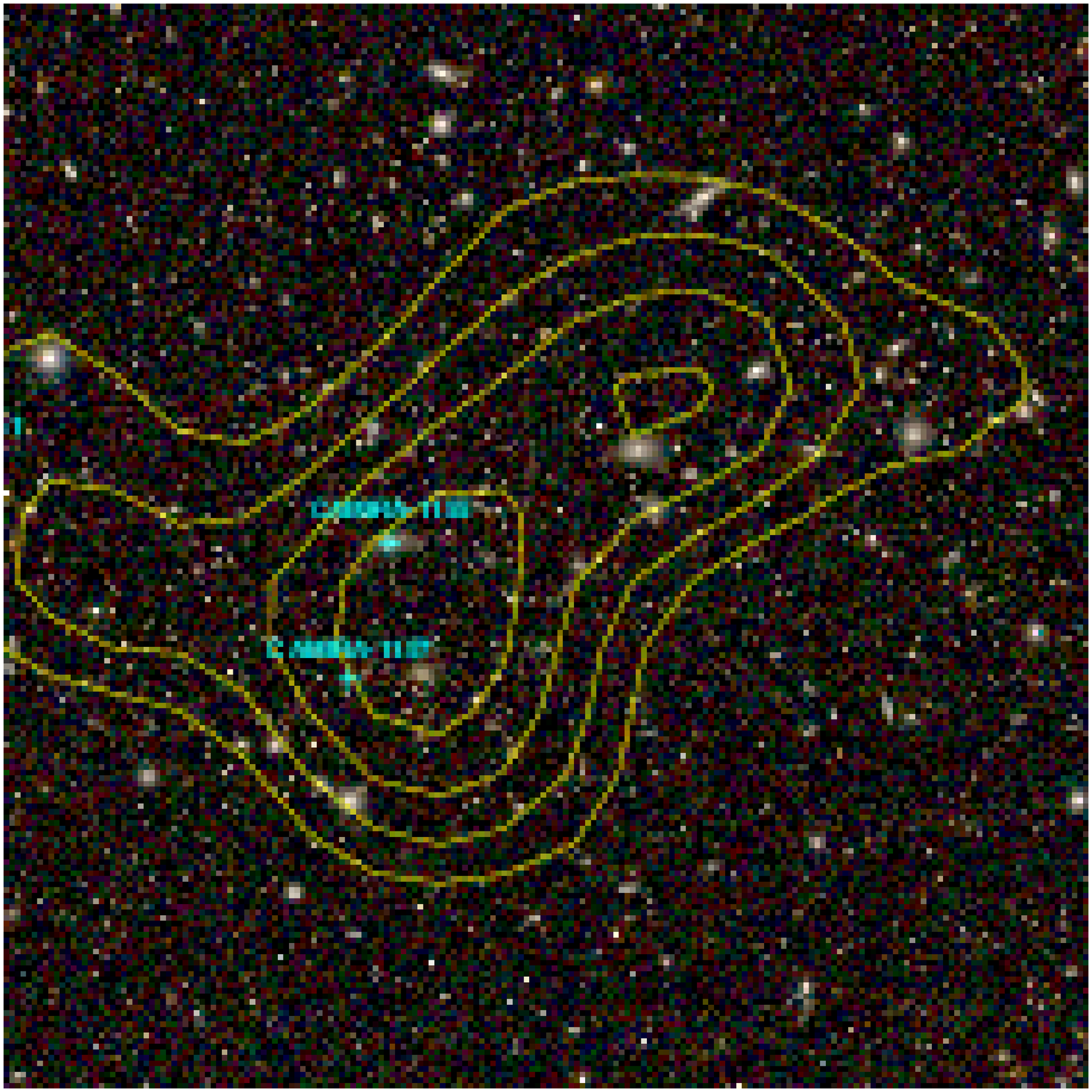}\\
          (l) HWL16a-068 (upper peak), and 069 (lower peak)
      \end{minipage}\\
      \vspace{0mm}\\
    \end{tabular}
  \end{center}
  \caption{(Continued)}
\end{figure*}

%
%
\addtocounter{figure}{-1}
\begin{figure*}
    \begin{tabular}{c}
      \begin{minipage}{0.5\hsize}
          \includegraphics[width=66mm]{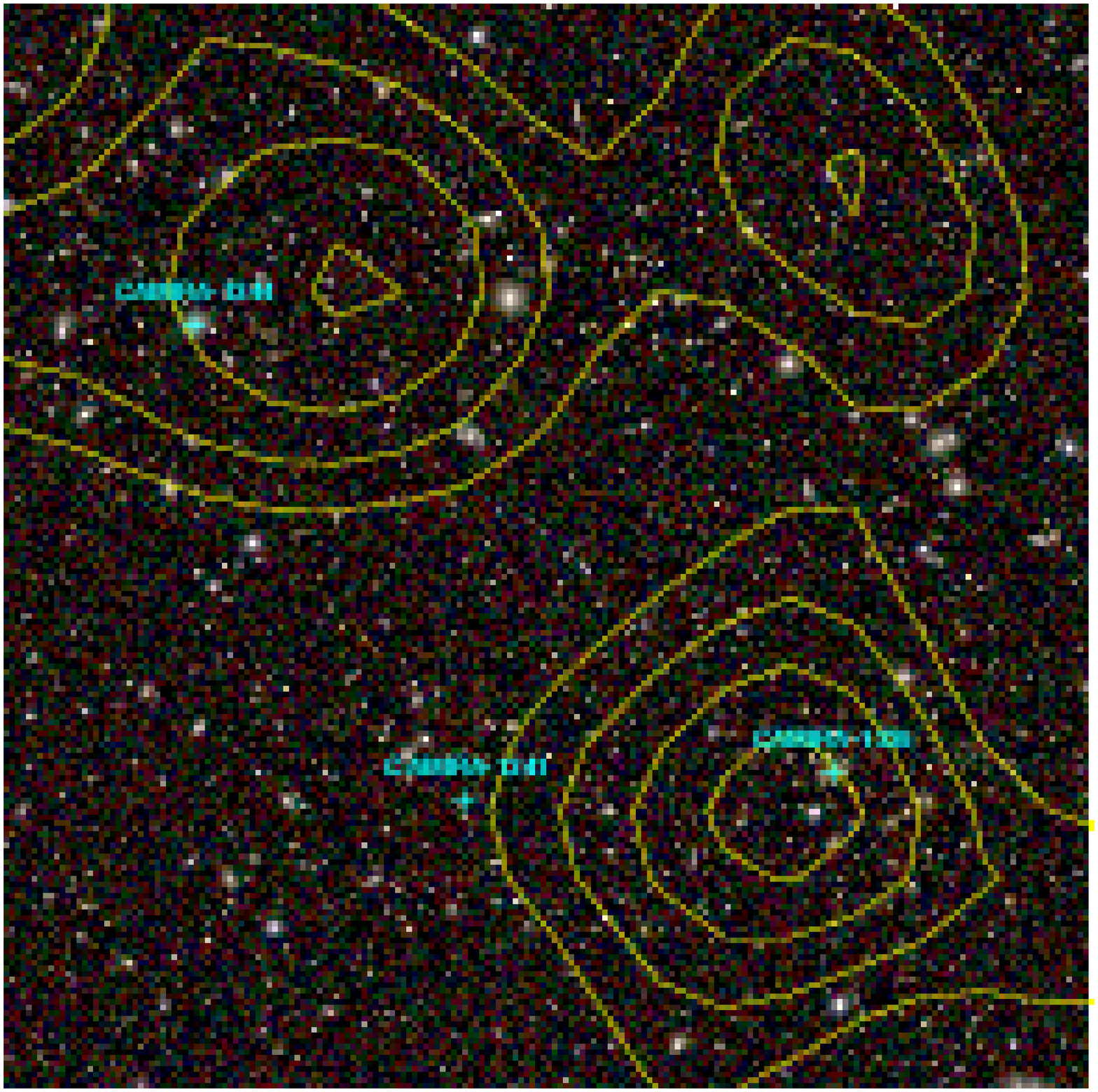}\\
          (m) HWL16a-085 (lower peak), and 086 (upper-left peak)
      \end{minipage}
      \begin{minipage}{0.5\hsize}
          \includegraphics[width=66mm]{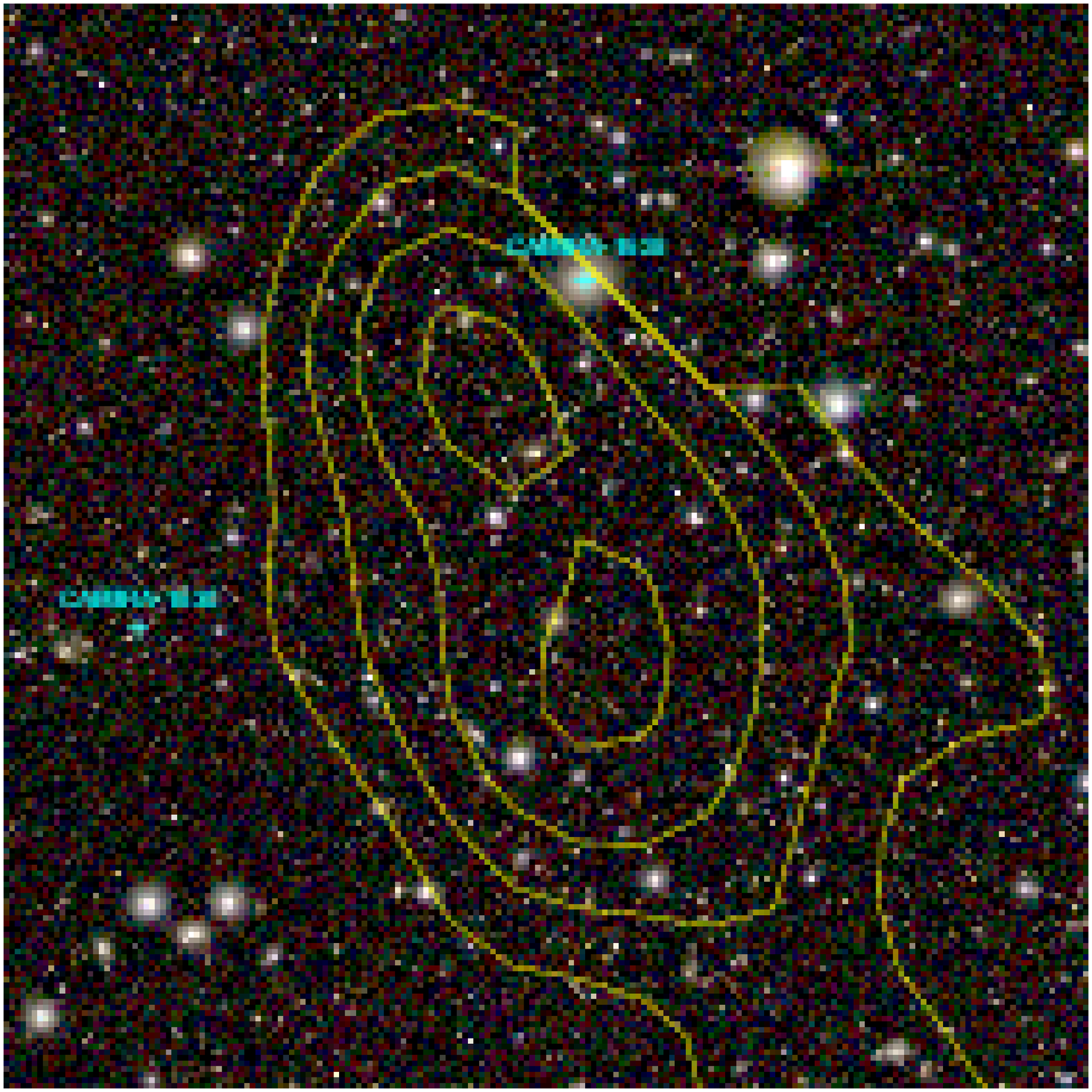}\\
          (n) HWL16a-105 (lower peak), and 106 (upper peak)
      \end{minipage}\\
      \vspace{0mm}\\
    \end{tabular}
  \caption{(Continued)}
\end{figure*}

\subsection{Cluster counterparts of weak lensing peaks with no CAMIRA-HSC
  cluster counterpart}
\label{sec:no_camira}

Among 124 weak lensing merged peaks, 17 peaks have no CAMIRA-HSC cluster
counterparts within a 5 arcmin radius from the peak positions.
However, note that one of them, HWL16a-121, is a known cluster
(Abell 2457) at $z=0.059$ which is outside the redshift coverage of
CAMIRA algorithm \citep{2018PASJ...70S..20O}. 

We search for cluster counterparts of those peaks in a known cluster
database taken from a compilation by {\tt NASA/IPAC
  Extragalactic Database} (NED).
Clusters matched within 5 arcmin radius from the peak positions are
summarized in Table \ref{table:no_camira}. 
In 10 out of 17 peaks, possible counterpart clusters are found.
For the remaining 7 peaks, we present HSC $riz$-band composite images in
Figure \ref{riz_image} [panels (a)--(g)].
In that Figure, we find apparent galaxy concentrations near the weak
lensing peaks of HWL16a-001, 011, 096, 118, and 122.
It follows from this that those peaks are not necessarily false
signals, but undiscovered counterpart clusters may exist.

In summary, combining results of cross-matching with CAMIRA-HSC cluster
catalog and with known cluster database, we have found
possible counterpart clusters for 117 out of 124 weak lensing peaks. 
However, since our matching is based on a simple positional
correlation, some of matches can be chance projections.
Future followup studies of individual peaks on a case-by-case basis are
required to reveal physical connections between weak lensing peaks and
matched clusters.

\subsection{Cross-matching with XXL clusters}
\label{sec:xxl}

\citet{2018A&A...620A...5A} have presented a sample of 365 clusters of
galaxies detected in the XXL Survey, which is a wide-field and deep
X-ray imaging survey conducted with {\it XMM-Newton} \citep{2016A&A...592A...1P}.
The XXL survey consists of two survey fields, each covering $\sim25$ deg$^2$
area, and its north field (XXL-N) largely overlaps with our XMM
field \citep[see Figure 1 of][]{2020ApJ...890..148U}. 
Since the selection function of XXL clusters with respect to the cluster
mass and redshift well covers that of our clusters \citep[see, e.g., Fig
12 of][]{2018PASJ...70S..27M}, the XXL cluster sample provides good
reference data to test the completeness of our weak lensing clusters.

Among 23 weak lensing merged peaks in XMM field (HWL16a-001--023), 14 peaks are located
on the XXL survey footprints \citep[see Figure~1 of ][]{2016A&A...592A...2P}.
11 out of 14 peaks have XXL cluster counterparts (see Table \ref{table:peaklist}).
The peaks with no XXL cluster counterpart are HWL16a-015, 018, and 019,
for which brief descriptions are given below, though future detail
investigations of each peak are required to reveal their real nature:
\begin{itemize}
\item HWL16a-015 matches with XLSSC~074 \citep{2014MNRAS.444.2723C}
  which is not contained in XXL 365 cluster sample
  \citep{2018A&A...620A...5A}.
  There are three known clusters within 5 arcmin from the peak
  position; see Appendix \ref{sec:no_camira} for details.
\item HWL16a-018 has no CAMIRA-HSC cluster counterpart.
  There are four known clusters within 5 arcmin from the peak position;
  see Appendix \ref{sec:no_camira} for details.
\item HWL16a-019 has one CAMIRA-HSC cluster counterpart
  (CAMIRA-ID 388, $z_{cl}=1.011$), but the separation between them is
  5.0 arcmin. Therefore the physical connection between the weak lensing
  peak and CAMIRA-HSC 388 is uncertain. There is one known
  cluster, CFHTLS~W1-2587 ($z_{cl}=0.30$,
  \citealp{2011A&A...535A..65D}), with the angular separation of 1.2 arcmin.
\end{itemize}

We note that both HWL16a-021 and HWL16a-022 match with the
same XXL cluster, XLSSC~151 at $z=0.189$.
Also, both peaks match with
the same CAMIRA-HSC cluster (CAMIRA-ID 355, the estimated redshift of
$z=0.276$).
In fact, those two are a close pair of peaks with their separation of
$\sim 3$ arcmin (see Figure \ref{riz_image} (i)), though they are identified as
two individual peaks under our peak identification criteria (described
in section \ref{sec:peak-finding}).
It is seen in Figure \ref{riz_image} (i) that the X-ray cluster XLSSC~151 is
at the peak position of HWL16a-021, whereas CAMIRA-HSC cluster 355
matches better with HWL16a-022.
Considering the difference in redshifts of those two clusters, it is
likely that the twin peaks of the weak lensing $SN$ map arise from a chance
line-of-sight projection of two physically separated clusters.
If this is the case, HWL16a-022 is another weak lensing peak having
no XXL cluster counterpart. 

\subsection{Cross-matching with weak lensing peaks in \citet{2018PASJ...70S..27M}}
\label{sec:M18}

\citet[][M18 hereafter]{2018PASJ...70S..27M} presented a sample of weak lensing peaks
detected in mass maps constructed from the HSC first-year shape catalog
\citep{2018PASJ...70S..25M} that we also used in this study.
Although their method of weak lensing mass map construction is very
similar to that of this study, differences in source galaxy selection
and criteria of peak identification result in different peak samples.
Their peak sample contains 65 peaks with $SN>4.7$.
We cross-match their peaks with our extended-sample (peaks with $SN\ge
4$ are included) by their peak positions to a
tolerance of $\theta_G=1.5$ arcmin.
Among their 65 peaks, 51 peaks match with our final merged peaks
($SN\ge 5$, see Table \ref{table:peaklist}).
The remaining 14 peaks fall into the following five categories:
\begin{enumerate}
\item {[M18~rank~51]}: A corresponding peak exists in our final merged
  sample (HWL16a-054), but their separation is 2.6 arcmin which
  exceeds the tolerance length.
\item {[M18~rank~49, 55, and 60]}: A matched peak
  exists in our extended-sample with $5 > SN_{\rm max}\ge 4$, but its $SN_{\rm max}$
  is below our threshold. 
\item {[M18~rank~7, 27, 31, 41 and 43]}: A corresponding peak exists in our
  extended-sample with $SN_{\rm max}\ge 4$, but is located in the edge-region. 
\item {[M18~rank~37 and 38]}: No corresponding peak exists in our
  extended-sample. Note that both the peaks are located in our edge-region.
\item {[M18~rank~15, 45 and 46]}: Those peaks are located in our
  masked-regions where we have not performed weak lensing analysis.
\end{enumerate}

In summary, among 65 peaks in M18 sample, 
all the 55 peaks located in our data-regions have counterpart peaks in our
extended-sample (including M18~rank~51--HWL16a-054).

%
%
\section{Systems of neighboring weak lensing peaks}
\label{sec:neighboring_peaks}

In weak lensing mass maps, there are systems of neighboring peaks;
those are either two isolated clusters or one cluster having
a significant substructure or under a merging process.
Distinguishing clusters' dynamical states with only the weak lensing
information is practically impossible.
Nevertheless, we have adopted a simple criterion that  
neighboring peaks with
their separation larger than $\sqrt{2}\times \theta_G \simeq 2.1$ arcmin
are regarded as two isolated peaks (see Section \ref{sec:peak-finding}).
Consequently, there are systems of neighboring
peaks, whose dynamical states are ambiguous, in our final peak catalog
(Table \ref{table:peaklist}).

Here we describe those systems of neighboring peaks having a common
CAMIRA-HSC cluster within 5 arcmin from both the peaks.
There are seven such systems, whose HSC $riz$ composite images are shown in
Figure \ref{riz_image} [panels (h)--(n)].
Below we give short descriptions of them:

\begin{itemize}
\item HWL16a-007 and 008 [Figure \ref{riz_image} (h)]:
  Although both peaks have a common CAMIRA-HSC counterpart (ID-149,
  $z_{cl}=0.287$), they match with the different XXL clusters, XLSSC~111
  ($z=0.299$) and XLSSC~117 ($z=0.300$). Thus those are likely two
  isolated clusters at very close redshifts.
\item HWL16a-021 and 022 [Figure \ref{riz_image} (i)]: 
  See Appendix \ref{sec:xxl}.
\item HWL16a-044 and 045 [Figure \ref{riz_image} (j)]: CAMIRA-HSC cluster (ID-870, $z_{cl}=0.260$) matches better
  with HWL16a-045. Not enough information is available to infer the
  physical connection between those two peaks.
\item HWL16a-062 and 063 [Figure \ref{riz_image} (k)]: CAMIRA-HSC cluster (ID-1103, $z_{cl}=0.144$)
  matches better with HWL16a-062. Another cluster (ID-1105,
  $z_{cl}=1.105$) is probably a non-related high-$z$ cluster as no
  associated lensing signal appears.
  Other than that, no known clusters can be associated with it.
  However, a good correlation between
  HWL16a-063 and an apparent concentration of bright
  galaxies is clearly seen.
\item HWL16a-068 and 069 [Figure \ref{riz_image} (l)]: Both the peaks
  match with CAMIRA-HSC clusters (ID-1155, $z_{cl}=0.322$) and
  (ID-1157, $z_{cl}=0.515$).
  Because of the difference in the cluster redshifts, the lensing
  signals likely originate from line-of-sight projections of the two
  clusters at different redshifts.
\item HWL16a-085 and 086 [Figure \ref{riz_image} (m)]: HWL16a-085 matches
  better with CAMIRA-HSC cluster (ID-1339, $z_{cl}=0.536$), whereas
  HWL16a-086 matches 
  better with CAMIRA-HSC cluster (ID-1344, $z_{cl}=0.149$).  Another cluster (ID-1341,
  $z_{cl}=0.884$) is probably a non-related high-$z$ cluster, as no
  associated lensing signal appears.
  Thus the two peaks originate from two isolated clusters.
\item HWL16a-105 and 106 [Figure \ref{riz_image} (n)]: CAMIRA-HSC
  cluster (ID-1628, $z_{cl}=0.100$) matches better
  with HWL16a-106. No apparent lensing signal associated with another
  cluster (ID-1680, $z_{cl}=0.357$) appears. Not enough information is
  available to infer the physical connection between those two peaks.
\end{itemize}

%
%
\section{Cluster mass estimate}
\label{sec:cluster_mass}

%
%
\begin{figure*}
  \includegraphics[height=152mm,angle=270]{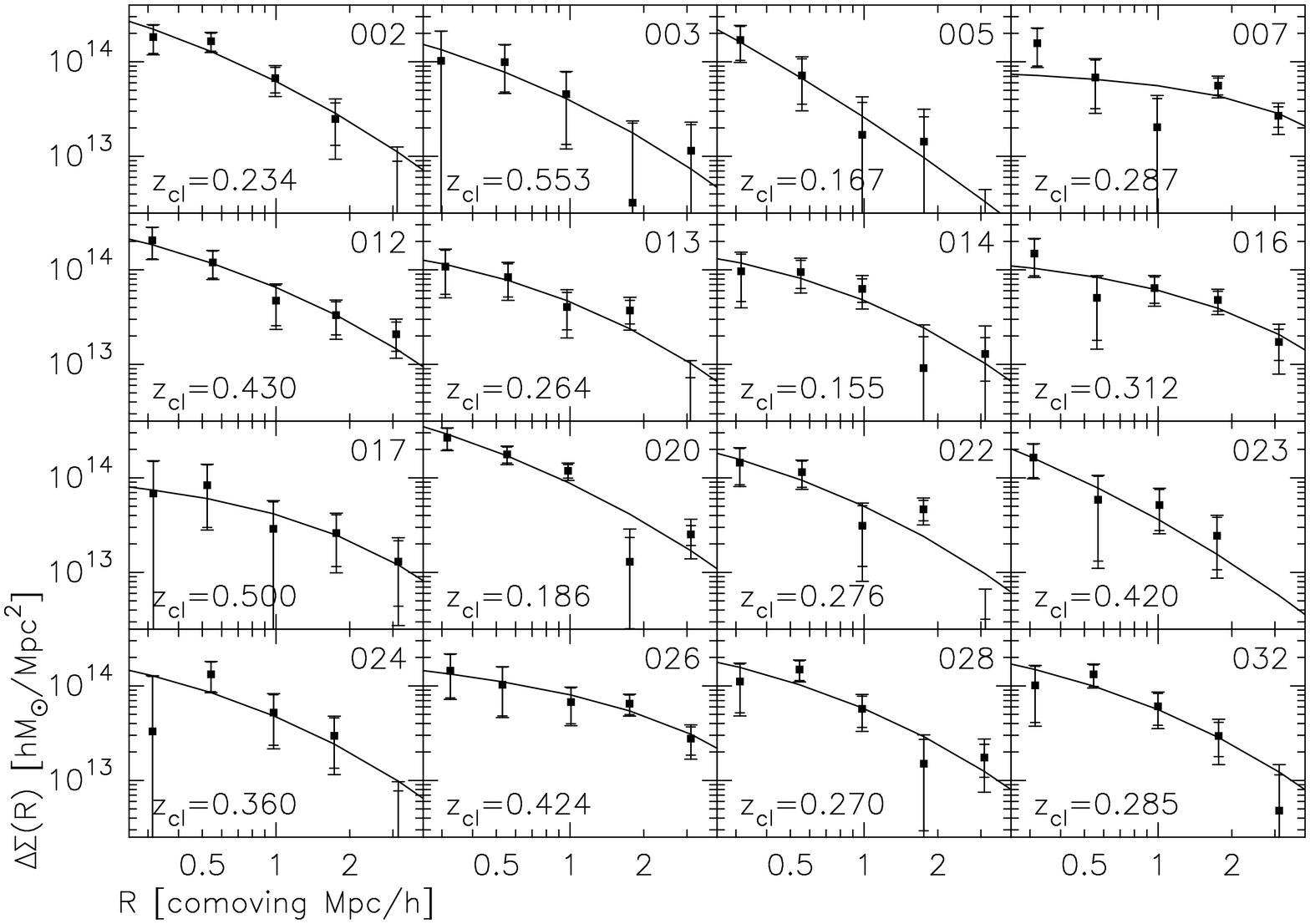}\\
  \includegraphics[height=152mm,angle=270]{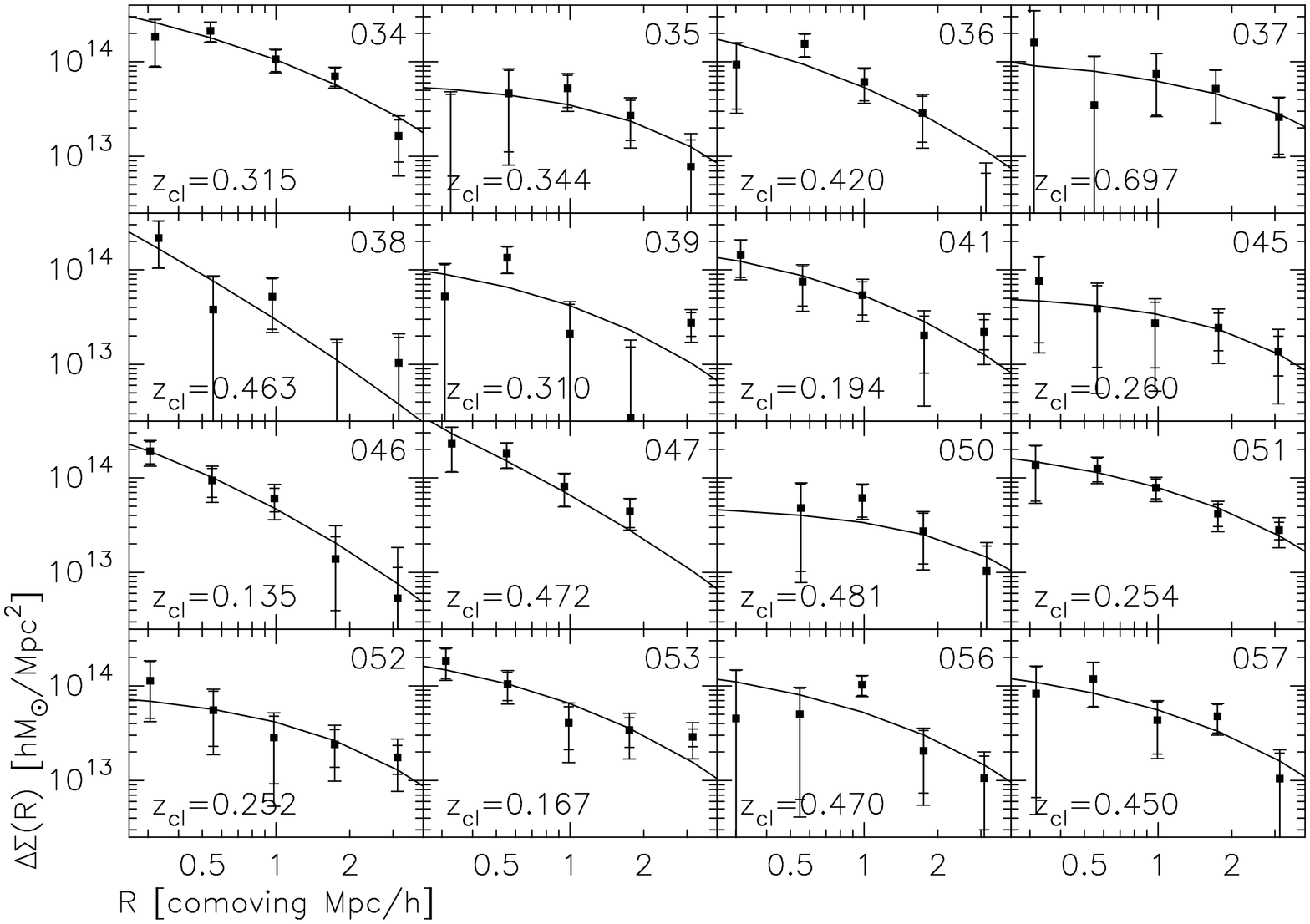}
  \caption{Radial profiles of the excess surface mass density 
    of individual clusters (the weak lensing peak ID is denoted in
    upper-right of each panel). Measurement results are plotted with filled
    squares with error bars; inner error bars show the shape
    noise components ($\sigma_{\rm shape}$) only,
    whereas outer error bars show the total errors being composed of
    shape noise, cosmic shear covariance due to large-scale
    structures, and intrinsic scatter components (the diagonal
    components of ${\rm Cov}^{\rm shape} + {\rm Cov}^{\rm lss} + {\rm
      Cov}^{\rm int}$, see Section
    3.3 of \citealp{2020ApJ...890..148U} and references therein).
    The best-fit NFW model in $M_{200c}$-$c_{200c}$ space is
    shown by the solid line.
    Here all the results are based on the
    cosmological parameters inferred
    from the WMAP 9-year results \citep{2013ApJS..208...19H}.
    \label{fig:sufmass_fit}}
\end{figure*}

%
%
\addtocounter{figure}{-1}
\begin{figure*}
  \includegraphics[height=152mm,angle=270]{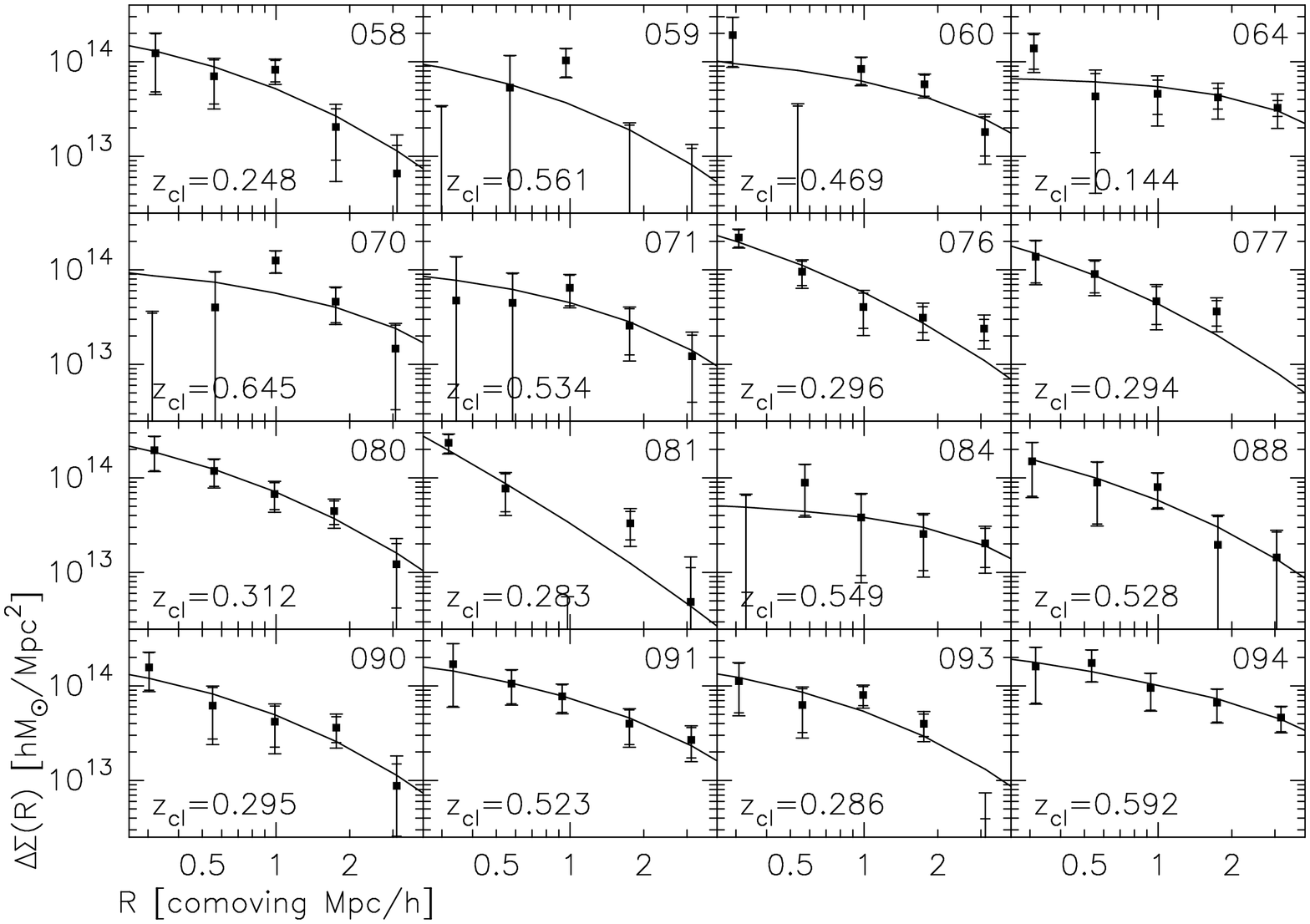}\\
  \includegraphics[height=152mm,angle=270]{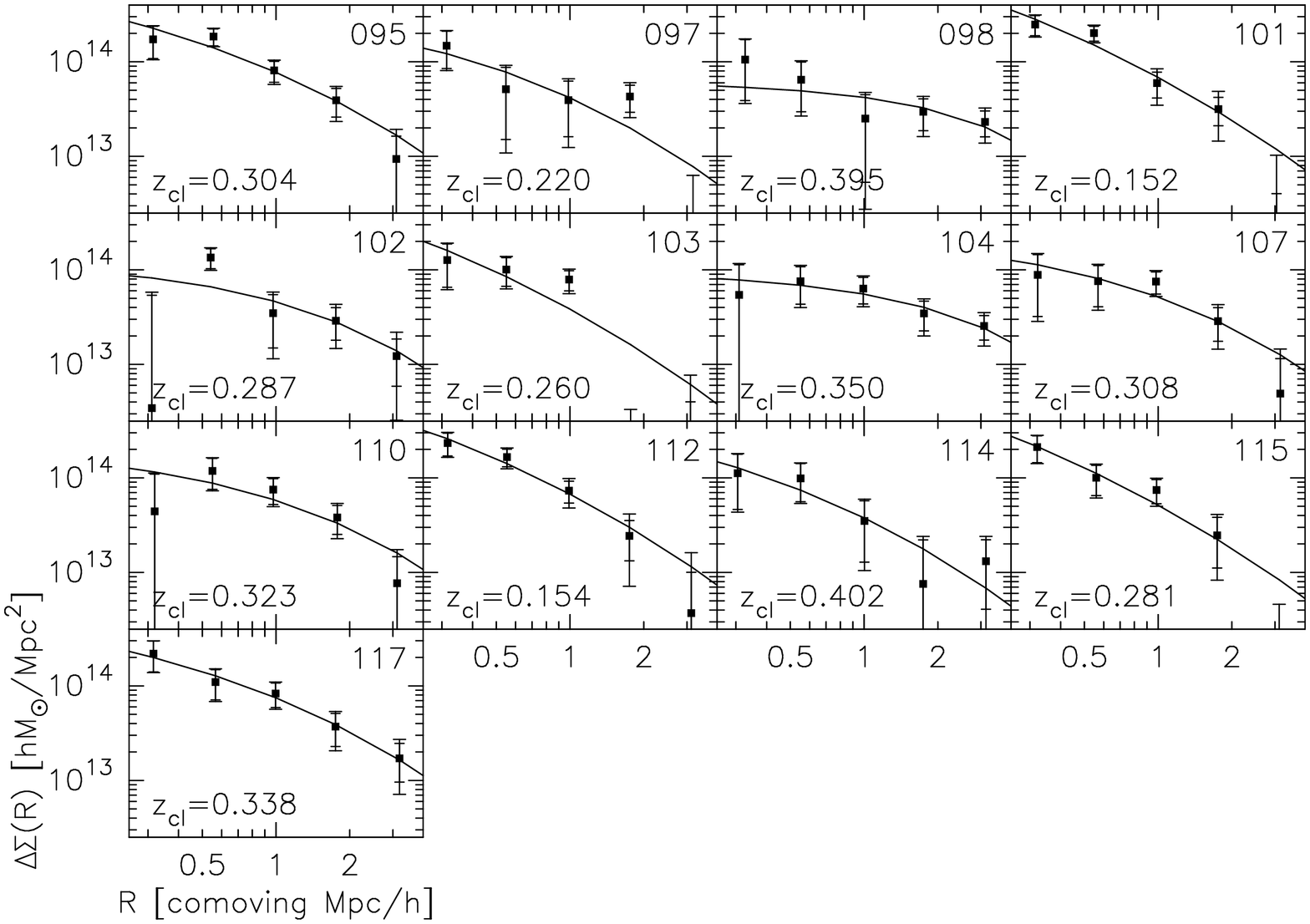}
  \caption{(Continued)}
\end{figure*}

%
%
%
\begin{longtable}{llcccccc}
  \caption{Results of the likelihood analysis of the surface mass density profile of individual clusters
  based on NFW model (see Appendix \ref{sec:cluster_mass}).
  Cluster redshifts ($z_{cl}$) are taken from the estimated redshift of matched
  CAMIRA clusters except for HWL16a-002 for which the redshift of matched XXL cluster (XLSSC 114) is taken
  (marked by $^\ast$). Peaks and 68.3\% confidence intervals of marginalized posterior distributions of $c_{200c}$,
  $M_{200c}$, $M_{500c}$ are summarized.
  The results based on the cosmological parameters from the WMAP 9-year
  \citep[][$\Omega_{\rm m}=0.279$, $\Omega_{\rm b}=0.046$, $\Omega_\Lambda=0.721$, $n_s=0.97$,
  $\sigma_8=0.82$, and $h=0.7$]{2013ApJS..208...19H},
  and Planck 2018 results 
  \citep[][$\Omega_{\rm m}=0.32$, $\Omega_{\rm b}=0.049$, $\Omega_\Lambda=0.68$, $n_s=0.96$,
  $\sigma_8=0.83$, and $h=0.67$]{2018arXiv180706209P} are presented.
  "N/A" in the results of $c_{200c}$ means either the upper/lower bound of 68.3\% confidence
  interval or the minimum of the marginalized log-likelihood function is not enclosed within
  the parameter range of $0.01\le c_\Delta\le30$.
  \label{table:clustermass}}
  \hline
  {} & {} & \multicolumn{3}{l}{WMAP9 cosmology} & \multicolumn{3}{l}{Planck2018 cosmology} \\
  ID & $z_{cl}$ & $c_{200c}$ & $M_{200c}$ & $M_{500c}$ & $c_{200c}$ & $M_{200c}$ & $M_{500c}$ \\
    &          &           & $[10^{14}h^{-1}M_{\odot}]$ & $[10^{14}h^{-1}M_{\odot}]$ &           & $[10^{14}h^{-1}M_{\odot}]$ & $[10^{14}h^{-1}M_{\odot}]$ \\
  \hline
  \endhead
HWL16a-002 & 0.234$^\ast$ & $5.34_{-2.23}^{+5.19}$ & $1.91_{-0.48}^{+0.74}$ & $1.51_{-0.36}^{+0.49}$ & $4.74_{-2.03}^{+4.74}$ & $1.85_{-0.49}^{+0.77}$ & $1.46_{-0.37}^{+0.50}$ \\
HWL16a-003 & 0.553 & $2.45_{-1.85}^{+5.94}$ & $1.10_{-0.54}^{+0.69}$ & $0.86_{-0.42}^{+0.49}$ & $2.76_{-2.00}^{+7.11}$ & $1.08_{-0.53}^{+0.66}$ & $0.85_{-0.41}^{+0.47}$ \\
HWL16a-005 & 0.167 & $9.89_{-7.45}^{+{\rm N/A}}$ & $0.89_{-0.36}^{+0.41}$ & $0.72_{-0.29}^{+0.32}$ & $8.64_{-6.48}^{+{\rm N/A}}$ & $0.90_{-0.37}^{+0.43}$ & $0.73_{-0.30}^{+0.33}$ \\
HWL16a-007 & 0.287 & $0.50_{-0.43}^{+0.70}$ & $3.99_{-1.66}^{+1.96}$ & $1.78_{-0.75}^{+0.81}$ & $0.43_{-0.41}^{+0.68}$ & $3.89_{-2.12}^{+2.16}$ & $1.55_{-0.68}^{+0.96}$ \\
HWL16a-012 & 0.430 & $3.01_{-1.54}^{+3.22}$ & $2.03_{-0.57}^{+1.19}$ & $1.60_{-0.43}^{+0.66}$ & $2.29_{-1.30}^{+2.74}$ & $1.89_{-0.56}^{+1.25}$ & $1.49_{-0.42}^{+0.66}$ \\
HWL16a-013 & 0.264 & $2.14_{-1.15}^{+2.43}$ & $1.11_{-0.39}^{+0.75}$ & $0.86_{-0.29}^{+0.41}$ & $2.46_{-1.34}^{+3.11}$ & $1.15_{-0.39}^{+0.62}$ & $0.90_{-0.29}^{+0.38}$ \\
HWL16a-014 & 0.155 & $2.07_{-1.15}^{+2.41}$ & $1.05_{-0.40}^{+0.69}$ & $0.81_{-0.29}^{+0.40}$ & $2.02_{-1.15}^{+2.46}$ & $1.05_{-0.39}^{+0.67}$ & $0.82_{-0.30}^{+0.40}$ \\
HWL16a-016 & 0.312 & $1.20_{-0.63}^{+1.03}$ & $3.22_{-1.39}^{+1.56}$ & $1.77_{-0.69}^{+0.72}$ & $1.19_{-0.64}^{+1.09}$ & $3.15_{-1.72}^{+1.73}$ & $1.70_{-0.71}^{+0.81}$ \\
HWL16a-017 & 0.500 & $0.98_{-0.88}^{+2.09}$ & $1.00_{-0.51}^{+0.81}$ & $0.76_{-0.38}^{+0.48}$ & $0.75_{-{\rm N/A}}^{+1.84}$ & $0.93_{-0.50}^{+0.79}$ & $0.70_{-0.37}^{+0.48}$ \\
HWL16a-020 & 0.186 & $5.39_{-1.95}^{+3.74}$ & $3.06_{-0.76}^{+1.29}$ & $2.41_{-0.55}^{+0.79}$ & $5.14_{-1.87}^{+3.64}$ & $3.06_{-0.79}^{+1.39}$ & $2.40_{-0.57}^{+0.85}$ \\
HWL16a-022 & 0.276 & $3.42_{-1.71}^{+4.07}$ & $1.38_{-0.42}^{+0.66}$ & $1.09_{-0.33}^{+0.41}$ & $3.06_{-1.55}^{+3.62}$ & $1.35_{-0.44}^{+0.69}$ & $1.05_{-0.32}^{+0.44}$ \\
HWL16a-023 & 0.420 & $5.19_{-3.00}^{+13.79}$ & $1.12_{-0.40}^{+0.50}$ & $0.89_{-0.31}^{+0.36}$ & $5.10_{-2.97}^{+14.04}$ & $1.13_{-0.40}^{+0.50}$ & $0.90_{-0.32}^{+0.36}$ \\
HWL16a-024 & 0.360 & $2.57_{-1.48}^{+3.43}$ & $1.29_{-0.55}^{+0.79}$ & $0.99_{-0.41}^{+0.51}$ & $2.24_{-1.34}^{+3.00}$ & $1.28_{-0.56}^{+0.83}$ & $0.98_{-0.42}^{+0.52}$ \\
HWL16a-026 & 0.424 & $1.20_{-0.63}^{+0.89}$ & $6.20_{-2.07}^{+2.60}$ & $3.34_{-0.98}^{+1.10}$ & $1.21_{-0.63}^{+0.90}$ & $6.04_{-2.09}^{+2.63}$ & $3.27_{-1.00}^{+1.12}$ \\
HWL16a-028 & 0.270 & $2.93_{-1.36}^{+2.59}$ & $1.63_{-0.54}^{+1.13}$ & $1.25_{-0.39}^{+0.60}$ & $2.95_{-1.37}^{+2.66}$ & $1.64_{-0.53}^{+1.13}$ & $1.27_{-0.39}^{+0.60}$ \\
HWL16a-032 & 0.285 & $2.87_{-1.27}^{+2.40}$ & $1.61_{-0.55}^{+1.15}$ & $1.22_{-0.39}^{+0.60}$ & $2.83_{-1.29}^{+2.52}$ & $1.55_{-0.52}^{+1.11}$ & $1.19_{-0.37}^{+0.59}$ \\
HWL16a-034 & 0.315 & $3.22_{-1.08}^{+1.67}$ & $5.69_{-1.62}^{+1.87}$ & $3.94_{-0.95}^{+1.03}$ & $3.10_{-1.06}^{+1.66}$ & $5.61_{-1.68}^{+1.96}$ & $3.87_{-0.99}^{+1.07}$ \\
HWL16a-035 & 0.344 & $0.73_{-0.55}^{+1.02}$ & $0.80_{-0.49}^{+1.24}$ & $0.54_{-0.32}^{+0.46}$ & $0.67_{-0.53}^{+0.96}$ & $0.74_{-0.48}^{+1.33}$ & $0.50_{-0.31}^{+0.48}$ \\
HWL16a-036 & 0.420 & $3.09_{-1.36}^{+2.55}$ & $1.55_{-0.56}^{+1.15}$ & $1.17_{-0.40}^{+0.64}$ & $2.97_{-1.32}^{+2.49}$ & $1.51_{-0.56}^{+1.17}$ & $1.15_{-0.40}^{+0.64}$ \\
HWL16a-037 & 0.697 & $0.32_{-{\rm N/A}}^{+1.35}$ & $2.08_{-1.10}^{+2.58}$ & $1.57_{-0.82}^{+1.22}$ & $0.40_{-{\rm N/A}}^{+1.39}$ & $1.97_{-1.06}^{+2.09}$ & $1.49_{-0.80}^{+1.11}$ \\
HWL16a-038 & 0.463 & $3.25_{-2.68}^{+{\rm N/A}}$ & $1.10_{-0.53}^{+0.62}$ & $0.89_{-0.42}^{+0.46}$ & $3.33_{-2.79}^{+{\rm N/A}}$ & $1.06_{-0.53}^{+0.61}$ & $0.86_{-0.42}^{+0.46}$ \\
HWL16a-039 & 0.310 & $0.95_{-{\rm N/A}}^{+2.51}$ & $0.90_{-0.41}^{+0.70}$ & $0.69_{-0.31}^{+0.40}$ & $1.52_{-{\rm N/A}}^{+3.43}$ & $0.98_{-0.40}^{+0.59}$ & $0.77_{-0.32}^{+0.38}$ \\
HWL16a-041 & 0.194 & $1.69_{-1.11}^{+2.47}$ & $1.24_{-0.44}^{+0.74}$ & $0.97_{-0.33}^{+0.44}$ & $1.74_{-1.20}^{+2.82}$ & $1.19_{-0.43}^{+0.65}$ & $0.93_{-0.32}^{+0.42}$ \\
HWL16a-045 & 0.260 & $0.55_{-0.52}^{+1.30}$ & $0.70_{-0.35}^{+0.51}$ & $0.54_{-0.27}^{+0.32}$ & $0.52_{-{\rm N/A}}^{+1.32}$ & $0.68_{-0.34}^{+0.49}$ & $0.53_{-0.27}^{+0.32}$ \\
HWL16a-046 & 0.135 & $4.88_{-2.49}^{+8.25}$ & $1.31_{-0.37}^{+0.50}$ & $1.05_{-0.29}^{+0.34}$ & $4.41_{-2.30}^{+7.59}$ & $1.29_{-0.37}^{+0.52}$ & $1.03_{-0.29}^{+0.36}$ \\
HWL16a-047 & 0.472 & $9.26_{-4.79}^{+{\rm N/A}}$ & $2.50_{-0.63}^{+0.74}$ & $2.01_{-0.49}^{+0.55}$ & $10.26_{-5.51}^{+{\rm N/A}}$ & $2.53_{-0.63}^{+0.73}$ & $2.04_{-0.50}^{+0.55}$ \\
HWL16a-050 & 0.481 & $0.50_{-0.46}^{+0.79}$ & $1.30_{-1.13}^{+1.60}$ & $0.46_{-0.35}^{+0.67}$ & $0.44_{-{\rm N/A}}^{+0.73}$ & $1.53_{-1.35}^{+1.63}$ & $0.48_{-0.37}^{+0.71}$ \\
HWL16a-051 & 0.254 & $1.67_{-0.78}^{+1.23}$ & $4.40_{-1.43}^{+1.73}$ & $2.66_{-0.73}^{+0.80}$ & $1.43_{-0.72}^{+1.16}$ & $4.28_{-1.54}^{+1.89}$ & $2.50_{-0.77}^{+0.85}$ \\
HWL16a-052 & 0.252 & $0.76_{-0.68}^{+1.83}$ & $0.92_{-0.39}^{+0.60}$ & $0.71_{-0.29}^{+0.37}$ & $0.67_{-0.63}^{+1.63}$ & $0.89_{-0.40}^{+0.59}$ & $0.69_{-0.30}^{+0.37}$ \\
HWL16a-053 & 0.167 & $1.66_{-1.12}^{+2.45}$ & $1.59_{-0.49}^{+1.07}$ & $1.25_{-0.37}^{+0.58}$ & $1.68_{-1.18}^{+2.72}$ & $1.53_{-0.48}^{+0.86}$ & $1.20_{-0.36}^{+0.53}$ \\
HWL16a-056 & 0.470 & $1.72_{-0.96}^{+1.70}$ & $1.92_{-0.90}^{+1.47}$ & $1.28_{-0.52}^{+0.72}$ & $1.71_{-0.95}^{+1.71}$ & $1.75_{-0.81}^{+1.48}$ & $1.20_{-0.50}^{+0.72}$ \\
HWL16a-057 & 0.450 & $1.50_{-0.93}^{+1.70}$ & $1.71_{-0.77}^{+1.93}$ & $1.23_{-0.51}^{+0.84}$ & $1.28_{-0.84}^{+1.57}$ & $1.63_{-0.74}^{+2.08}$ & $1.19_{-0.50}^{+0.84}$ \\
HWL16a-058 & 0.248 & $2.42_{-1.28}^{+2.84}$ & $1.43_{-0.50}^{+0.80}$ & $1.11_{-0.37}^{+0.47}$ & $2.23_{-1.21}^{+2.69}$ & $1.37_{-0.49}^{+0.81}$ & $1.06_{-0.37}^{+0.48}$ \\
HWL16a-059 & 0.561 & $1.76_{-1.33}^{+3.61}$ & $0.96_{-0.67}^{+0.87}$ & $0.72_{-0.52}^{+0.61}$ & $2.24_{-1.60}^{+4.66}$ & $1.05_{-0.68}^{+0.85}$ & $0.80_{-0.53}^{+0.61}$ \\
HWL16a-060 & 0.469 & $0.96_{-0.59}^{+0.91}$ & $4.29_{-1.72}^{+2.06}$ & $2.25_{-0.89}^{+0.95}$ & $0.97_{-0.61}^{+0.96}$ & $4.09_{-1.80}^{+2.10}$ & $2.14_{-0.91}^{+0.98}$ \\
HWL16a-064 & 0.144 & $0.32_{-{\rm N/A}}^{+0.69}$ & $3.03_{-2.23}^{+2.19}$ & $0.99_{-0.41}^{+0.95}$ & $0.26_{-{\rm N/A}}^{+0.71}$ & $1.10_{-0.45}^{+3.58}$ & $0.88_{-0.36}^{+0.77}$ \\
HWL16a-070 & 0.645 & $0.82_{-0.60}^{+0.83}$ & $4.10_{-1.95}^{+2.57}$ & $2.07_{-0.99}^{+1.10}$ & $0.87_{-0.60}^{+0.85}$ & $3.94_{-1.92}^{+2.51}$ & $2.01_{-0.99}^{+1.10}$ \\
HWL16a-071 & 0.534 & $1.10_{-0.82}^{+1.65}$ & $1.43_{-0.65}^{+1.24}$ & $1.05_{-0.46}^{+0.59}$ & $1.08_{-0.79}^{+1.58}$ & $1.47_{-0.66}^{+1.34}$ & $1.07_{-0.46}^{+0.62}$ \\
HWL16a-076 & 0.296 & $4.09_{-1.93}^{+4.52}$ & $1.61_{-0.37}^{+0.69}$ & $1.29_{-0.29}^{+0.44}$ & $4.10_{-1.98}^{+4.95}$ & $1.58_{-0.36}^{+0.63}$ & $1.27_{-0.29}^{+0.40}$ \\
HWL16a-077 & 0.294 & $3.84_{-2.02}^{+5.92}$ & $1.25_{-0.40}^{+0.53}$ & $0.99_{-0.31}^{+0.36}$ & $3.80_{-2.01}^{+5.93}$ & $1.28_{-0.40}^{+0.55}$ & $1.02_{-0.32}^{+0.37}$ \\
HWL16a-080 & 0.312 & $3.04_{-1.39}^{+2.74}$ & $2.32_{-0.69}^{+1.43}$ & $1.80_{-0.49}^{+0.74}$ & $2.99_{-1.41}^{+2.92}$ & $2.23_{-0.63}^{+1.35}$ & $1.75_{-0.46}^{+0.72}$ \\
HWL16a-081 & 0.283 & ${\rm N/A}$ & $1.19_{-0.32}^{+0.37}$ & $0.97_{-0.26}^{+0.28}$ & ${\rm N/A}$ & $1.20_{-0.33}^{+0.36}$ & $0.98_{-0.27}^{+0.28}$ \\
HWL16a-084 & 0.549 & $0.33_{-{\rm N/A}}^{+1.02}$ & $1.15_{-0.70}^{+2.41}$ & $0.78_{-0.45}^{+0.75}$ & $0.31_{-{\rm N/A}}^{+1.04}$ & $1.05_{-0.64}^{+2.24}$ & $0.73_{-0.43}^{+0.71}$ \\
HWL16a-088 & 0.528 & $2.35_{-1.44}^{+3.01}$ & $1.57_{-0.61}^{+1.05}$ & $1.22_{-0.47}^{+0.64}$ & $2.42_{-1.43}^{+3.07}$ & $1.54_{-0.60}^{+1.00}$ & $1.19_{-0.46}^{+0.63}$ \\
HWL16a-090 & 0.295 & $1.98_{-1.18}^{+2.76}$ & $1.25_{-0.43}^{+0.72}$ & $0.98_{-0.33}^{+0.42}$ & $1.90_{-1.17}^{+2.76}$ & $1.25_{-0.44}^{+0.71}$ & $0.98_{-0.33}^{+0.42}$ \\
HWL16a-091 & 0.523 & $1.51_{-0.86}^{+1.54}$ & $3.31_{-1.43}^{+2.78}$ & $2.23_{-0.72}^{+1.16}$ & $1.47_{-0.82}^{+1.44}$ & $3.57_{-1.70}^{+2.66}$ & $2.23_{-0.75}^{+1.18}$ \\
HWL16a-093 & 0.286 & $2.13_{-0.95}^{+1.70}$ & $1.82_{-0.80}^{+1.17}$ & $1.20_{-0.42}^{+0.61}$ & $2.12_{-0.97}^{+1.75}$ & $1.69_{-0.71}^{+1.32}$ & $1.18_{-0.41}^{+0.64}$ \\
HWL16a-094 & 0.592 & $0.95_{-0.57}^{+0.78}$ & $10.90_{-4.25}^{+5.99}$ & $5.59_{-1.76}^{+2.00}$ & $0.85_{-0.56}^{+0.78}$ & $10.51_{-4.19}^{+5.91}$ & $5.26_{-1.71}^{+1.95}$ \\
HWL16a-095 & 0.304 & $3.97_{-1.42}^{+2.49}$ & $3.07_{-1.03}^{+1.30}$ & $2.26_{-0.65}^{+0.76}$ & $3.81_{-1.38}^{+2.41}$ & $3.07_{-1.08}^{+1.37}$ & $2.24_{-0.67}^{+0.81}$ \\
HWL16a-097 & 0.220 & $2.30_{-1.40}^{+4.30}$ & $1.00_{-0.38}^{+0.52}$ & $0.79_{-0.30}^{+0.36}$ & $1.89_{-1.17}^{+3.24}$ & $0.95_{-0.39}^{+0.54}$ & $0.75_{-0.30}^{+0.37}$ \\
HWL16a-098 & 0.395 & $0.38_{-{\rm N/A}}^{+1.03}$ & $1.27_{-0.53}^{+1.89}$ & $0.94_{-0.37}^{+0.56}$ & $0.55_{-{\rm N/A}}^{+1.31}$ & $1.24_{-0.49}^{+1.30}$ & $0.95_{-0.36}^{+0.51}$ \\
HWL16a-0101 & 0.152 & $7.60_{-3.20}^{+8.72}$ & $2.22_{-0.48}^{+0.67}$ & $1.77_{-0.37}^{+0.47}$ & $6.87_{-2.91}^{+7.78}$ & $2.19_{-0.50}^{+0.72}$ & $1.75_{-0.39}^{+0.49}$ \\
HWL16a-0102 & 0.287 & $1.20_{-0.71}^{+1.14}$ & $1.76_{-1.04}^{+1.18}$ & $0.91_{-0.44}^{+0.62}$ & $1.10_{-0.68}^{+1.12}$ & $1.76_{-1.11}^{+1.31}$ & $0.86_{-0.41}^{+0.68}$ \\
HWL16a-0103 & 0.260 & $5.71_{-3.02}^{+11.68}$ & $1.15_{-0.37}^{+0.45}$ & $0.92_{-0.29}^{+0.33}$ & $5.21_{-2.71}^{+9.56}$ & $1.22_{-0.39}^{+0.48}$ & $0.97_{-0.30}^{+0.34}$ \\
HWL16a-0104 & 0.350 & $0.71_{-0.51}^{+0.74}$ & $3.67_{-1.41}^{+1.79}$ & $1.75_{-0.65}^{+0.73}$ & $0.75_{-0.52}^{+0.76}$ & $3.49_{-1.43}^{+1.83}$ & $1.69_{-0.67}^{+0.75}$ \\
HWL16a-0107 & 0.308 & $1.99_{-0.97}^{+1.75}$ & $1.55_{-0.65}^{+1.28}$ & $1.09_{-0.39}^{+0.61}$ & $1.90_{-0.95}^{+1.70}$ & $1.50_{-0.63}^{+1.39}$ & $1.08_{-0.40}^{+0.63}$ \\
HWL16a-0110 & 0.323 & $1.76_{-0.84}^{+1.36}$ & $2.52_{-1.20}^{+1.35}$ & $1.51_{-0.63}^{+0.74}$ & $1.55_{-0.77}^{+1.24}$ & $2.70_{-1.33}^{+1.49}$ & $1.57_{-0.69}^{+0.78}$ \\
HWL16a-0112 & 0.154 & $6.34_{-2.75}^{+7.26}$ & $2.12_{-0.49}^{+0.71}$ & $1.69_{-0.38}^{+0.48}$ & $6.10_{-2.65}^{+6.96}$ & $2.17_{-0.52}^{+0.75}$ & $1.72_{-0.39}^{+0.51}$ \\
HWL16a-0114 & 0.402 & $2.80_{-1.90}^{+5.43}$ & $0.99_{-0.40}^{+0.53}$ & $0.78_{-0.31}^{+0.37}$ & $2.23_{-1.60}^{+4.37}$ & $0.93_{-0.41}^{+0.53}$ & $0.72_{-0.31}^{+0.37}$ \\
HWL16a-0115 & 0.281 & $6.52_{-3.29}^{+12.45}$ & $1.68_{-0.44}^{+0.55}$ & $1.35_{-0.35}^{+0.39}$ & $6.99_{-3.62}^{+15.38}$ & $1.71_{-0.44}^{+0.56}$ & $1.38_{-0.35}^{+0.39}$ \\
HWL16a-0117 & 0.338 & $3.06_{-1.36}^{+2.65}$ & $2.48_{-0.76}^{+1.67}$ & $1.92_{-0.54}^{+0.87}$ & $2.79_{-1.27}^{+2.43}$ & $2.49_{-0.80}^{+1.84}$ & $1.91_{-0.55}^{+0.93}$ \\
\hline
\end{longtable}

Here we present the method and results of cluster mass estimate of the weak lensing peaks
that meet the following two conditions:
\begin{enumerate}
\item Those peaks should be classified as weak lensing secure clusters
(see Section \ref{sec:cross-matching}) to ensure the presence of 
a secure cluster counterpart, and to avoid systems with line-of-sight
projection. The latter is required as our cluster model (described
below) assumes a single dark matter halo.
\item Cluster redshift should be lower than 0.7 to have a sufficient
  number density of background galaxies for the measurement of weak
  lensing shear profile (described below). 
\end{enumerate}
61 weak lensing secure clusters meet those conditions (see Table
\ref{table:clustermass}).

We derive cluster masses by fitting the NFW model
\citep{1997ApJ...490..493N} to measured weak
lensing shear profiles based on the standard likelihood analysis.
We employ the weak lensing mass estimate procedure of
\citet{2020ApJ...890..148U} who 
used the same HSC first-year shear catalog as one used in this study,
allowing us to closely follow their procedure.
Since details of the procedure are described in
\citet[][and see the references therein]{2020ApJ...890..148U}, below we
describe those aspects that are directly relevant to this study.

For each cluster, we select background galaxies using the $P$-cut method
(see Section 3.4 of \citealp{2020ApJ...890..148U}, and see also 
\citealp{2018PASJ...70...30M}) with the cluster 
redshift taken from the estimated redshift of matched CAMIRA-HSC
clusters\footnote{For HWL16a-002, the redshift of matched
  XXL cluster (XLSSC 114) is taken as it is based on spectroscopic
  redshifts \citep{2018A&A...620A...5A}.}, 
and we measure the azimuthally averaged tangential shear
($\gamma_t$) 
which relates to the excess surface mass density $\Delta \Sigma$ as
\citep{1995ApJ...439L...1K}
\begin{equation}
  \label{eq:gamma2Sigma}
  \gamma_t(R)={{\bar{\Sigma}(<R)-\Sigma(R)} \over
    {\Sigma_{cr}(z_{cl},z_s)}}
  \equiv {{\Delta \Sigma(R)} \over
    {\Sigma_{cr}(z_{cl},z_s)}},
\end{equation}
where $\Sigma(R)$ is the azimuthally averaged surface mass density at
$R$, ${\bar{\Sigma}}(<R)$ denotes the average surface mass density
interior to $R$, and $\Sigma_{cr}(z_{cl},z_s)$ is the critical surface
mass density.
We take the peak positions as the cluster centers, and we measure
$\gamma_t(R)$ in 5 radial bins of equal logarithmic spacing of
$\Delta \log R = 0.25$ with bin centers of
$R_c(i)=0.3\times 10^{i\Delta \log R}[h^{-1}$Mpc] where $i$ runs from
0 to 4.
We use the photo-$z$ probability distribution functions (PDFs) of
background galaxies to evaluate $\Sigma_{cr}(z_{cl},z_s)$ following 
\citet[][Section 3.2]{2020ApJ...890..148U}.
The resulting $\Delta \Sigma(R)$ signals are shown in Figure \ref{fig:sufmass_fit}.

We adopt the NFW model to make the 
model prediction of the weak lensing shear profile by a cluster.
The spherical NFW density profile is specified by two parameters, the characteristic
density parameter ($\rho_s$), and the scale radius ($r_s$),
as $\rho_{\rm NFW}(r)=\rho_s/[r/r_s(1+r/r_s)^2]$.
We define the halo mass by the over-density mass ($M_\Delta$) which is
given by integrating the halo density profile out to the corresponding
over-density radius ($r_\Delta$) at which the mean interior density is
$\Delta \times \rho_{cr}(z)$.
The corresponding concentration parameter is defined by $c_\Delta=r_\Delta/r_s$.
For a given set of ($M_\Delta, c_\Delta$), which is of our primary
interest, the NFW parameters ($\rho_s, r_s$) are uniquely determined,
and thus $\Delta \Sigma(R)$ is as well.
Therefore we take ($M_\Delta, c_\Delta$) as fitting parameters in the
likelihood analysis.
We consider two cases, $\Delta=200$ and $\Delta=500$.

We employ the standard likelihood analysis for deriving constraints on
the model parameters.
The log-likelihood is given by,
\begin{equation}
  \label{eq:lnL}
  -2\ln {\cal{L}}(\bm{p})=\sum_{i,j}[d_i-m_i(\bm{p})]
  {\rm Cov}_{ij}^{-1}[d_j-m_j(\bm{p})], 
\end{equation}
where the data vector $d_i=\Delta\Sigma(R_i)$, and $m_i(\bm{p})$ is the
model prediction with the model parameters $\bm{p}=(M_\Delta, c_\Delta)$.
The covariance matrix (${\rm Cov}$) is composed of the three
components \citep[see][and references therein for detailed
  descriptions]{2020ApJ...890..148U}:
The statistical uncertainty due to the galaxy shape noise
(${\rm Cov}^{\rm shape}$), the cosmic shear covariance due to
uncorrelated large-scale structures projected along the line of sight
\citep{2003MNRAS.339.1155H} (${\rm Cov}^{\rm lss}$), and the intrinsic
variation of the cluster lensing signals at the fixed model parameters
due to e.g., cluster asphericity, and
the presence of correlated halos (${\rm Cov}^{\rm int}$) \citep{2015MNRAS.449.4264G,2019ApJ...875...63M}.

We compute the log-likelihood function over the two-parameter
space in the ranges of $0.01<M_\Delta[\times 10^{14}h^{-1}M_\odot]<30$ and
$0.01<c_\Delta<30$, and marginalize it to derive one-parameter posterior
distributions.
Peaks and 68.3\% confidence intervals of marginalized posterior distributions of $c_{200c}$,
$M_{200c}$, and $M_{500c}$ are summarized in Table \ref{table:clustermass}.
We present two sets of results based on the cosmological parameters
from the WMAP 9-year results
\citep{2013ApJS..208...19H}, 
and from the Planck 2018 results 
\citep[][$\Omega_{\rm m}=0.32$, $\Omega_{\rm b}=0.049$, $\Omega_\Lambda=0.68$, $n_s=0.96$,
  $\sigma_8=0.83$, and $h=0.67$]{2018arXiv180706209P}.
The differences in the derived values between two cosmological models
are much smaller than the derived 68.3\% confidence intervals.
Note that those differences are not systematic but rather random.
The reason for this is that since we take the comoving angular
distance, which depends on the cosmology, for binning of $\gamma_t(R)$
measurement, the corresponding angular ranges of bins vary
between two analyses and thus measured signals and errors do as well.
Note that "N/A" in the results of $c_{200c}$ means either the upper/lower bound of
68.3\% confidence interval or the minimum of the marginalized likelihood
function is not enclosed within the parameter range of $c_{200c}$.
This is due to the limited coverage in $R$ with relatively large
error bars.
In Figure \ref{fig:sufmass_fit}, we compare the measured excess
surface mass density profiles with the best-fit NFW 
model (based on WMAP 9-year cosmological model) in the
$M_{200c}$-$c_{200c}$ space, from which
the reader can judge the goodness of fits.

%
%
\section{The locally normalized $SN$ estimator}
\label{sec:local_estimator}

In this study, we have adopted the globally normalized $SN$ estimator defined
by equation~(\ref{eq:sn}) with equations (\ref{eq:shear2kap_sum}) and
(\ref{eq:sigma_shape}). 
In some studies \citep[for
  example,][]{2015PASJ...67...34H}, however,
the peak $SN(\bm{\theta})$ is defined by the locally normalized
estimators, for which $\cal{K}(\bm{\theta})$ and
$\sigma_{\rm shape}^2(\bm{\theta})$
are normalized by the local galaxy number density, $n_g(\bm{\theta})$,
instead of the mean density $\bar{n}_g$.
Here we compare those two estimators using a simple model, and using
the actual weak lensing data.
See \citet{2011ApJ...735..119S} for a related study on those estimators.

Following the same manner as introduced
in Section~\ref{sec:global-kap},
the local estimators are given by, 
\begin{equation}
\label{eq:shear2kap_sum_L}
{\cal K }_L(\bm{\theta}) 
={1\over {n_{fg}+n_{bg}+n_{cl}(\bm{\theta})}}
\sum_{i\in bg} \hat{\gamma}_{t,i} Q_i,
\end{equation}
and
\begin{eqnarray}
\label{eq:sigma_shape_L}
\sigma_{{\rm shape},L}^2(\bm{\theta}) 
&=&{1\over {2 [n_{fg}+n_{bg}+n_{cl}(\bm{\theta})]^2}}\nonumber \\
&&\times \left(
\sum_{i\in bg} \hat{e}_i^2 Q_i^2
+\sum_{i\in fg} \hat{e}_i^2 Q_i^2
+\sum_{i\in cl} \hat{e}_i^2 Q_i^2
\right).
\end{eqnarray}
Notice that contributions from cluster member
population can not be ignored at the cluster central
regions where we are interested in.
Thus we have,
\begin{eqnarray}
\label{eq:sn_L}
SN_L(\bm{\theta})=
{\sqrt{2}{\sum_{bg} \hat{\gamma}_{t,i} Q_i}
  \over
      {
        \left(
        \sum_{bg} \hat{e}_i^2 Q_i^2
        + \sum_{fg} \hat{e}_i^2 Q_i^2
        + \sum_{cl} \hat{e}_i^2 Q_i^2
        \right)^{1/2}}}.
\end{eqnarray}
Therefore, the locally defined $SN$ is affected by the cluster
member population and can be smaller than the globally defined $SN$ [see
equation (\ref{eq:sn_G})], 
though it depends on the local proportion of cluster member galaxies to
background and foreground galaxies.

%
%
\begin{figure}
\begin{center}
  \includegraphics[width=82mm]{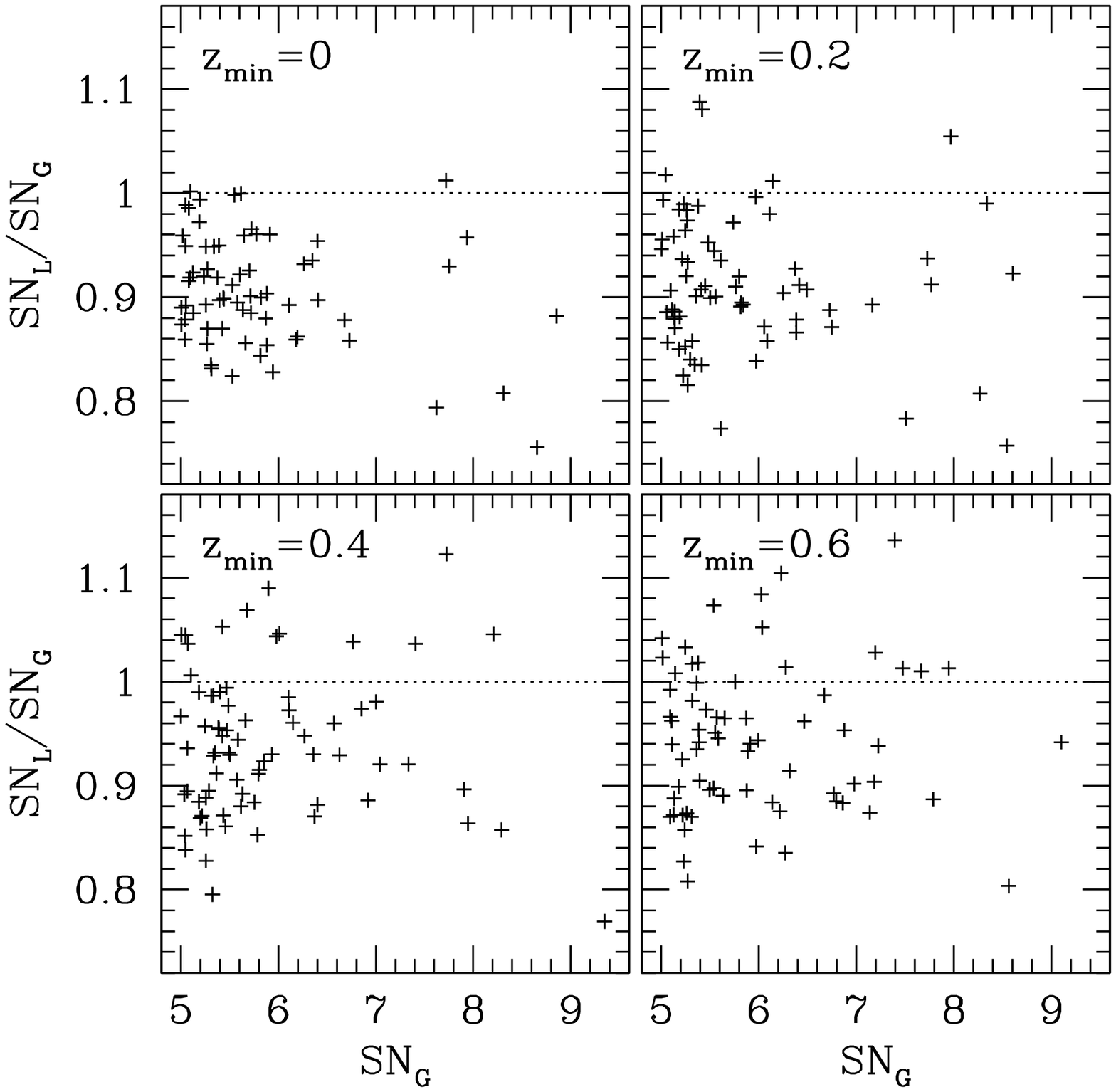}
\end{center}
\caption{Peak $SN_G$ values in the globally normalized $SN$ maps are
  compared with $SN_L$ values at the same positions in the locally
  normalized $SN$ maps. 
  Plus marks are for high peaks ($SN_G\ge 5$) located in the globally
  normalized $SN$ maps.
  Different panels are for different source samples with $z_{\rm min}$
  being shown in each panel.
  \label{gsn2lsn}}
\end{figure}

\begin{figure}
\begin{center}
  \includegraphics[width=82mm]{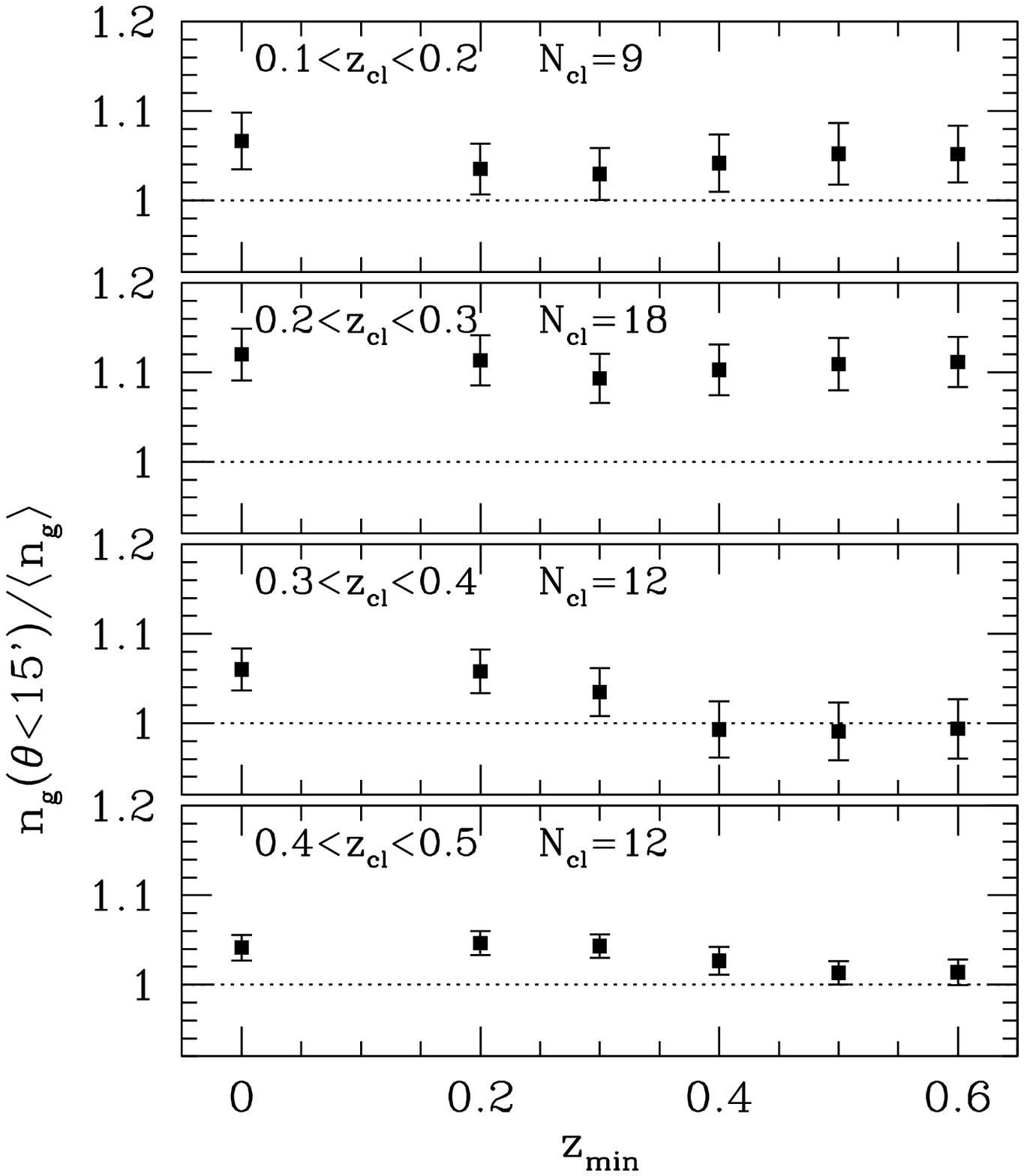}
\end{center}
\caption{Shown is the local galaxy number density at cluster regions,
  which is defined by 
  the mean number density within an angular radius of 15 arcmin from
  peak positions, normalized by the global mean galaxy number density. 
  Weak lensing secure clusters are used, and are divided into four
  sub-samples based on the cluster redshifts (denoted in panels).
  The horizontal axis is $z_{\rm min}$ of source galaxy samples.
  For each sub-sample and each source galaxy sample, the mean and
  its 1-$\sigma$ error among the clusters (the number of clusters in each
  sub-sample is given in each panel) are plotted.
  \label{avgdens_zmim_stats}}
\end{figure}

We examine the actual differences between the globally normalized and the
locally normalized  $SN$ values using our source galaxy samples.
We have generated the locally normalized $SN$ maps for the six source
samples used in this study.
We evaluate locally normalized $SN_L$ values at positions of high peaks
($SN_G\ge 5$) located in the globally normalized $SN$ maps.
This $SN_G$--$SN_L$ comparison is done for six sets of $SN$ maps.
Results are shown in Figure \ref{avgdens_zmim_stats}, in which we find
that $SN_L$ tends to be smaller than $SN_G$, and that this trend is more
clearly seen in lower $z_{\rm min}$ cases as expected.
We find that $SN_L$ is smaller than about 10 percent on average than
$SN_G$ for weak lensing maps used in this study.

%
%

We note that one may take an averaged local shape noise
(that is $\langle \sigma_{{\rm shape},L}^2 \rangle$) to define the $SN$,
instead of the locally defined one.
In this case, deriving its expression using the above manner is not
straightforward, because it is necessary to take into account the
covariance between numerator and denominator in
equation~(\ref{eq:sigma_shape_L}) (see \citealp{2011ApJ...735..119S} for
an approximative approach to this).  
Instead, we evaluate $\langle \sigma_{{\rm shape},L}^2 \rangle$ with
actual weak lensing data used in this study and compare it with 
$\langle \sigma_{{\rm shape},G}^2 \rangle$.
We find that the two are very close;
$\langle \sigma_{{\rm shape},L}^2 \rangle^{1/2}$ is only slightly smaller than
$\langle \sigma_{{\rm shape},G}^2 \rangle^{1/2}$ (to be specific, the
fractional difference is smaller than 0.5 percent).
Therefore, replacing $\sigma_{{\rm shape},L}^2(\bm{\theta})$ with
$\langle \sigma_{{\rm shape},L}^2 \rangle$ does not
mitigate the dilution effect, but an additional $n_{cl}(\bm{\theta})$
term in the normalization of ${\cal K }_L(\bm{\theta})$
suppresses the peak signal [compare
equations~(\ref{eq:shear2kap_sum_G}) with (\ref{eq:shear2kap_sum_L})].
We measure $n_{cl}$ from our data;
in doing this, we have defined the local galaxy number density at cluster
regions by the mean number density within a circular area with an angular
radius of 15 arcmin from peak positions.\footnote{We note that the {\it
    local galaxy number density} is not uniquely defined, because it is
  necessary to define a {\it local scale}, or an {\it averaging scheme}.
  Thus the estimated values given there are not general but are specific
  to our definition of the local galaxy number density.} 
Results are shown in Figure
\ref{avgdens_zmim_stats} for four cluster redshift ranges and six source
samples, in which we find that
for $z_{\rm min}$ from 0 to 0.3, the galaxy 
density excess is 5-10 percent; while for higher $z_{\rm min}$, it is consistent
with zero for higher redshift clusters ($0.3<z_{cl}<0.5$), but the
excess is still 5-10 percent for lower redshift clusters.
It follows from these results that for low-$z_{\rm min}$ source samples, a
peak $SN$ from the globally normalized estimator can be 5-10
percent larger than one from the locally normalized estimator.

We note that the decreasing trend of the number excess at higher
$z_{\rm min }$ seen in higher redshift clusters is expected, as
$z_{\rm  min}$-cut may exclude cluster member galaxies of clusters at 
$z_{cl} < z_{\rm min}$.
However, the trend is not seen in the lower redshift clusters. 
The reason for this is not understood well; possible causes are the
line-of-sight projection of undiscovered clusters at higher redshifts, and
errors in photo-$z$ (cluster member galaxies at a low-$z$
are mis-estimated as higher-$z$ galaxies).
We are not going into this issue in this study but leave it for a future
study. 

\end{document}